\pdfoutput=1
\documentclass[fleqn,usenatbib,usedcolumn]{mnras}
\usepackage{adjustbox}
\usepackage{amsmath} 
\usepackage{amssymb} 
\usepackage{graphics}
\usepackage{graphicx}
\usepackage{grffile}
\usepackage{epsfig}
\usepackage{epstopdf}
\usepackage{times}
\usepackage[T1]{fontenc} 
\usepackage{aecompl}
\usepackage{grffile}
\usepackage{subfig}
\usepackage{multirow}
\usepackage{array}
\usepackage{tikz}
\usepackage{times}
\usepackage{tabularx}
\usepackage{adjustbox}
\usepackage{blindtext}
\usetikzlibrary{positioning}
\usepackage{marvosym}
\usepackage{hyperref}
\usepackage{float}
\usepackage{pdflscape}
\usepackage{afterpage}
\usepackage{rotating}
\usepackage{cite}
\usepackage{color}
\usepackage{longtable}

\hypersetup{pdfauthor={J. Mejia-Restrepo},
            pdftitle={lack Hole Mass  estimation of quasars by means of broad emission lines},
            pdfkeywords={galaxies: active  quasars:general quasars:supermassive black holes quasars: emission lines},
            bookmarksnumbered=true}
\volume{{\rm in press}}

\def\be{\begin{equation}}
\def\ee{\end{equation}}
\def\ba{\begin{eqnarray}}
\def\ea{\end{eqnarray}}

\newcommand{\todo}{\ifmmode \text{\Huge{\(\bullet\)}} \else {\Huge$\bullet$}\fi}
\newcommand{\tido}{\ifmmode {\bullet} \else $\bullet$\fi}

\newcommand{\Ntot}{39}
\newcommand{\Nbright}{30}

\newcommand{\ergs}	{\ifmmode {\text{erg\,s}}^{-1} \else erg s$^{-1}$\fi}
\newcommand{\kms}	{\ifmmode {\text{km\,s} }^{-1} \else km\,s$^{-1}$\fi}

\newcommand{\LCDM}{$\Lambda$CDM~}
\newcommand{\beq}{\begin{eqnarray}}  
\newcommand{\eeq}{\end{eqnarray}}

\newcommand{\ly}{{\ifmmode{{\text{Ly} }\alpha}\else{Ly$\alpha$}\fi}}

\newcommand{\hMpc  }{{\ifmmode{h^{-1}{\text{Mpc} }}\else{$h^{-1}$Mpc }\fi}}  
\newcommand{\hGpc  }{{\ifmmode{h^{-1}{\text{Gpc} }}\else{$h^{-1}$Gpc }\fi}}  
\newcommand{\hmpc  }{{\ifmmode{h^{-1}{\text{Mpc}  }}\else{$h^{-1}$Mpc }\fi}}  
\newcommand{\hkpc  }{{\ifmmode{h^{-1}{\text{kpc}  }}\else{$h^{-1}$kpc }\fi}}  
\newcommand{\hMsun }{{\ifmmode{h^{-1}{\text{M_{\rm \odot}}}}\else{$h^{-1}{\text{M_{\rm \odot}}}$}\fi}}   
\newcommand{\hmsun }{{\ifmmode{h^{-1}{\text{M_{\odot}}}}\else{$h^{-1}{\text{M_{\odot}}}$}\fi}}   
\newcommand{\Msun  }{\ifmmode M_{\rm \odot} \else $M_{\rm \odot}$\fi}  
\newcommand{\msun  }{\ifmmode M_{\rm \odot} \else $M_{\rm \odot}$\fi}

\newcommand{\rand  }{\ifmmode{{\mathcal{R}}}\else{${\mathcal{R}}$}\fi}

\newcommand{\Mbh   }{\ifmmode M_{\text{BH} } \else $M_{\text{BH} }$\fi}
\newcommand{\mbh   }{\ifmmode M_{\text{BH} } \else $M_{\text{BH} }$\fi}
\newcommand{\Mdot  }{\ifmmode \dot{M} \else $\dot{M}$\fi}
\newcommand{\LLedd }{\ifmmode L/L_{\text{Edd} } \else $L/L_{\text{Edd} }$\fi}
\newcommand{\lledd }{\ifmmode L/L_{\text{Edd} } \else $L/L_{\text{Edd} }$\fi}

\newcommand{\astar }{\ifmmode a_{*} \else  $a_{*}$\fi}
\newcommand{\RBLR  }{\ifmmode R_{\text{BLR} } \else $R_{\text{BLR}}$\fi}
\newcommand\abs[1]{\left|#1\right|}

\newcommand*{\rom}[1]{\expandafter\@slowromancap\romannumeral #1@}
\def\spose#1{\hbox to 0pt{#1\hss}}
\def\simlt{\mathrel{\spose{\lower 3pt\hbox{$\mathchar"218$}}
     \raise 2.0pt\hbox{$\mathchar"13C$}}}
\def\simgt{\mathrel{\spose{\lower 3pt\hbox{$\mathchar"218$}}
     \raise 2.0pt\hbox{$\mathchar"13E$}}}

\newcommand{  \Halpha   }{\ifmmode {\text{H} }\alpha \else H$\alpha$\fi}
\newcommand{  \ha   	}{\ifmmode {\text{H}}\alpha \else H$\alpha$\fi}
\newcommand{  \Hbeta    }{\ifmmode {\text{H} }\beta \else H$\beta$\fi}
\newcommand{  \hb    	}{\ifmmode {\text{H} }\beta \else H$\beta$\fi}
\newcommand{  \mgii     }{\ifmmode {\text{Mg} }\,\textsc{ii} \else Mg\,\textsc{ii}\fi}
\newcommand{  \MgII    }{\ifmmode {\text{Mg} }\,\textsc{ii}\,\lambda2798 \else Mg\,\textsc{ii}\,$\lambda2798$\fi}
\newcommand{  \HeII    }{\ifmmode {\text{He} }\,\textsc{ii}\,\lambda1640 \else He\,\textsc{ii}\,$\lambda1640$\fi}
\newcommand{  \NIII     }{\ifmmode {\text{N} }\,\textsc{iii}]\,\lambda1750 \else N\,\textsc{iii}]\,$\lambda1750$\fi}
\newcommand{  \NIV     }{\ifmmode {\text{N} }\,\textsc{iv}\,\lambda1718 \else N\,\textsc{iv}\,$\lambda1718$\fi}
\newcommand{  \NIVo     }{\ifmmode {\text{N} }\,\textsc{iv}]\,\lambda1486 \else N\,\textsc{iv}]\,$\lambda1486$\fi}
\newcommand{  \OIII     }{\ifmmode {\text{O} }\,\textsc{iii}]\,\lambda1663 \else O\,\textsc{iii}]\,$\lambda1663$\fi}
\newcommand{  \oiii     }{\ifmmode {\text{O} }\,\textsc{iii}]\,\lambda5007 \else O\,\textsc{iii}]\,$\lambda5007$\fi}
\newcommand{  \civ      }{\ifmmode {\text{C} }\,\textsc{iv}  \else C\,\textsc{iv}\fi}
\newcommand{  \ciii      }{\ifmmode {\text{C} }\,\textsc{iii}]  \else C\,\textsc{iii}]\fi}
\newcommand{  \CIV      }{\ifmmode {\text{C} }\,\textsc{iv}\,\lambda1549 \else C\,\textsc{iv}\,$\lambda1549$\fi}
\newcommand{  \CIII      }{\ifmmode {\text{C} }\,\textsc{iii}]\,\lambda1909 \else C\,\textsc{iii}]\,$\lambda1909$\fi}
\newcommand{  \SiOIV      }{\ifmmode {\text{Si} }\,\textsc{IV}+{\text{O} }\,\textsc{IV}]\, \lambda1400 \else Si\,\textsc{iv}+O\,\textsc{iv}]  \,$\lambda1400$\fi}
\newcommand{  \sioiv      }{\ifmmode {\text{Si} }\,\textsc{IV}+{\text{O} }\,\textsc{IV}]\ \else Si\,\textsc{iv}+O\,\textsc{iv}]\fi}
\newcommand{  \feii      }{\ifmmode {\text{Fe} }\,\textsc{ii}  \else Fe\,\textsc{ii}\fi}

\newcommand{  \FeII      }{\ifmmode {\text{Fe} }\,\textsc{ii}  \else Fe\,\textsc{ii}\fi}

\newcommand{  \FeIII      }{\ifmmode {\text{Fe} }\,\textsc{iii}  \else Fe\,\textsc{iii}\fi}
\newcommand{ \Lhb   }{\ifmmode L\left(\hb\right) \else $L\left(\hb\right)$\fi}

\newcommand{ \fwhb  }{\ifmmode {\text{FWHM} }\left(\hb\right) \else FWHM(\hb)\fi}
\newcommand{ \fwhbloc  }{\ifmmode {\text{FWHM} }\left(\hb\right)_{\text{local} } \else FWHM(\hb)_{\text{local} } \fi}
\newcommand{ \fwhbglob  }{\ifmmode {\text{FWHM} }\left(\hb\right)_{\text{global} } \else FWHM(\hb)_{\text{global} } \fi}

\newcommand{ \Lha   }{\ifmmode L\left(\ha\right) \else $L\left(\ha\right)$\fi}
\newcommand{ \fwha  }{\ifmmode {\text{FWHM} }\left(\ha\right) \else FWHM(\ha)\fi}

\newcommand{ \Lmg   }{\ifmmode L\left(\mgii\right) \else $L\left(\mgii\right)$\fi}
\newcommand{ \fwmg  }{\ifmmode {\text{FWHM} }\left(\mgii\right) \else FWHM(\mgii)\fi}
\newcommand{ \Lciv  }{\ifmmode L\left(\civ\right) \else $L\left(\civ\right)$\fi}
\newcommand{ \fwciv }{\ifmmode {\text{FWHM} }\left(\civ\right) \else FWHM(\civ)\fi}
\newcommand{ \fwhm  }{\ifmmode {\text{FWHM} } \else \text{FWHM}\fi} 
\newcommand{ \voff  }{\ifmmode v_{\text{off} } \else $v_{\text{off} }$\fi} 

\newcommand{ \sigline  }{\ifmmode \sigma_{\text{line}} \else $\sigma_{\text{line}}$\fi}

\newcommand{ \sigmamg  }{\ifmmode {\sigma }\left(\mgii\right) \else $\sigma$(\mgii)\fi}
\newcommand{ \sigmaciv  }{\ifmmode {\sigma }\left(\civ\right) \else $\sigma$(\civ)\fi}
\newcommand{ \sigmahb  }{\ifmmode {\sigma }\left(\Hbeta\right) \else $\sigma$(\Hbeta)\fi}
\newcommand{ \sigmaha  }{\ifmmode {\sigma }\left(\Halpha\right) \else $\sigma$(\Halpha)\fi}

\newcommand{\MbhHb   }{\ifmmode M_{\text{BH}  } \left( \Hbeta \right) \else $M_{\text{BH} } \left( \Hbeta \right)$\fi}
\newcommand{\MbhHa   }{\ifmmode M_{\text{BH} }  \left( \Halpha \right) \else $M_{\text{BH} } \left( \Halpha \right)$\fi}%
\newcommand{\MbhMg   }{\ifmmode M_{\text{BH}  } \left( \mgii \right) \else $M_{\text{BH} } \left( \mgii \right)$\fi}
\newcommand{\MbhC   }{\ifmmode M_{\text{BH}  } \left( \civ \right) \else $M_{\text{BH} } \left( \civ \right)$\fi}

\newcommand{\Mbhfw   }{\ifmmode M_{\text{BH}}\left(\text{FWHM}\right)    \else $M_{\text{BH}}\left(\text{FWHM}\right)$\fi}
\newcommand{\Mbhsig   }{\ifmmode M_{\text{BH}}\left(\sigma_{\text{line}}\right)    \else $M_{\text{BH}}\left(\sigma_{\text{line}}\right)$\fi}

\newcommand{\local}{ \textit{local\ } }
\newcommand{\glob}{ \textit{global\ } }
\newcommand{\globapp}{ \glob approach\ }
\newcommand{\localapp}{ \local approach\ }

\newcommand{\fw}{\ifmmode \fwhm_{\text{local}} \else $\fwhm_{\text{local}}$\fi}
\newcommand{\fwtdc   }{\ifmmode \fwhm_{\text{global}} \else $\fwhm_{\text{global}}$\fi}
\newcommand{\Llocal   }{\ifmmode L_{\text{local}} \else $L_{\text{local}}$\fi}
\newcommand{\Ltdc   }{\ifmmode L_{\text{global}} \else $L_{\text{global}}$\fi}

\newcommand{ \mumg  }{\ifmmode \mu\left(\mgii\right) \else $\mu\left(\mgii\right)$\fi}
\newcommand{ \fmg   }{\ifmmode f\left(\mgii\right) \else $f\left(\mgii\right)$\fi}
\newcommand{ \muciv }{\ifmmode \mu\left(\civ\right) \else $\mu\left(\civ\right)$\fi}
\newcommand{ \fciv  }{\ifmmode f\left(\civ\right) \else $f\left(\civ\right)$\fi}

\newcommand{  \Luv      }{\ifmmode L_{1450} \else $L_{1450}$\fi}
\newcommand{  \Lop      }{\ifmmode L_{5100} \else $L_{5100}$\fi}
\newcommand{  \Loploc      }{\ifmmode L_{5100}^{\text{local}} \else $L_{5100}^{\text{local}}$\fi}
\newcommand{  \Lopglob      }{\ifmmode L_{5100}^{\text{global}} \else $L_{5100}^{\text{global}}$\fi}
\newcommand{  \Lsix      }{\ifmmode L_{6200} \else $L_{6200}$\fi}
\newcommand{  \Lthree   }{\ifmmode L_{3000} \else $L_{3000}$\fi}
\begin{document}
\title[Black Hole Mass estimation in quasars]
{Active galactic nuclei  at \boldmath$z\sim 1.5$: \textsc{II}.\ Black Hole Mass  estimation by means of broad emission lines}

\author[ J. E. Mejia-Restrepo et al]{J. E. Mejia-Restrepo,$^{1}$\thanks{Email: jemejia@das.uchile.cl}, B. Trakhtenbrot,$^{2}$ \thanks{Zwicky postdoctoral fellow} P. Lira$^{1}$, H. Netzer,$^{3}$ D. M. Capellupo$^{3,4}$ \vspace*{6pt}\\ $^{1}$ Departamento de Astronom\'{i}a, Universidad de Chile, Camino el Observatorio 1515, Santiago, Chile \\ $^{2}$ Institute for Astronomy, Dept. of Physics, ETH Zurich, Wolfgang-Pauli-Strasse 27, CH-8093 Zurich, Switzerland\\ $^{3}$ School of Physics and Astronomy, Tel Aviv University, Tel Aviv 69978, Israel \\$^{4}$ Department of Physics, McGill University, Montreal, Quebec, H3A 2T8, Canada} 

\maketitle

\begin{abstract}
This is the second in a series of papers aiming to test how the mass (\Mbh), accretion rate (\Mdot) and spin (\astar) of super massive black holes (SMBHs) determine the observed properties of type-I active galactic nuclei (AGN). 
Our project utilizes a sample of 39 unobscured AGN at $z\simeq1.55$ observed by VLT/X-shooter, selected to map a large range in \Mbh\ and \LLedd\ and covers the most prominent UV-optical (broad) emission lines, including \Halpha, \Hbeta, \MgII, and \CIV. This paper focuses on single-epoch, ``virial''  \Mbh\ determinations  from broad emission lines and examines the implications of different continuum modeling approaches in line width measurements.
We find that using a \local power-law continuum instead of a physically-motivated thin disk continuum leads to only slight underestimation of the \fwhm\ of the lines and the associated \Mbhfw. However, the line dispersion \sigline\ and associated \Mbhsig\ are strongly affected by the continuum placement and provides less reliable mass estimates than FWHM-based methods. Our analysis shows that \Halpha, \Hbeta\ and \mgii\ can be  safely used for virial \Mbh\ estimation. The \civ\ line, on the other hand, is not reliable in the majority of the cases, this may indicate that the gas emitting this line is not virialized. While \Halpha\ and \Hbeta\ show very similar line widths, the mean \fwmg\ is about 30\% narrower than \fwhb. We confirm several recent suggestions to improve the accuracy in \civ-based mass estimates, relying on other UV emission lines. Such improvements do not reduce the scatter between \civ-based and Balmer-line-based mass estimates.
\end{abstract}


\section{Introduction}
\label{sec:intro}

The mass (\Mbh) of Super Massive Black Holes (SMBHs), along with the SMBH spin (\astar) and accretion rate (\Mdot), are the fundamental parameters that drive the physical, geometric and kinematic properties of the SMBH  environment \citep[e.g.][]{Kaspi2005, SloneNetzer2012, Capellupo2015}. \Mbh\ is also known to be correlated with several properties of the host galaxy, suggesting a so-called ``co-evolutionary'' scenario for the SMBH and stellar component of the host \citep[e.g.][]{Ferrarese2000,HaringRix2004,Gultekin2009,Xiao2011}.  
Therefore, accurate and precise determinations of \Mbh, across cosmic epochs, are crucial for our understanding
 of SMBH physics and evolution.

For un-obscured, type-I actively growing SMBHs (active galactic nuclei - AGN), \Mbh\ can be estimated from single epoch spectra of several broad emission lines.
{ 
The method, which was used for many large samples of AGN across cosmic epochs \citep[e.g.,][]{Croom2004,MclureDunlop2004,Onken2004,Fine2006,Shen2008,RafieeHall2011,TrakhtenbrotNetzer2012}, is based on a combination of two basic ingredients \citep{Vestergaard2002,Peterson2004}.
First, reverberation mapping (RM) experiments provide an empirical relation between the BLR size and the AGN continuum luminosity ($\RBLR  = K' 
(\lambda L_{\lambda} )^{\alpha} $, with $\alpha\sim0.5-0.7$; see \citealt{Kaspi2000,Kaspi2005,Bentz2009,Bentz2013}, and references therein). 
Second, the gas in the broad line region  (BLR) is assumed to be virialized \citep[as suggested by several empirical studies, e.g.,][]{PetersonWandel1999,Onken2004} .
}
After taking the line width of the BLR lines as a natural estimation of the virial velocity of the gas in the BLR ($V_\text{BLR}$), one may obtain the mass from the virial relation:
 \begin{equation}
 \Mbh =f G^{ - 1} R_{\text{BLR} } V_{\text{BLR}}^{2} = K (\lambda L_{\lambda} )^{\alpha}\fwhm^{2}
  \label{eqn:virial_eqn}
\end{equation}
where $K=K'G^{ - 1}f$ and $f$ is a general geometrical function which correct for the unknown structure and inclination to the line of sight. $f$ can be determined experimentally by requiring RM-\Mbh\ estimations to be consistent, on average, with those predicted from the \Mbh-bulge stellar velocity dispersion (\Mbh-$\sigma_{*}$) relation of local  galaxies where \Mbh\ have been dynamically estimated  \citep[e.g.][]{Onken2004,Woo2010,Graham2011, Graham2015, Woo2015}.  In this paper,  we  assume $f=1$, which is appropriate for the \fwhm\ \MbhHb\ estimates \citep{Woo2015}. However, in addition to the  still  large uncentainty in this value (50\%), $f$ can also be different for different  lines and could even depend on luminosity and/or  line properties \citep[e.g. equivalent widths, line offsets \fwhm][]{Shen2013}.

Among the RM-based $\RBLR-L$ relations, the most reliable one is the $\RBLR\left(\Hbeta\right)-\Lop$ relation, which is the only one based on a large number of sources, with $\Lop\lesssim 10^{46}\,\ergs$.  
Thus, the \Mbh\ determination based on other lines and luminosities at other wavelengths needs to be re-calibrated to match \Mbh\ measurements based on \Hbeta\ and \Lop. 
Particularly, \CIV, hereafter \civ, \citep[e.g.][]{VestergaardPeterson2006,Park2013}, \MgII, hereafter \mgii, \citep[e.g.][]{MclureJarvis2002,VestergaardOsmer2009,Wang2009,Trakhtenbrot2011,ShenLiu2012,TrakhtenbrotNetzer2012} and \Halpha\ \citep[e.g.][]{GreeneHo2005,Xiao2011,ShenLiu2012} have been re-calibrated accordingly, and are widely used lines to determine \Mbh\ at high redshifts.  

Earlier \Mbh\ recalibrations based on \mgii\ and \Halpha\ have showed good agreement and low scatter with \Hbeta-based \Mbh\ calibration \citep{GreeneHo2005,Xiao2011,TrakhtenbrotNetzer2012}. 
{ 
However, \Mbh\ recalibrations using the \civ\ line  are more problematic, compared with those based on lower-ionization lines. 
First, the correlation between the widths of \civ\ and the other lines was shown to be weak, or indeed insignificant, and to present a large scatter, in many AGN samples \citep[e.g.,][]{BaskinLaor2005,Netzer2007,Shang2007,Shen2008,Fine2010,Ho2012,ShenLiu2012,Tilton2013}.
Moreover, about 40\% of the objects have $\fwciv\lesssim\fwhb$, in contrast to the expectations from RM experiments and the virial assumption, that suggest $\fwciv\simeq 2\times \fwhb$ (see detailed discussion in TN12, and additional samples in \citet{Ho2012,ShenLiu2012,Tilton2013}).  
Second, significant blueshifts of the \emph{entire} \civ\ profile (i.e., not necessarily a specific sub-component of the line), reaching several 1000s \kms, are ubiquitously measured in the vast majority of AGN \citep{Richards2002,BaskinLaor2005,Shang2007,Richards2011,TrakhtenbrotNetzer2012}. 
Some of these findings were explained either by a disc outflow  wind \citep[e.g.][]{Gaskell1982,Sulentic2007,Richards2011} or, alternatively, by scattering off an in-falling medium in the innermost \civ-emitting regions, which would produce the \civ\ blueshifts \citep[e.g.][]{KallmanKrolik1986,GoosmannGaskell2007,Gaskell2009,GaskellGoosmann2013}.
Finally, the detailed re-analysis of the RM data for \civ\ performed by \citet{Denney2012} found that the (narrowest) core of the broad \civ\ line does not reverberate in response to continuum variability. 
This implies that the \emph{outermost} \civ\ emitting regions may not be virialized, either. 
All this leads to the conclusion that the simplified models and prescriptions discussed above may be incorrect, or at least incomplete, for some lines.
}

The \Mbh\ determination is also subjected to several uncertainties, related to the limitations of spectral analysis, and/or the need to make several assumptions regarding the universality of some AGN properties. 
The former includes 
the blending  of neighboring emission and/or absorption features; 
incorrect determination of the continuum emission \citep[][hereafter S07]{Shang2007};
poor statistics due to non-homogeneous or small nature of the sample under study  \citep[e.g.][]{Ho2012}; 
poor data quality \citep[e.g.,][]{Denney2013,Tilton2013};
and measurements obtained from non-simultaneous data \citep[see e.g.][]{ShenLiu2012,Marziani2013}. 
The latter, somewhat more fundamental uncertainties, include 
non virial gas motion;
{ 
the orientation of the (generally non-spherical) BLR with respect to the line-of-sight \citep{Runnoe2014,ShenHo2014,BrothertonSinghRunnoe2015}};
and the extrapolation of the $\RBLR-L$ relations to luminosities which are well beyond the range probed by RM experiments.

There have been many efforts to improve single-epoch \Mbh\ determinations, addressing some of the aforementioned  limitations \citep[e.g.][]{GreeneHo2005,VestergaardPeterson2006,Fine2008,Wang2009,Fine2010,Xiao2011,ShenLiu2012,TrakhtenbrotNetzer2012,Marziani2013, Park2013,Runnoe2013, Brotherton2015, Zuo2015}.   \citet[][hereafter TN12]{TrakhtenbrotNetzer2012} combined Sloan digital sky survey  archival data \citep[SDSS;][]{Abazajian2009} with smaller surveys and samples to improve earlier \mgii-based \Mbh\ prescriptions  \citep[e.g.,][]{MclureJarvis2002,MclureDunlop2004,Wang2009}, by assuming virialization of the \mgii\ emitting clouds. 
As mentioned above, the TN12 study emphasized the fact that a large fraction of AGN show $\fwciv<\fwhb$.  
\citet{Marziani2013} (hereafter M13) also used SDSS data to perform an Eigenvector 1 analysis \citep{BorosonGreen1992}, and to separate the population into ``population A'' ($\fwhb<4000\,\kms$) and ``population B'' ($\fwhb>4000\,\kms$) sources. 
They suggested that \Hbeta- and \mgii-based \Mbh\ estimates in population B sources could be systematically overestimated due to a red-shifted, extremely broad emission component. 
The study of\citet{ShenLiu2012} combined SDSS optical observations of high-z objects (1.5$\lesssim$z$\lesssim$2.2) with follow up  FIRE-IR observations, which allowed them to compare and recalibrate the \civ, \mgii, \Hbeta\ and \Halpha\ \Mbh\ relations as well as contrast them with previous calibrations. While they found that   \fwmg\ correlates well with the Balmer lines, the \fwciv\ does not show such correlations and is not a reliable viral mass estimator. 
The \citet{ShenLiu2012} results are however subjected to  low quality SDSS data, non homogeneous sample selection and non simultaneous observations. \citet{Ho2012} obtained simultaneous UV, optical and infrared X-Shooter spectra for 7 objects at $1.3\lesssim z \lesssim 1.6$, resulting in similar conclusions regarding the usability of \mgii-based \Mbh\ estimates, and the limitations associated with \civ.

The studies of \citep[][hereafter D13]{Denney2013} and \citet{Tilton2013} claimed that in spectra of limited S/N and/or spectral resolution, \fwciv\ measurements are underestimating the ``real'' line widths, in objects with strong intrinsic absorption features that cannot be deblended from the emission lines. 
This would partially explains the TN12 finding that about 40\% of the objects shows \fwciv$<$\fwhb. 
{ 
However, objects with no evidence of absorption features, and yet ``intrinsic'' line widths with \fwciv$<$\fwhb\ are known to exist \citep[e.g.,][]{CorbinBoroson1996}.}
After correcting for intrinsic \civ\ absorption, D13 claimed  that although \fwciv\ still does not correlate well with \fwhb, \sigmaciv\ shows a strong correlation with \sigmahb\ and can safely be used for \civ\ based \Mbh\ determinations.  Based on these results, \citep{Park2013} obtained high quality data in 39 out of 45 objects of the RM experiments campaign  and  improved the \citet{VestergaardPeterson2006}  \civ-based \Mbh\ estimator based on the \sigmaciv. Both D13 and \citep{Park2013} used non homogeneous and multi-epoch samples  that could affect their results. In addition,  \sigline\ measurements are highly dependent on the continuum determination method (see discussion in \citep{Peterson2004}).

{  Recently, \citet{Runnoe2013} (hereafter R13) and \citet{Brotherton2015} used a sample of 85  low-redshift ($0.03<z<1.4$) and low-luminosity ($43.37<\log{\Lop}<46.45$) AGN with quasi-simultaneous UV and optical spectra to propose a method to rehabilitate \civ\ for \Mbh\ determination, based on a correlation that they found between the   \sioiv$-$\civ\  line peak intensity ratio and the \Hbeta$-$\civ\ \fwhm\ ratio.  
This allowed these authors to predict \fwhb\ from measurements of the \sioiv\ emission. These studies suggested that this correlation may be driven by the so-called Eigenvector 1.}

 In this work, we use  X-shooter high-quality observations that  combines simultaneous UV, optical and infrared spectroscopy of  a unique sample of AGN at z$\sim$1.55, selected
  by both their \Mbh\ and Eddington ratio, \LLedd\ as described in \citet{Capellupo2015}  
(hereafter paper I).  Selecting objects at this redshift allows simultaneous observations of  \Halpha, \Hbeta, \mgii\  and \civ\ which is optimal for comparing the various mass determination methods. In Paper I, we showed   that the accretion-disk continuum of most of the objects (25 out of 30)  can be successfully modeled by a geometrically thin, optically thick Shakura-Sunyaev  accretion disks \citep[][hereafter SS73]{ShakuraSunyaev1973} . The  models were taken from \citet{SloneNetzer2012}  who include several improvements upon the SS73 model, such as GR effects and a detailed treatment of the Comptonization in the disc atmosphere.	Paper I shows that most earlier attempts to fit accretion disk (AD) spectra to AGNs failed because of the limited wavelength coverage and/or non-simultaneous observations. The continuation of this work, that includes 9 more sources and a more comprehensive analysis, is described in Capellupo et al. (2016, in prep.), hereafter Paper III) which is published in this volume.

 The purpose of the present paper is to evaluate BH mass measurements based on different emission lines, as derived from our unique sample of X-shooter spectra. We also aim to provide to the community \Mbh\ correction factors that do not depend on the exact shape of the underlying continuum. The paper is structured as follows. In  section \ref{sec:data} we describe the sample. In section \ref{sec:fit} we first  introduce  the \local and  \glob thin disk continuum approaches  and describe the fitting procedures we follow to model the continuum, emission lines, iron pseudo continuum and Balmer continuum.  In section \ref{sec:results} we present and discuss  the main  results and in section \ref{sec:conclusions} we list the main  conclusions of our work.
Throughout this paper we assume a flat \LCDM\ cosmology with the following values for the cosmological parameters: $\Omega_{\Lambda}= 0.7$, $\Omega_{\text{M} }= 0.3$ and $H_{0}=70\,\kms\,\rm{Mpc}^{-1}$.

\begin{figure}
\centering
\includegraphics[scale=0.42]{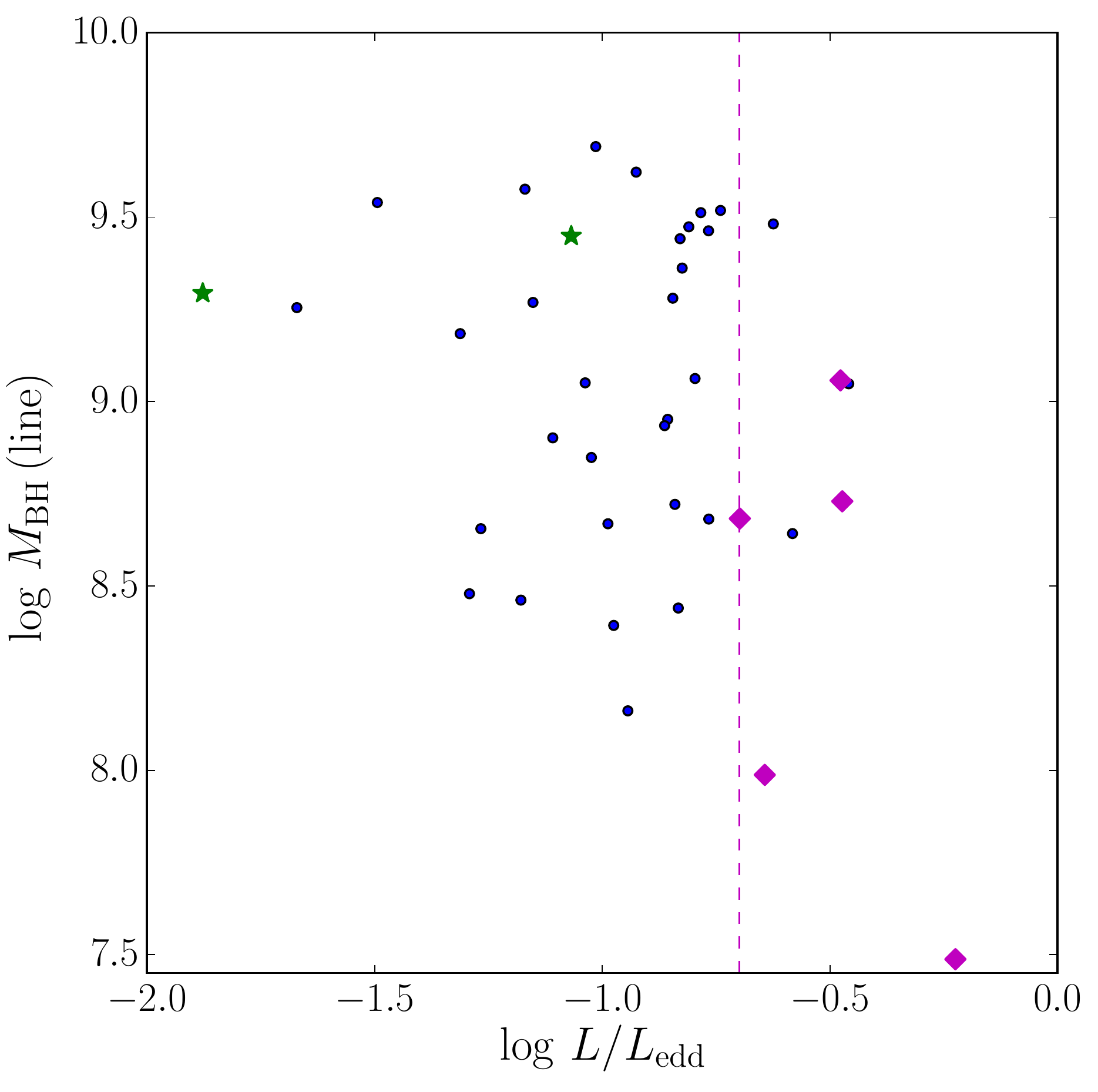}

\caption{{  \Mbh\ vs \LLedd\ using the values we obtained in this paper. Green stars and magenta diamonds represent the broad absorption line quasars (BALQSO)  and the broad-\mgii\ respectively (as defined in \S \ref{subsec:broadmg}) . The magenta dashed vertical line  represents \LLedd$=$0.2.}}
\label{fig:MLLedd}
\end{figure}

\begin{figure*}
\centering
\includegraphics[width=0.49\textwidth,angle=0]{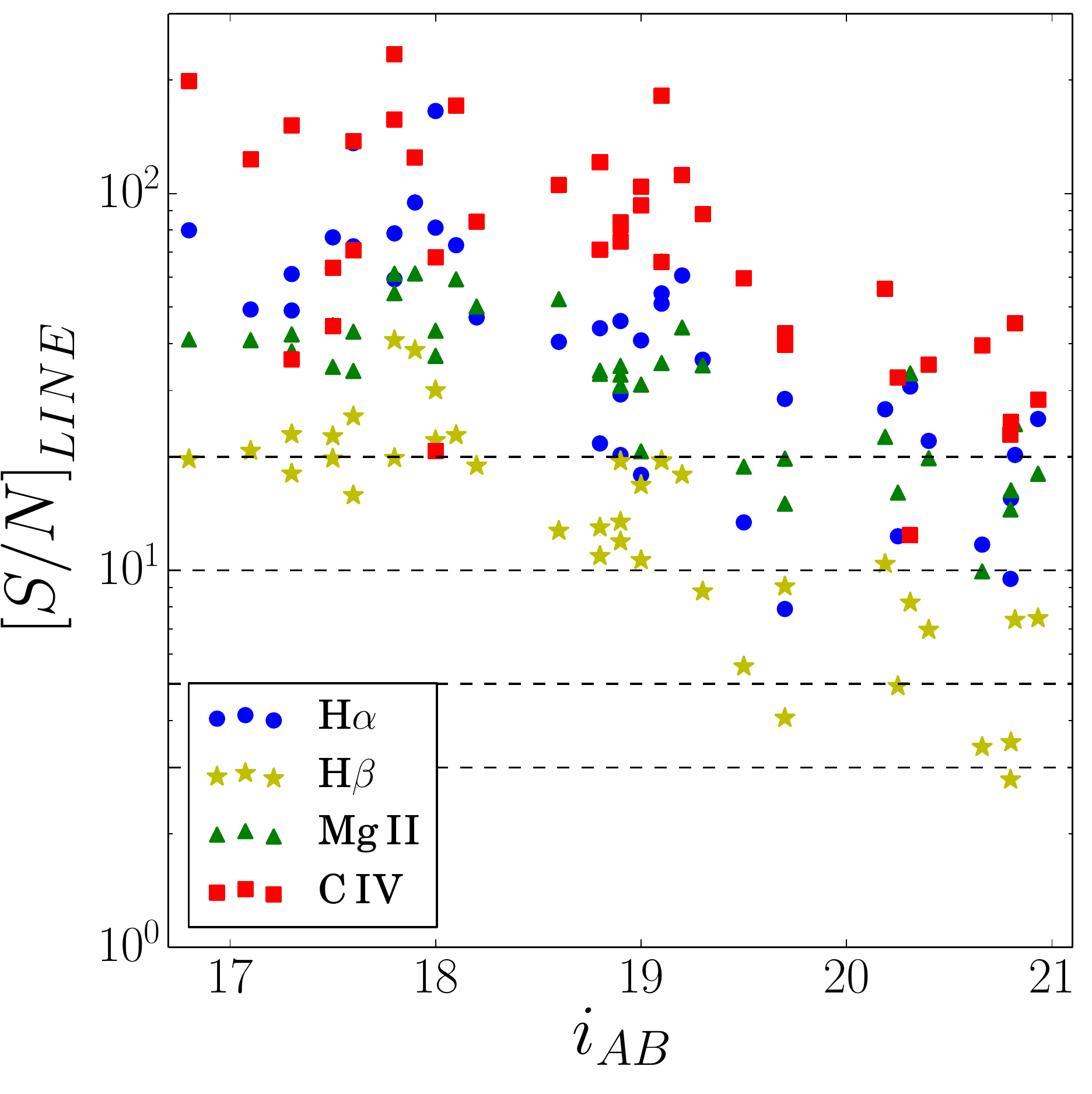}
\includegraphics[width=0.49\textwidth,angle=0]{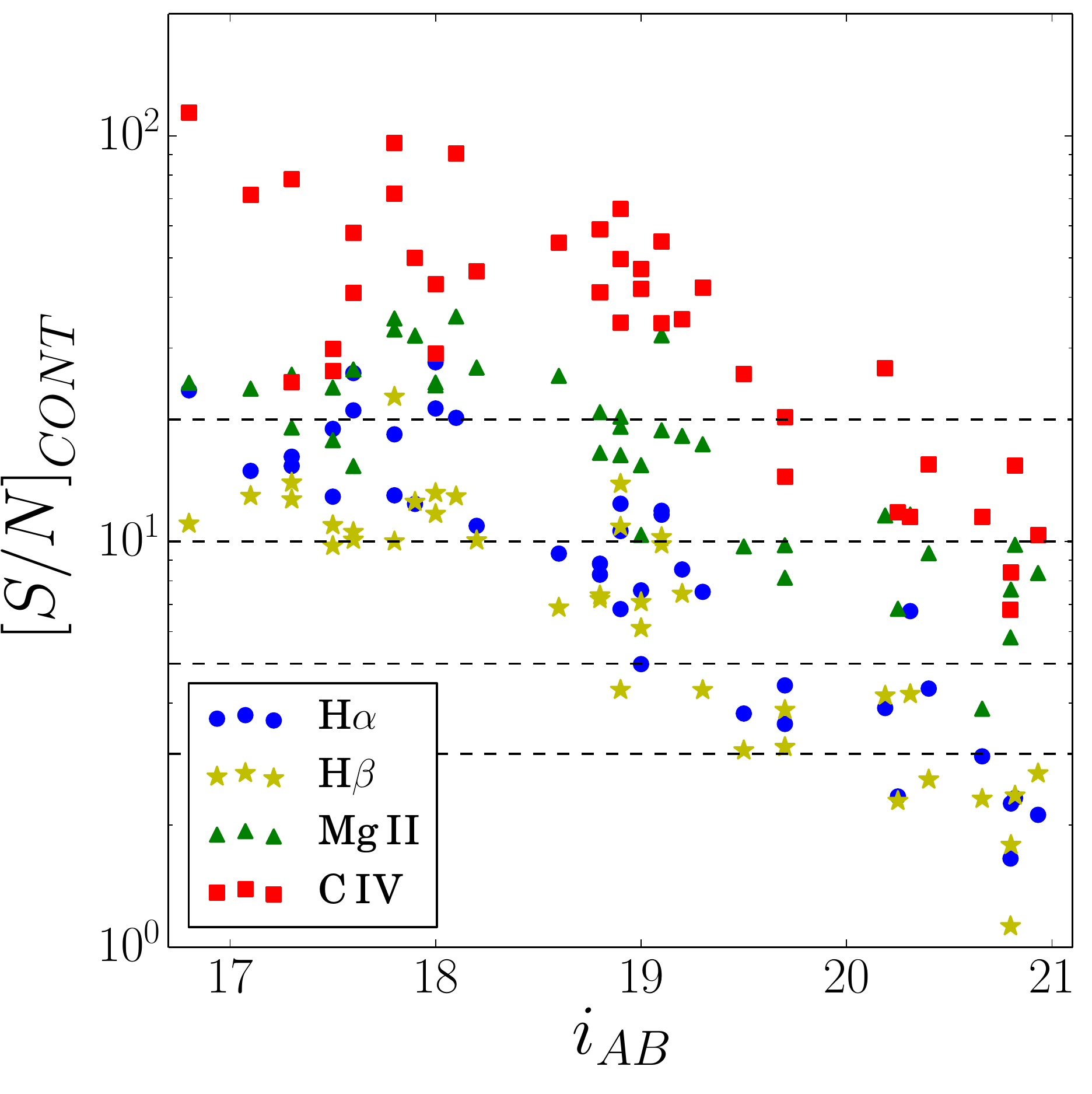}

\caption{ Signal to noise ratios (S/N) measured at  the peaks of each of the main broad emission lines (left), and over the corresponding nearby continuum bands (right), plotted against optical brightness, $i_{\text{AB} }$. 
The black dashed lines  represent, from bottom top,  S/N equal to 3, 5, 10 and 20.}
\label{fig:SN}
\end{figure*}

\section{Sample, Data and Analysis}
\label{sec:data}

The analysis presented in this paper is based on a sample of luminous, type-I  AGN  in a narrow redshift range around  $z\simeq1.55$,  for which we have obtained high signal to noise ($S/N$) single epoch spectroscopic observations using the X-Shooter instrument on the Very Large Telescope.  The \Ntot\ sources span a range in brightness of $i_{\text{AB}}\sim16.8-20.9$.
The sample selection, data acquisition and reduction for the \Nbright\ brightest sources were described in detail in paper I, and information about 9 other sources, obtained in ESO program 092.B-0613,  is provided in Paper III. 
Here we only briefly highlight a few essential aspects.

The sample has been selected from the seventh data release of the SDSS \citep{Abazajian2009} to homogeneously map the parameter space of \Mbh\ and \LLedd.
For the purposes of target selection, these quantities were initially obtained by spectral fitting of the \mgii\ broad emission line in the SDSS spectra as part of the large compilation described in TN12. 
{  In Figure. \ref{fig:MLLedd}  we show \Mbh\ vs \LLedd\ using  updated values  calculated in this paper based on the \Halpha\ broad emission line and following the procedure that we describe in section \ref{subsec:mass}. }

At the chosen redshift range of the sample, X-Shooter covers the rest-frame wavelength from about 1200\AA\ to  9200\AA. 
This broad spectral coverage has allowed us, after correction for Galactic extinction, to successfully model and constrain the observed  Spectral Energy distributions (SEDs). As shown in Papers I and III, we obtain satisfactory thin AD model fits to 37 sources, 6 of which require an intrinsic reddening correction for a satisfactory fit.  The  wide wavelength coverage,  together with the homogeneous selection of the sample  in the $\Mbh-\lledd$ plane, enables us to test the performance of the single epoch black hole mass  estimators for the \Halpha, \Hbeta,  \mgii\ and \civ\   lines and estimate the systemic bias induced when the physical SED is unknown.  

In Figure~\ref{fig:SN} we show the the signal to noise ratios ($S/N$) for our X-Shooter spectra, measured at the peaks of each of the main emission lines under study, as well as at the corresponding continuum bands, as a function of $i_{\text{AB}}$. 
{
We note that, even in the spectral region which overlaps with the available SDSS spectra, the X-Shooter data provide a significant improvement in terms of $S/N$ and spectral resolution (see an example in Fig.~\ref{fig:XshvsSDSS}, described in appendix \S\ref{app:xs_sdss_example}).
}
All the sources have fairly high $S/N$ ($\gtrsim 20$)  at the peaks of the \mgii\ and \civ\ lines and the adjacent continuum bands. 
However, this is not the case for \Halpha\ and \Hbeta. The continuum bands adjacent to \Halpha\ are much noisier. Most of the objects have $S/N<20$, and for those with $i_{\text{AB}}>18.5$, the ratio is below 10. Nevertheless, we are able to obtain reliable \Halpha\ line measurement because most objects have fairly high $S/N$ at their \Halpha\ line peak (34 out of 39 object have $S/N\gtrsim 20$ and all of them have $S/N\gtrsim 8	$ ). 
Moreover, the relevant continuum bands around \Halpha\ have low levels of contamination from iron or other, unresolved spectral features. 
Consequently, even a moderate continuum $S/N$ (i.e., $\gtrsim$3) is enough to have reliable  \Halpha\ fits. There are however 4 objects where  $S/N<3$ and their line measurements, especially their  \fwha\ are somewhat uncertain. 

Unfortunately, the \Hbeta\ line measurements are more problematic. 
In addition to the fact that \Hbeta\ is the weakest of the lines of interest, we can also see from Fig. \ref{fig:SN} that the relevant continuum band in 21 out of 39  objects have $S/N\lesssim 10$, and 14 of them are actually below $S/N\sim 5$. 
Near infrared (NIR) telluric absorption is another issue that could also crucially affect \Hbeta\ line measurements.  
The spectral regions with known low atmospheric transmission in the NIR, between \AA\ and between 13000\AA\ to 15000\AA\ typically translate to rest-frame bands at 4200-4500\AA\ and 5300-5800\AA\ at the redshift of the sample. 
These bands  are  known to show strong iron  emission which are  suppressed by such telluric absorption (see the example spectrum in Fig. \ref{fig:global} around 4400 and 5500\AA). 
The combined effect of the telluric absorption and the limited $S/N$ achieved for the fainter sources severely affects the correct  modeling  of their iron emission around  \Hbeta.  This, in turn, significantly increases the measurement uncertainties related to  \Hbeta, ultimately making \Hbeta\ measurements of faint objects less reliable. 

Fortunately, the \Halpha\ line shows very similar profiles to \Hbeta\ \citep[e.g.][]{GreeneHo2005} which  is in accordance with the expected radial ionization stratification of the BLR \citep{Kaspi2000}. 
Based on these results, we can probe several aspects related to the \Hbeta\ line using the more reliable \Halpha\ measurements.

\section{Spectral Decomposition}
\label{sec:fit}
In this section we describe the analysis procedures we used to model the X-Shooter spectra and to obtain  continuum and line emission measurements.
We discuss separately the analysis of emission corresponding to the continuum, the blended iron features, and the emission line components.
All the spectral modeling is done by employing the Levenberg-Marquardt algorithm for  $\chi^2$ minimization, using the \texttt{python} based spectroscopic analysis package \texttt{pyspeckit} \citep{GinsburgMirocha2011}.  
The fitting is preformed in the rest frame, after shifting the spectra using the improved SDSS redshifts provided by \citet{HewettWild2010}. 
{  We chose to use these redshifts, instead of using the \oiii\ line observed within the X-Shooter data, because of the limited quality of the relevant data and modeling of the \Hbeta-\oiii\ spectral region (see \S \ref{sec:data}) and the weak or absent \oiii\ emission in many of our sources.}

\subsection{Continuum Emission}
\label{subsec:cont}

We adopt here two different approaches to account for the continuum emission of the AGN,which we refer to as the \local  and \glob (thin disk)  continuum approaches. 
The \local continuum attempts to account for the usual approximation of the continuum emission by a single power law when the observed spectrum is limited to a narrow wavelength range.
The \glob thin disk  continuum, on the other hand, corresponds to the more physically-motivated AD model, that was obtained through a  Bayesian analysis taking advantage of our wide spectral wavelength coverage (see paper I).
A comparison of the measurements obtained with both approaches will allow us to quantify the possible bias imposed by ignoring the real SED shape, when wide-enough wavelength coverage is not available. 

\begin{figure*}
\centering
\includegraphics[width=1.00\textwidth]{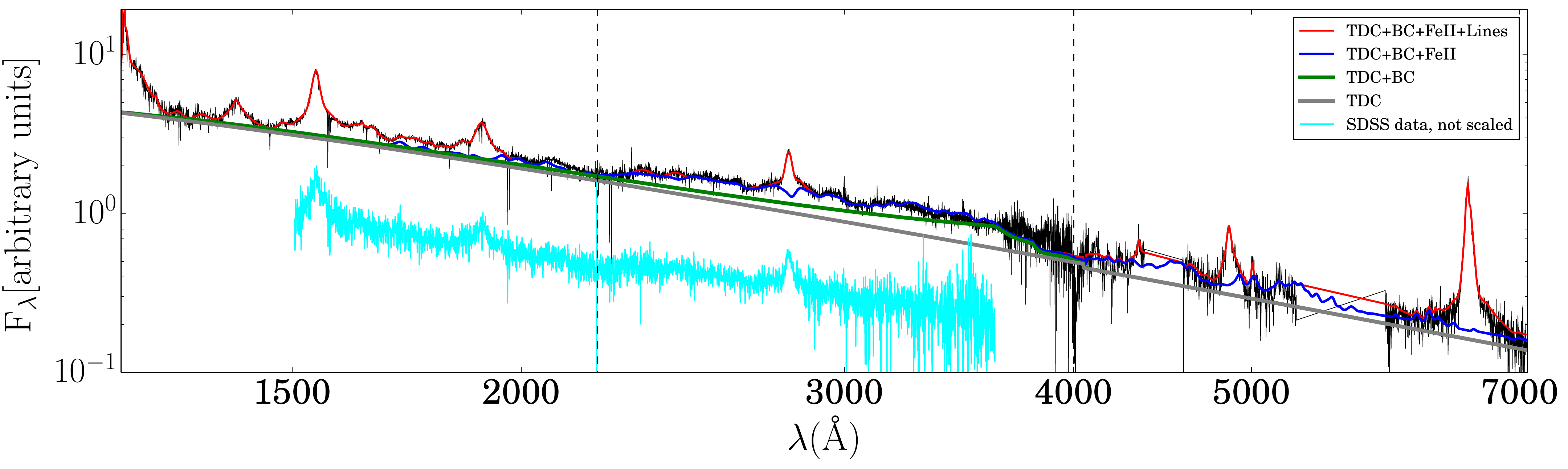}
\includegraphics[width=1.0\textwidth]{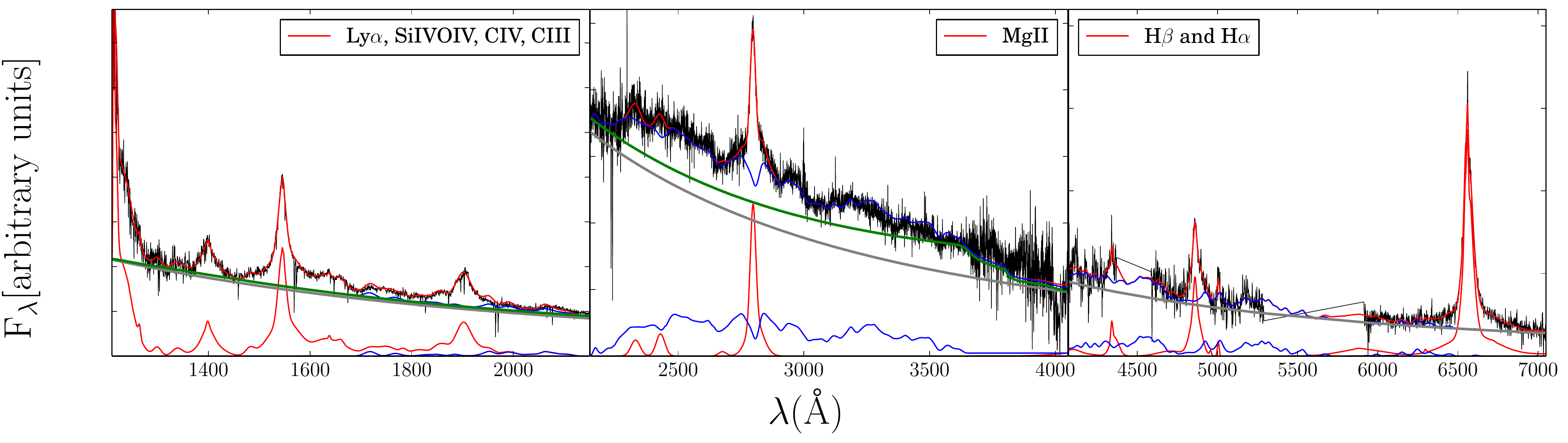}
\caption{The rest-frame X-Shooter spectrum (top) and main emission line complexes (bottom), over the three X-Shooter bands (UV:left, Optical:middle, Infrared:right), for the source J0143-0056 and the corresponding line fitting using the { \bf \glob thin disk continuum approach}.  
The observed spectrum is shown in black.
The best-fitting continuum is shown in gray. The blue lines represent the additional best-fit iron  emission. The red lines represent the additional best-fit models for the emission lines. For the sake of comparison we show the SDSS of the source in cyan.}
\label{fig:global}
\end{figure*}

\begin{figure*}
\centering
\includegraphics[width=1.00\textwidth]{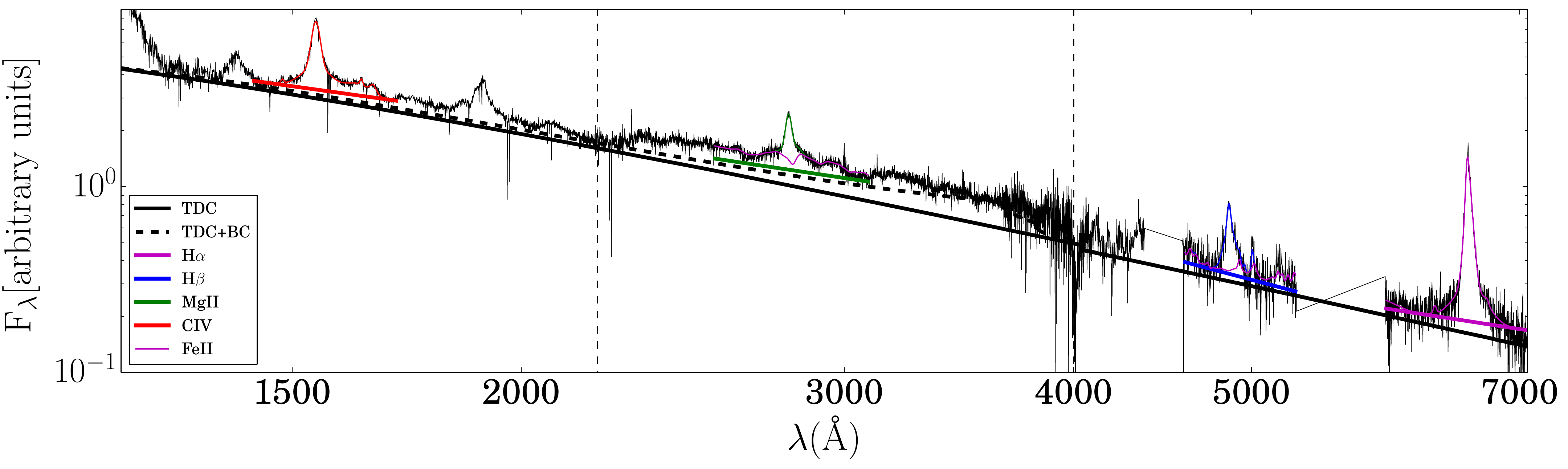} 
\includegraphics[width=1.0\textwidth]{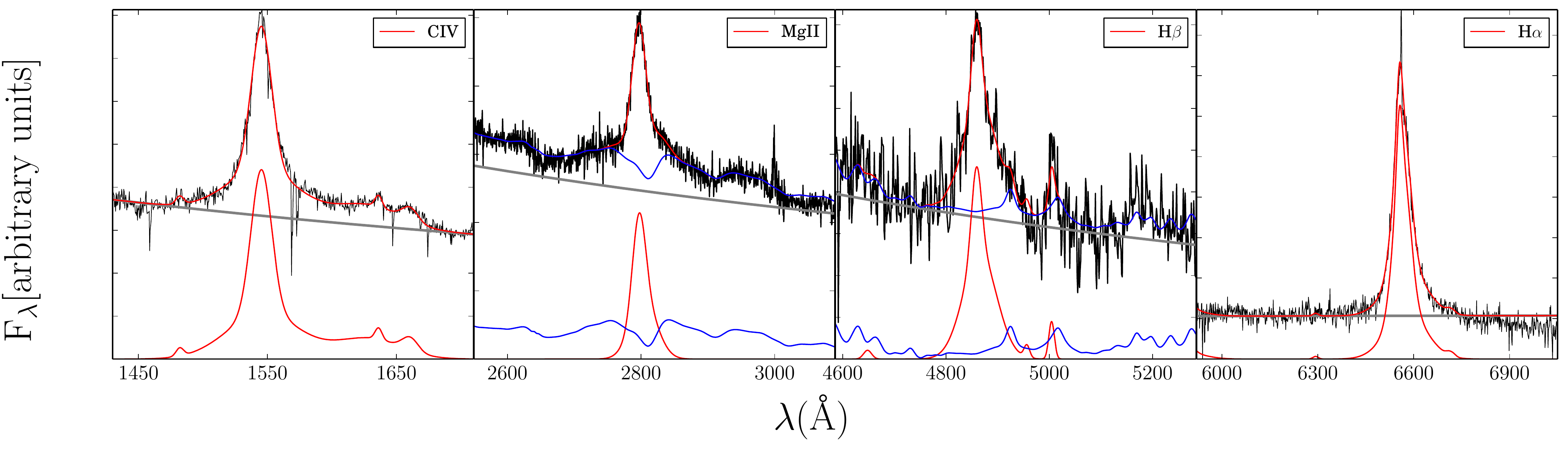} 
\caption{An example spectrum and spectral decomposition of one of the sources in our sample,  J0143-0056 using the \local approach. 
The top panel presents the rest-frame X-Shooter spectrum and the overall decomposition.
The solid black solid line corresponds to the ``thin disk'' continuum, while the dashed black line illustrates the addition of the Balmer continuum. We  highlight the spectral regions surrounding the most prominent broad emission lines (from left to right: \civ\ in red, \mgii\ in green, \Hbeta\ in blue, and \Halpha\ in magenta). The bottom panels show in detail the highlighted regions of the top panel as well as the individual \emph{\local} continuum determinations. 
Continuum fitting is in gray. Continuum plus iron emission fitting is in blue. Continuum plus, iron plus line fitting in red. Observed spectrum is in black.}
\label{fig:local}
\end{figure*}

\subsubsection{Local continuum approach and Biases}
 \label{subsubsec:LCA}
The \local continuum approach consists of separately fitting the continuum emission surrounding each of the lines
 of interest. For every source in the sample, each of these continua is approximated by a single power law, which connects neighboring spectral windows known to have little line contamination.
Our specific choice of such line-free continuum bands rely on several similar works (S07,TN12), and are listed in Table ~ \ref{tab:cwindows}. 

\begin{table}
\tabcolsep=0.5cm
\centering
\begin{tabularx}{0.47\textwidth}{  l l l  }
\hline
	Line Complex  & \multicolumn{2}{c}{--------- Continuum windows$^{1}$ ----------}   \\ 
  \hline
	\sioiv & 1340-1360\AA & 1420-1460\AA  \\ 
	\civ & 1420-1460\AA & 1680-1720\AA \\  
	\ciii & 1680-1720\AA & 1960-2020\AA \\ 
	\mgii & 2650-2670\AA & 3030-3070\AA \\ 
	\Hbeta & 4670-4730\AA & 5080-5120\AA \\ 
	\Halpha & 6150-6250\AA & 6950-7150\AA \\
\hline
\end{tabularx}
\caption{ Spectral pseudo-continuum windows used for our line fitting procedure under the \local continuum approach. $^{1}$For each object, we manually adjusted the continuum bands, using the listed wavelength ranges as a reference. }
\label{tab:cwindows}
\end{table}

The most important bias in the \localapp is that it commonly uses non real continuum windows that are affected by either 
 (1) weak line emission flux such as the continuum window at  1700\AA\  that is used for \civ\  line fitting, 
 (2)  iron continuum emission that affects continuum windows around 2600\AA\ and 3000\AA, as well as those  around  4650\AA\ and   5100\AA\ that are respectively used for \mgii\ and \Hbeta\  line fitting and finally, 
 (3) the Balmer continuum (BC) emission, at $\lambda<3647$ \AA, which can significantly affect \mgii\ measurements, and to a lesser extent even \civ\ measurements. 
 All these  biases are in the direction of an overestimation of the continuum emission when the \localapp is used which will translate into \fwhm\ and line flux underestimation.

An additional bias comes from the shape of the SED,  particularly at the turn over of most spectra at around 1000-1500\AA\ (exact wavelengths depend on BM mass, spin and accretion rate, see Papers I and III). The simple power-law approximation to the SED does not remain valid over this range and may lead to  measurement biases of the line profile properties of  \civ\  and \SiOIV\ (hereafter \sioiv). In this paper we use our AD SED fittings to quantify these biases. 

\subsubsection{Thin Disk continuum approach}
\label{subsubsec:TDCA}

The \glob AD  approach  is based on the best fits from the  thin-accretion-disk continuum models obtained for each  of the sources in Papers I and III. For the analysis in this paper we do not consider the two objects with no satisfactory 
fits to the thin disk continuum model.
 
As explained in paper I and III, the SEDs of the AD models used in this work are  determined by \Mbh, the accretion rate (\Mdot), the spin (\astar) and  the inclination of the disc with respect to the line-of-sight ($\theta$).
We adopted a Bayesian procedure to fit the thin AD model spectra to the observed X-shooter SEDs. \Mbh\ and \Mdot\ were taken as priors with Gaussian distributions centered on the observed values, obtained from \Halpha\ and \Lsix\ measurements (following the  procedures described in this paper), and with standard deviations of 0.3 and 0.2 dex, respectively.

Within the \glob continuum approach we also consider the BC emission that peaks near the Balmer edge (3647\AA) and gradually decreases towards shorter wavelengths. 
The Balmer continuum model we used is based on calculations of the photo-ionization code ION \citep{Netzer2006} with an H-atom containing 40  levels, solar abundances, hydrogen density of $10^{10} \rm{cm}^{-3}$, column density of $10^{23}\rm{cm}^{-2}$ and  ionization parameter of $10^{-1.5}$. The exact shape is insensitive to the exact value of these parameters and the normalization is done by direct fits to the observations.

An additional contribution to the continuum emission is due to starlight, mostly at wavelengths longer than about 6000\AA. 
For our AGN sample such a contribution is marginal in 32 out of 39 objects and does not severely affect the continuum
level and shape of the AGN SED as discussed in paper I and III ({  $<$3\% at 6200\AA} ).
For the 7 fainter objects we used the method described in paper III which assumes a template from an 11 Gyr old  stellar population  to model the host galaxy emission. 
The scale factor of the template is determined  from the ratio of the measured $\rm{EW}\left(\Halpha \right)$  and the median value of the  $\rm{EW}\left(\Halpha \right)$ distribution of the 29 brightest objects, as discussed in paper III.  
The host galaxy contribution is subtracted before the thin disk continuum fitting for those objects which require this correction.{  We find that in this sub-sample the host galaxy contribution  is between 6\% and 50\% at 6200\AA\ and smaller than 3\% at 3000\AA}. We  also tested several stellar populations in the age range from 1 to 11 Gyr, but we find no significant changes in the corrected spectrum (see paper III for details).

Finally, combining the X-Shooter spectra obtained by three different arms (UV, Optical and  NIR) may introduce additional uncertainties. 
As explained in Paper I, in most cases, the overlap and connection between the VIS and UVB arms are satisfactory, with no need for further adjustments but this is not the case for the VIS-NIR joint, as can be seen for J0043 in Fig. \ref{fig:global}. For several objects, the slope of the VIS arm was adjusted based on comparison to SDSS (see Paper I for more details). We therefore allow our fitting to rescale the \glob continuum   up to 10\%  in each 
of the regions covered by each arm   (1200-2200\AA, 2200-4000\AA, 4000-9000\AA) to take into account  the arm calibration  uncertainties. 

\subsection{Blended iron lines}

For an adequate modeling of  \Hbeta\ and \mgii\ line profiles it is crucial to first subtract the iron line emission, originating from a large number of blended features of \feii\ and Fe \textsc{iii}.
Generally, this is done by choosing the best-fit broadened,  shifted and scaled empirical iron line template. 
We constrain line center shifts to be smaller than 1000\,\kms\ and broadening is constrained to the range 1000-20000\,\kms. 
For the optical region around \Hbeta\ (4000-7000\AA) we used the iron template from \citet{BorosonGreen1992}. 
For the UV region around \mgii\  (1700-3647\AA) we initially used the \citet{Tsuzuki2006} template (hereafter T06).  
However, the fits obtained using this template was not  satisfactory, mainly due to an over-estimation of the continuum emission. We therefore built a new iron template (see Apppendix \ref{app:fe_template} and figure \ref{fig:feIItemplate})  based on the spectrum of I~Zw~1 reported by T06, which is a composite of their UV (HST) observation and the optical (KPNO) observation reported by \citet{Laor1997}. 
 
One of the main differences between the \local and \glob approaches is that under the \local approach different scaling factors for the UV iron template at each side of the \mgii\ line are required in order  to guarantee an acceptable match  to the observed spectrum. 
The scale factor in the red side of \mgii\ is found to be always larger than the one for the blue side, but by no more than 10\%. 
This type of correction is not needed in the \glob approach, when the complete continuum model (AD+BC) is considered. 
Given that  under the \localapp the BC cannot be accounted for directly and that the BC is monotonically increasing  from 2200 to 3647\AA, we suspect that the larger scale factor 
in the red side of \mgii\ might be due to the BC and not to intrinsic changes in iron line emission.
 
\subsection{Emission Line Measurements}

For the emission line modeling we have followed a procedure  similar to the one described in TN12 (see their appendix C) and \citet{Shang2007} 
In short, after removing the continuum emission (following either the \local or \glob approaches) and the iron template, we model the prominent broad emission lines with two broad Gaussian components. 
We allow for a range of line widths and shifts for each component, where the \fwhm\ ranges  between 1000\,\kms\ and 10000\,\kms and the line shifts are limited to $+/-1000\,\kms$ for the \Halpha, \Hbeta\  and \mgii\ lines, while for the \civ\ line we allowed blue-shifts of up to -3000\,\kms. 
These different choices are motivated by the findings of several earlier studies (e.g., \citet{VestergaardPeterson2006}, S07, R13, \citet{Park2013}).   
In the case of doublet lines (\civ\  and \mgii), we used 4 Gaussians, forcing the two broad and two narrower components to have the same profiles and intensity, and the theoretical wavelength separation. 
We fixed the \mgii\ and \civ\ doublet intensity ratios to 1:1, suitable for optically thick line emission.  
For each of the \Halpha, \Hbeta\  and \civ\ lines we have also included a third Gaussian component when needed to account for the  additional emission originating from the narrow line region (NLR). 
Each of these narrow components are modeled by a single Gaussian profile, their \fwhm\ is constrained not to exceed % 1000\kms
{  1300 \kms, and their line centers are tied to each other, with shifts of 400\kms, at most.}
We chose not to include a narrow component in the modeling of the \mgii\ and \civ\ lines  \citep[as in, e.g.,][]{Wills1993,Sulentic2007}, since we found no significant difference in the \mgii\ measurements (or fit quality) when trying to include it.\footnote{For example, for \civ\ we find that a narrow component typically contributes $\sim3\%$, and at most $6\%$, of the total line luminosity.}  
For other, weaker emission lines  (including He\,\textsc{ii}1640, N\,\textsc{iv}1718, Si\,\textsc{iii}]1892) we used only a single Gaussian component.  These lines are not necessary for the purpose of the present work except for limiting the continuum placement. More accurate modeling of these lines will be a topic of the fourth paper in this series.

All the Gaussian components we used are symmetric and defined by three parameters: peak flux density, \fwhm, and central wavelength.  We have made several simple, physically motivated simplifying assumptions, in order to minimize the number of free parameters: the Gaussian components of lines of the same species  were forced to share an identical width;
we have also tied together the relative shifts in the central wavelengths of some lines, based on their laboratory wavelengths; and assume line-intensity ratios for some lines based on their statistical weights (See Appendix ~\ref{app:fitting_params} and Table~\ref{tab:line_const} for further details on the different emission line parameters, their assumed ranges, inter-connections and delimitation of the emission line regions). Our line fitting procedure runs separately on each of the main emission line regions, while all the lines in each line region are fitted simultaneously. 

Generally, the \glob (see Figure \ref{fig:global}) and \local (see Figure \ref{fig:local}) continuum approaches follow the same line fitting procedures in terms of the number of components per emission line and the way they are tied together. 
One important difference is that in the \glob approach, the  \ciii\ and \civ\  line regions are considered a 
single region and are therefore fitted simultaneously. The reason for this is that under the \local approach we take as continuum windows the region around 1700\AA\, following the same procedure of previous works (e.g., S07, TN12, and references therein). However, this region is usually contaminated by weak emission lines like  N\,\textsc{iv}1718, and consequently the thin disk continuum fit does not allow us to fit \civ\ and \ciii\ independently.

In order to account for possible uncertainties in our spectral measurements, we performed 100 Monte-Carlo realizations for each of the spectra. 
In each of these realizations, the flux density at each spectral pixel was altered from the observed value by a random, normally distributed value, assuming the corresponding level of noise (i.e., using the noise spectrum).
From these sets of best-fit models we extracted, for each emission line, the line width \fwhm, the velocity dispersion (\sigline;  following \citet{Peterson2004}), integrated luminosity ($L$), rest-frame equivalent width (EW), the luminosity at the peak of the fitted profile ($L_{\text{P}}$) as well as its corresponding wavelength  ($\lambda_{\text{P}}$) and the offset of the line center (relative to the laboratory wavelength; $\Delta v$). 
{
The  line offsets were calculated using the flux-weighted central wavelength of the broad line profile:
\begin{equation}
\Delta v=\left(\int\lambda f_{\lambda}\left(\rm{line}\right)d\lambda/F\left(\rm{line}\right) - \lambda_{0}\right)c/\lambda_{0} 
\label{eqn:dlambda}
\end{equation}

where $f_{\lambda}\left(\rm{line}\right)$ is the flux density of the broad line profile at $\lambda$; $F\left(\rm{line}\right)$ is the integrated broad line flux, $F\left(\rm{line}\right)=\int f_{\lambda}\left(\rm{line}\right) d\lambda$ ; $\lambda_0$ is the laboratory wavelength of the line; and $c$ is the speed of light.
}

The best-fit values for all these parameters were taken from the medians of the parameter distribution, and the corresponding  uncertainties were estimated from the central 68\%\, percentiles. 
{  This ``re-sampling'' approach for the estimation of measurement-related uncertainties was used in several recent studies of spectral decomposition of AGN UV-optical spectra \citep[e.g.,][]{ShenLiu2012}.
Based on our experience, such errors reflect the true uncertainties related to measuring emission line profiles, while those  provided by the (statistical) spectral fitting procedure itself tend to underestimate the ``real'' uncertainties.
}

The measured parameters, and uncertainties, for the most prominent emission lines under the \localapp are summarized in Tables \ref{tab:line_params} and \ref{tab:line_params1}\,.\ {  The data is also available at \url{http://www.das.uchile.cl/~jemejia/big_table_mass_paper.tar.gz} which contains the plain text tables with these quantities   in the \local and  \glob approaches as well as the \fwhm s and \sigline s that we measured using the archival SDSS spectroscopy that  covers both the \civ\ and \mgii\ lines.}

\afterpage{\clearpage}{

\begin{landscape}
\begin{table}
\tabcolsep=0.09cm
\caption{Monochromatic continuum luminosities ($\lambda L\left[\lambda \right]$), line peak luminosity densities ($L_{\rm P}$) and line offsets ($\Delta v$), under \localapp}
\label{tab:line_params}
\begin{tabular}{lcccccccccccccccccccccccc}
\hline 
\multicolumn{1}{c}{Object} & \multicolumn{8}{c}{---------------$\log\left(\lambda L_{\lambda}/\ergs\right)$---------------} & \multicolumn{2}{c}{$\log\left(L_{line}/\ergs\right)$} &\multicolumn{6}{c}{--------------$\log\left(L_{\text{P}}/\ergs \ \text{\AA}  \right)$--------------} & \multicolumn{8}{c}{-------------------------------------$\Delta\, v\left[\kms \right]$-------------------------------------} \\ \hline
\multicolumn{1}{c}{name}   & \multicolumn{1}{c}{1450} & \multicolumn{1}{c}{$\Delta$} & \multicolumn{1}{c}{3000} & \multicolumn{1}{c}{$\Delta$} & \multicolumn{1}{c}{5100} & \multicolumn{1}{c}{$\Delta$} & \multicolumn{1}{c}{6200} & \multicolumn{1}{c}{$\Delta$} & \multicolumn{1}{c}{\Halpha} & \multicolumn{1}{c}{$\Delta$} & \multicolumn{1}{c}{\sioiv} & \multicolumn{1}{c}{$\Delta$} & \multicolumn{1}{c}{\civ} & \multicolumn{1}{c}{$\Delta$} & \multicolumn{1}{c}{\ciii} & \multicolumn{1}{c}{$\Delta$} & \multicolumn{1}{c}{\civ} & \multicolumn{1}{c}{$\Delta$} & \multicolumn{1}{c}{\mgii} & \multicolumn{1}{c}{$\Delta$} & \multicolumn{1}{c}{\Hbeta} & \multicolumn{1}{c}{$\Delta$} & \multicolumn{1}{c}{\Halpha} & \multicolumn{1}{c}{$\Delta$} \\ \hline 
J1152+0702  &       46.81  &         0.01  &       46.55  &         0.01  &       46.11  &        0.01  &       46.17  &        0.01  &        44.8  &        0.01  &       43.15  &        0.02  &       43.43  &         0.01  &       42.94  &        0.16  &    -2452.68  &      104.71  &      -97.53  &      101.35  &     1042.62  &      148.83  &      -79.15  &       70.63 \\
J0155-1023  &       46.62  &         0.01  &       46.41  &        0.01  &       46.13  &        0.01  &       46.07  &         0.01  &       44.87  &        0.01  &       43.17  &        0.01  &       43.29  &         0.01  &       42.88  &        0.07  &    -2294.86  &      123.19  &     -139.62  &       75.73  &      510.34  &       86.22  &     -172.03  &       90.25 \\
J0303+0027  &       46.53  &         0.01  &       46.36  &        0.01  &       46.03  &        0.01  &       45.99  &        0.01  &       44.79  &        0.02  &       43.01  &        0.02  &       43.21  &        0.01  &       42.49  &        0.27  &     -484.41  &       61.98  &       157.4  &      123.36  &     1094.57  &        53.3  &     -386.05  &      193.83 \\
J1158-0322  &       46.52  &         0.01  &       46.35  &        0.01  &       46.08  &        0.01  &       45.99  &        0.01  &       44.82  &        0.01  &       43.02  &        0.03  &       43.22  &        0.01  &       42.83  &        0.05  &    -1775.53  &       40.87  &       -4.04  &       66.53  &    -1067.76  &      168.86  &     -534.82  &       64.61 \\
J0043+0114  &       46.46  &         0.01  &       46.26  &        0.01  &       45.93  &        0.01  &       45.89  &         0.01  &       44.68  &         0.01  &        42.9  &        0.03  &       43.11  &         0.01  &       42.87  &        0.06  &    -2823.52  &      102.51  &     -340.23  &       67.91  &     -583.57  &        94.8  &     -330.63  &       51.76 \\
J0209-0947  &       46.56  &         0.01  &       46.38  &         0.01  &       46.09  &        0.01  &       46.01  &         0.01  &       44.86  &         0.01  &       43.05  &        0.01  &       43.47  &         0.01  &       42.95  &        0.03  &    -1534.54  &       19.99  &     -109.47  &       74.22  &      668.06  &      296.98  &     -328.73  &      105.64 \\
J0842+0151  &       46.39  &         0.01  &       46.21  &        0.01  &       45.79  &        0.04  &       45.78  &        0.02  &       44.74  &        0.02  &       42.97  &        0.05  &       43.24  &         0.01  &       42.81  &        0.08  &    -2393.53  &       53.53  &      -249.3  &       65.66  &      488.33  &      346.79  &      -545.4  &      222.08 \\
J1002+0331  &       46.55  &         0.01  &       46.29  &        0.01  &       45.99  &        0.01  &       45.97  &        0.01  &       44.83  &        0.01  &       42.02  &        0.39  &       43.34  &        0.03  &       43.27  &        0.02  &    -1425.91  &        67.7  &     -224.11  &       84.88  &      771.98  &      121.85  &      127.18  &       72.02 \\
J0323-0029  &       46.51  &         0.01  &       46.27  &        0.01  &       45.95  &        0.01  &       45.92  &        0.01  &       44.81  &        0.01  &       43.03  &        0.02  &       43.37  &         0.01  &       43.14  &        0.02  &      278.52  &       85.01  &     -477.62  &       92.79  &      674.22  &       97.41  &     -141.79  &      107.94 \\
J0152-0839  &       46.31  &         0.01  &       46.11  &        0.01  &       45.83  &        0.01  &       45.69  &        0.01  &       44.56  &        0.01  &        42.7  &        0.02  &       43.02  &         0.01  &       42.57  &        0.09  &    -2152.61  &       48.73  &     -245.03  &      100.87  &     -523.93  &      142.87  &      -512.4  &      104.87 \\
J0941+0443  &       46.27  &         0.01  &       46.08  &        0.01  &       45.79  &        0.02  &       45.74  &        0.01  &       44.68  &        0.01  &       42.73  &        0.02  &       43.11  &         0.01  &        42.6  &        0.09  &      306.62  &       92.72  &      -27.67  &      138.44  &     -174.21  &      401.45  &     -191.83  &       45.64 \\
J0148+0003  &        46.4  &        0.01  &       46.38  &        0.01  &       46.11  &        0.02  &       46.07  &        0.01  &       44.97  &        0.01  &       42.85  &        0.07  &       43.05  &        0.08  &       42.92  &        0.06  &     -933.05  &      104.45  &      -60.68  &       95.14  &      799.06  &       63.93  &     -360.83  &       64.41 \\
J0934+0005  &       46.15  &         0.01  &       45.92  &        0.01  &       45.68  &        0.01  &       45.62  &        0.01  &       44.43  &        0.01  &        42.7  &        0.02  &       42.77  &        0.02  &       42.62  &        0.06  &    -2156.93  &       71.57  &     -155.35  &       80.23  &       556.3  &       140.5  &     -449.29  &      174.13 \\
J0019-1053  &       45.89  &         0.01  &       45.78  &        0.01  &        45.4  &        0.01  &       45.39  &        0.01  &       44.26  &        0.01  &       42.33  &        0.03  &       42.79  &         0.01  &       42.35  &        0.09  &     -571.41  &       38.73  &      269.01  &       71.92  &      789.89  &      108.23  &      725.37  &        59.2 \\
J0850+0022  &        45.8  &         0.01  &       45.73  &        0.01  &       45.52  &        0.01  &       45.48  &        0.01  &       44.14  &        0.01  &        42.8  &        0.02  &       42.85  &        0.02  &       42.53  &        0.12  &    -2754.99  &      225.22  &      -41.26  &      125.73  &      316.41  &      196.98  &    -1044.97  &      142.74 \\
J0404-0446  &        45.9  &         0.01  &       45.72  &        0.01  &       45.62  &        0.03  &       45.46  &        0.02  &       43.92  &        0.05  &       42.79  &        0.02  &       42.71  &        0.01  &       42.51  &        0.12  &    -3440.87  &        99.7  &      -214.1  &       86.61  &     -435.09  &      451.73  &     -560.96  &      145.65 \\
J1052+0236  &       45.78  &         0.01  &        45.7  &        0.02  &       45.44  &        0.02  &       45.39  &        0.02  &       44.21  &        0.03  &       42.14  &        0.14  &       42.59  &        0.06  &       42.07  &        0.07  &      -70.84  &      116.54  &      263.36  &      147.25  &      1059.5  &      215.22  &      475.95  &       166.7 \\
J0223-0007  &       45.86  &         0.01  &       45.67  &        0.01  &       45.32  &        0.06  &       45.28  &        0.05  &       44.19  &        0.04  &        42.4  &        0.02  &        42.7  &        0.01  &       42.36  &        0.03  &     -1568.3  &       56.19  &        7.56  &        99.2  &     -481.33  &      692.18  &     -404.17  &      271.04 \\
J0240-0758  &       45.89  &         0.01  &       45.76  &        0.01  &       45.43  &        0.02  &       45.36  &        0.01  &       44.32  &        0.01  &       42.51  &        0.02  &       42.98  &        0.01  &       42.53  &        0.03  &     -293.37  &       33.52  &      185.33  &       57.24  &      478.62  &      212.48  &      710.93  &       26.89 \\
J0136-0015  &        45.80  &         0.01  &       45.64  &        0.01  &       45.28  &        0.03  &        45.2  &        0.01  &       44.12  &        0.02  &       42.48  &        0.13  &       42.59  &        0.01  &       42.33  &        0.04  &    -2756.36  &       79.89  &     -318.27  &       89.48  &      656.58  &      232.12  &     -648.21  &      114.67 \\
J0213-1003  &        46.2  &        0.01  &       45.93  &        0.01  &       45.64  &        0.02  &       45.58  &        0.01  &       44.39  &        0.01  &       42.98  &        0.05  &       42.98  &        0.01  &       42.65  &        0.04  &    -1739.83  &      261.66  &      -27.74  &      144.11  &     -360.87  &      278.85  &     -666.62  &       72.89 \\
J0341-0037  &       45.73  &         0.01  &       45.56  &        0.01  &        45.3  &        0.03  &       45.22  &        0.03  &       44.12  &        0.03  &       42.39  &        0.02  &        42.6  &        0.01  &       42.33  &        0.03  &    -2330.08  &      129.88  &     -222.25  &      102.02  &     -840.53  &      688.41  &     -536.48  &       261.4 \\
J0143-0056  &       45.72  &         0.01  &       45.52  &        0.01  &       45.18  &        0.02  &        45.1  &        0.04  &       43.99  &        0.04  &       42.35  &        0.03  &       42.73  &         0.01  &       42.24  &        0.06  &     -524.69  &        40.00  &      264.87  &       77.58  &      396.95  &      199.42  &      587.99  &      149.68 \\
J0927+0004  &       45.74  &        0.01  &       45.51  &        0.01  &       45.18  &        0.02  &       45.18  &        0.01  &       44.04  &        0.02  &       42.33  &        0.03  &       42.71  &         0.01  &       42.17  &        0.09  &      486.01  &      220.26  &       72.03  &       72.29  &     1010.25  &      278.06  &      345.72  &       94.13 \\
J0213-0036  &       45.64  &         0.01  &       45.46  &        0.01  &       45.15  &        0.02  &       45.12  &        0.02  &       44.11  &        0.01  &       42.29  &        0.05  &       42.86  &        0.01  &       42.29  &        0.04  &     -643.71  &       19.59  &      137.23  &      103.97  &      174.78  &       261.3  &       27.18  &      206.44 \\
J1050+0207  &       45.61  &        0.01  &       45.43  &        0.02  &       45.12  &        0.06  &       44.94  &        0.03  &       43.89  &        0.03  &       41.89  &        0.07  &       42.36  &        0.03  &       41.92  &        0.07  &     -591.26  &      218.74  &      443.37  &      152.79  &      122.92  &       566.2  &      526.65  &      337.22 \\
J0948+0137  &       45.43  &        0.01  &        45.3  &        0.02  &       45.03  &        0.04  &       45.02  &        0.05  &       43.96  &        0.03  &       42.09  &        0.11  &       42.53  &        0.02  &       42.13  &        0.08  &     -961.26  &      106.85  &      176.82  &      190.95  &      -62. &       414.8  &     -424.22  &      354.05 \\
J0042+0008  &       45.42  &        0.01  &       45.21  &        0.01  &       44.87  &        0.05  &       44.76  &        0.03  &       43.67  &        0.04  &       41.95  &         0.2  &       42.21  &        0.08  &       41.81  &        0.02  &    -1683.76  &      138.32  &     -312.02  &       99.56  &      697.42  &      205.71  &       26.51  &      160.33 \\
J1013+0245  &       45.38  &        0.01  &       45.18  &        0.02  &       45.08  &        0.04  &       45.01  &        0.03  &       43.59  &        0.05  &       42.03  &        0.07  &       42.19  &        0.01  &       41.95  &        0.09  &    -1110.04  &      244.02  &     -113.76  &      240.13  &    -1167.31  &     1020.54  &     -197.79  &      213.94 \\
J1021-0027  &       44.96  &        0.02  &       44.97  &        0.01  &       44.78  &        0.05  &       44.89  &        0.02  &       43.88  &        0.01  &       42.02  &        0.04  &       42.35  &        0.25  &       41.83  &        0.01  &      206.99  &      527.72  &     -219.52  &       49.35  &     1100.26  &      105.14  &     -925.29  &      101.62 \\
J0038-0019  &        45.1  &        0.01  &       44.97  &        0.01  &       44.78  &        0.04  &       44.81  &        0.02  &       43.45  &        0.02  &       41.51  &        0.12  &       42.21  &        0.05  &       41.65  &        0.04  &       51.63  &       92.82  &       18.95  &      162.72  &      -647.1  &      338.03  &      445.16  &       83.86 \\
J0912-0040  &       45.11  &        0.01  &       44.93  &        0.01  &       44.71  &        0.12  &       44.73  &        0.05  &       43.44  &        0.04  &       41.67  &        0.14  &       42.17  &        0.38  &       41.58  &        0.09  &    -1396.33  &      110.18  &       -79.1  &       162.0  &       600.7  &      567.47  &     -116.43  &      117.77 \\
J1048-0019  &       45.09  &        0.02  &       44.88  &        0.01  &       44.57  &        0.08  &       44.32  &         1.9  &        43.2  &        0.25  &       41.69  &         0.1  &       42.22  &        0.21  &       41.53  &        0.06  &     -322.86  &       79.87  &     -190.12  &      165.63  &      241.99  &      335.73  &      270.64  &      155.81 \\
J1045-0047  &       44.92  &        0.03  &       44.79  &        0.02  &       44.31  &        0.06  &       44.43  &        0.05  &       43.37  &        0.03  &       41.14  &        0.37  &       42.02  &        0.23  &       41.36  &        0.07  &     -990.37  &      115.71  &      -37.63  &      155.99  &     -158.89  &      564.79  &       66.88  &        89.9 \\
J0042-0011  &       44.87  &        0.02  &       44.77  &        0.01  &       44.63  &        0.07  &       44.47  &        0.05  &       43.06  &        0.04  &       41.62  &        0.09  &       41.92  &        0.17  &       41.37  &        0.02  &     -2104.9  &       83.54  &     -341.65  &      209.76  &      327.91  &      322.67  &     -169.64  &       73.31 \\
J1046+0025  &       44.97  &        0.01  &        44.7  &        0.02  &       44.42  &         0.1  &       44.41  &         0.1  &       42.48  &        0.12  &       41.66  &        0.09  &       42.07  &        0.34  &       41.43  &        0.05  &    -1255.21  &       69.88  &     -791.01  &      175.69  &      -801.5  &      452.36  &    -1299.09  &      125.12 \\
J0930-0018  &       44.81  &        0.02  &       44.66  &        0.02  &       44.36  &        0.97  &       44.11  &        0.11  &       42.95  &        0.26  &        41.4  &        0.25  &       41.98  &        0.28  &       41.38  &        0.03  &    -1115.71  &      371.44  &     -878.01  &      115.31  &    -1023.75  &      423.79  &     -996.61  &      147.11 \\
J1108+0141  &       46.53  &        0.01  &       46.47  &        0.01  &        46.2  &         0.01  &       46.09  &         0.01  &       44.82  &        0.01  &        43.1  &        0.03  &       42.95  &        0.02  &       42.93  &        0.06  &     -523.43  &       73.63  &     -729.27  &       47.84  &     1298.13  &       19.36  &     -354.62  &       45.75 \\
J1005+0245  &       45.96  &        0.01  &       46.05  &        0.01  &       45.87  &        0.01  &       45.89  &        0.01  &       44.74  &        0.01  &       42.57  &        0.13  &       42.75  &        0.04  &       42.79  &        0.03  &    -1202.56  &      828.92  &       11.81  &       79.42  &     1257.99  &       96.71  &      126.41  &      100.23 \\
\hline
\end{tabular}
\end{table}
 
\end{landscape}

\begin{landscape}
\begin{table}
\tabcolsep=0.035cm
\caption{Broad line widths and corresponding mass estimates, under the \localapp}
\begin{tabular}{lcccccccccccccccccccccccccc}
\hline
\multicolumn{1}{c}{Object} & \multicolumn{8}{c}{--------------------------$\fwhm \left[\kms \right]$--------------------------} & \multicolumn{8}{c}{--------------------------$\sigline \left[\kms \right]$--------------------------}   & \multicolumn{10}{c}{--------------------$\log\left(\Mbh\left(\fwhm\right)/\text{M}_{\odot}\right)$--------------------}                                                                                                                                                                                                                                                      \\ \hline
\multicolumn{1}{c}{name}   & \multicolumn{1}{c}{\civ} & \multicolumn{1}{c}{$\Delta$} & \multicolumn{1}{c}{\mgii} & \multicolumn{1}{c}{$\Delta$} & \multicolumn{1}{c}{\Hbeta} & \multicolumn{1}{c}{$\Delta$} & \multicolumn{1}{c}{\Halpha} & \multicolumn{1}{c}{$\Delta$}  & \multicolumn{1}{c}{\civ} & \multicolumn{1}{c}{$\Delta$} & \multicolumn{1}{c}{\mgii} & \multicolumn{1}{c}{$\Delta$} & \multicolumn{1}{c}{\Hbeta} & \multicolumn{1}{c}{$\Delta$} & \multicolumn{1}{c}{\Halpha} & \multicolumn{1}{c}{$\Delta$} & \multicolumn{1}{c}{\civ} & \multicolumn{1}{c}{$\Delta$} & \multicolumn{1}{c}{\mgii} & \multicolumn{1}{c}{$\Delta$} & \multicolumn{1}{c}{\Hbeta} & \multicolumn{1}{c}{$\Delta$} & \multicolumn{1}{c}{$\Halpha_{6200}$ $^{\rm a}$} & \multicolumn{1}{c}{$\Delta$} & \multicolumn{1}{c}{$\Halpha_{\rm line}$ $^{\rm b}$} & \multicolumn{1}{l}{$\Delta$} \\ \hline 
J1152+0702  &     6573.17  &      133.06  &     3202.24  &      135.98  &     4729.64  &      270.58  &     4283.31  &      240.51  &     3870.09  &      351.63  &     2165.29  &       179.2  &     4611.16  &       79.61  &     3186.29  &      246.89  &        9.65  &        0.02  &        9.47  &        0.04  &        9.44  &        0.06  &        9.48  &        0.05  &        9.32  &        0.06 \\
J0155-1023  &     6581.92  &       92.02  &     3468.25  &       66.93  &     5458.28  &      139.76  &     4785.95  &      115.95  &     5553.35  &      131.83  &     2113.16  &      129.33  &     2601.71  &       50.29  &     4010.01  &      147.36  &        9.54  &        0.01  &        9.46  &        0.02  &        9.58  &        0.03  &        9.51  &        0.02  &        9.46  &        0.03 \\
J0303+0027  &     5913.74  &      185.57  &     4790.24  &       136.5  &     7342.45  &      165.96  &     6229.92  &       99.96  &     3856.09  &      381.76  &     2519.43  &       219.8  &     3883.75  &       61.91  &     4897.97  &      308.97  &        9.39  &        0.03  &         9.7  &        0.03  &        9.77  &        0.02  &        9.69  &        0.02  &        9.64  &        0.02 \\
J1158-0322  &      4836.3  &       72.18  &     3375.87  &      132.76  &     5192.01  &      408.19  &      4854.6  &       169.2  &     2678.57  &       52.06  &     1900.87  &      106.72  &     4516.99  &      348.21  &     4114.39  &      261.77  &        9.21  &        0.01  &        9.39  &        0.04  &         9.5  &        0.08  &        9.47  &        0.03  &        9.44  &        0.04 \\
J0043+0114  &     7323.78  &      120.88  &     3379.59  &       93.25  &     4057.87  &       492.7  &     3227.27  &      123.33  &     3928.23  &         0.0  &     2508.89  &      204.13  &      5328.8  &       93.02  &     3633.45  &       83.24  &        9.54  &        0.02  &        9.34  &        0.03  &        9.19  &        0.12  &        9.06  &        0.04  &        9.01  &        0.04 \\
J0209-0947  &     5032.08  &       56.01  &     2900.08  &       64.16  &     5124.13  &      278.41  &      4722.6  &      176.26  &     4030.67  &       60.19  &      2298.9  &      145.81  &     4313.91  &      330.76  &     3973.72  &       146.8  &        9.26  &        0.01  &        9.28  &        0.02  &         9.5  &        0.06  &        9.46  &        0.04  &        9.44  &        0.04 \\
J0842+0151  &     6524.91  &      109.83  &     3951.92  &      128.07  &     5231.26  &      563.34  &     4955.88  &       55.58  &     3625.84  &      233.43  &      2044.7  &      165.81  &     4408.65  &      196.49  &     5900.66  &      136.71  &        9.39  &        0.02  &        9.45  &        0.03  &        9.32  &        0.12  &        9.36  &        0.02  &        9.42  &        0.02 \\
J1002+0331  &     5017.31  &       439.6  &     2563.88  &       69.75  &      5464.7  &      596.11  &     4738.24  &        81.6  &     2445.63  &        42.2  &     1854.53  &      235.07  &     4000.48  &      125.61  &     3315.12  &      276.48  &        9.26  &        0.08  &        9.12  &        0.03  &        9.49  &        0.11  &        9.44  &        0.02  &        9.43  &        0.02 \\
J0323-0029  &      5189.8  &       68.08  &     1889.87  &       69.24  &      2990.6  &      446.07  &      3127.0  &      668.27  &     3283.82  &       57.48  &     3206.57  &       47.72  &     3441.97  &      173.99  &      2955.2  &      188.82  &        9.26  &        0.01  &        8.84  &        0.04  &        8.94  &        0.14  &        9.05  &        0.22  &        9.05  &        0.21 \\
J0152-0839  &     6481.41  &       69.84  &     3116.09  &       158.2  &     4306.36  &      340.16  &     4813.74  &      164.25  &     3754.43  &      348.65  &     2110.44  &      186.45  &     3871.83  &       228.1  &     5716.66  &      259.13  &        9.34  &        0.01  &        9.18  &        0.05  &        9.18  &        0.08  &        9.28  &        0.03  &        9.29  &        0.04 \\
J0941+0443  &     7721.23  &       63.08  &     4218.07  &      286.42  &     6151.72  &      308.41  &     6520.48  &      120.97  &     3811.46  &       34.47  &     2948.63  &        94.8  &     3823.26  &      430.01  &     4161.47  &      141.57  &        9.46  &        0.01  &        9.42  &        0.07  &        9.46  &        0.06  &        9.58  &        0.02  &        9.62  &        0.02 \\
J0148+0003  &     6662.44  &     1461.18  &     4542.74  &       183.5  &     6474.99  &      340.44  &     5411.48  &       67.74  &     3844.55  &      258.27  &     2278.39  &      233.47  &     3458.09  &      114.54  &      4510.4  &      185.17  &        9.42  &        0.22  &        9.67  &        0.04  &        9.72  &        0.06  &        9.62  &        0.01  &        9.62  &        0.01 \\
J0934+0005  &     6536.46  &      282.31  &     2806.89  &       57.93  &     2880.02  &      144.81  &     2694.27  &      163.45  &     2702.42  &         0.0  &     1635.09  &      235.95  &     2573.13  &      254.74  &     2995.19  &       596.9  &        9.25  &        0.04  &        8.98  &        0.02  &        8.73  &        0.05  &        8.73  &        0.06  &        8.71  &        0.06 \\
J0019-1053  &     5219.79  &       89.76  &     4425.79  &       76.59  &      5708.5  &      376.28  &     5908.82  &      506.23  &     3484.23  &      195.22  &     2818.44  &      115.48  &     2849.73  &      479.65  &     4694.24  &      467.45  &         8.9  &        0.02  &        9.28  &        0.02  &        9.15  &        0.07  &        9.27  &        0.09  &         9.3  &        0.08 \\
J0850+0022  &     6115.68  &      272.63  &     2415.05  &      123.67  &     3503.46  &      960.02  &     3763.47  &      419.19  &      2905.3  &      123.92  &     1749.98  &      437.53  &     4858.84  &      322.37  &     3435.86  &      135.78  &        8.98  &        0.04  &        8.73  &        0.05  &         8.8  &        0.29  &        8.93  &        0.11  &        8.84  &        0.11 \\
J0404-0446  &     4341.32  &      304.12  &     1965.55  &       50.96  &     2788.42  &     1363.96  &     2731.63  &      140.01  &     1841.37  &       99.93  &     1666.83  &       264.0  &      867.64  &      230.43  &     6567.38  &      769.42  &        8.74  &        0.06  &        8.54  &        0.03  &        8.66  &         0.6  &        8.64  &        0.06  &        8.43  &        0.08 \\
J1052+0236  &     9627.85  &      852.07  &      5412.1  &      422.04  &    10110.01  &      1132.0  &     8117.99  &      338.71  &     5192.56  &      272.53  &      2701.9  &      339.79  &     5726.97  &      567.53  &     4513.15  &      408.93  &        9.37  &        0.08  &        9.41  &        0.08  &        9.66  &        0.11  &        9.54  &        0.05  &        9.55  &        0.05 \\
J0223-0007  &      5239.8  &      174.63  &     3077.91  &       80.99  &     3416.23  &      549.13  &     5009.03  &      271.49  &     3500.23  &      374.37  &     2425.53  &      288.24  &     5199.39  &      403.63  &     5461.09  &     1392.14  &        8.88  &        0.03  &         8.9  &        0.03  &        8.65  &        0.19  &        9.05  &        0.08  &        9.12  &        0.07 \\
J0240-0758  &     5858.64  &      135.99  &     2827.82  &       94.14  &     3542.36  &      445.25  &      4189.8  &      158.88  &      3013.8  &       32.58  &     2184.61  &      133.16  &      5569.2  &      275.57  &     4136.64  &      445.19  &         9.0  &        0.02  &        8.88  &        0.03  &        8.75  &        0.13  &        8.95  &        0.04  &        9.03  &        0.04 \\
J0136-0015  &     7051.94  &      114.59  &     3553.55  &       86.89  &     2730.93  &     1000.53  &     3467.62  &      276.33  &     4489.88  &      434.54  &     2248.54  &      192.31  &     3648.03  &      195.22  &     5068.74  &      586.32  &         9.1  &        0.02  &        9.01  &        0.03  &        8.43  &        0.42  &        8.68  &        0.08  &        8.76  &        0.08 \\
J0213-1003  &      5093.2  &      167.27  &     2741.59  &      251.38  &     3392.86  &      279.21  &     4054.31  &      192.65  &      3523.3  &      932.25  &     1939.26  &      287.79  &      4174.0  &      871.34  &     4579.36  &      117.79  &        9.06  &        0.03  &        8.96  &        0.09  &        8.85  &        0.08  &        9.06  &        0.05  &        9.05  &        0.05 \\
J0341-0037  &     5246.89  &      130.51  &     2571.38  &       106.3  &     3218.83  &     2612.53  &     3405.73  &      610.87  &      3982.5  &      398.66  &     1901.38  &      277.54  &     3567.48  &      736.67  &     4447.75  &      907.79  &         8.8  &        0.02  &        8.68  &        0.04  &        8.58  &        1.47  &        8.68  &        0.19  &        8.74  &        0.19 \\
J0143-0056  &     3890.11  &       76.91  &     2856.85  &      107.36  &     5509.54  &      1371.7  &     3894.47  &      570.51  &      3277.0  &      107.64  &     1400.35  &      212.49  &     1876.82  &      121.42  &     4417.88  &      1204.3  &        8.54  &        0.02  &        8.75  &        0.04  &        8.97  &        0.26  &        8.72  &        0.16  &        8.78  &        0.16 \\
J0927+0004  &     8307.68  &      152.96  &     5663.93  &      202.87  &     7418.97  &      915.99  &     6255.58  &       148.7  &      4555.5  &      291.75  &     2477.82  &      189.86  &     4265.65  &      355.71  &     4497.04  &       557.7  &        9.21  &        0.02  &        9.34  &        0.04  &        9.23  &        0.13  &        9.18  &        0.03  &        9.23  &        0.03 \\
J0213-0036  &     4063.84  &      101.99  &     3462.69  &       155.4  &     5697.36  &     1018.59  &     4458.74  &       171.6  &     2719.59  &       41.51  &     1811.81  &      159.25  &      2680.4  &      569.02  &     4997.53  &       672.0  &        8.53  &        0.02  &        8.88  &        0.05  &        8.98  &        0.19  &        8.85  &        0.04  &        8.97  &        0.04 \\
J1050+0207  &     7608.29  &      816.82  &     4888.94  &      762.65  &     5204.91  &     1493.81  &     5402.67  &      504.92  &     4416.38  &      210.67  &     2409.78  &      196.71  &     2442.51  &       345.5  &     5260.54  &      706.67  &        9.06  &         0.1  &        9.15  &        0.16  &        8.88  &        0.33  &         8.9  &         0.1  &        9.01  &         0.1 \\
J0948+0137  &     5115.62  &      251.64  &     3363.61  &      331.15  &     3766.11  &      728.57  &     3880.89  &      392.36  &     3292.57  &      178.85  &     1997.33  &      367.65  &     3373.17  &      769.02  &     2879.06  &     1025.66  &         8.6  &        0.05  &        8.75  &         0.1  &        8.54  &        0.21  &        8.67  &        0.13  &        8.77  &        0.11 \\
J0042+0008  &     5362.14  &      347.78  &     2943.94  &      120.79  &     4020.34  &       345.5  &     3607.54  &      160.39  &     4970.52  &      292.54  &     1843.64  &      224.51  &      4028.2  &      473.32  &     5323.33  &      540.12  &        8.64  &        0.06  &        8.58  &        0.04  &         8.5  &        0.11  &        8.44  &        0.06  &        8.54  &        0.06 \\
J1013+0245  &     8990.08  &      304.21  &     4603.49  &      709.46  &     8395.63  &     1030.61  &     7716.39  &      872.62  &     5124.06  &      469.07  &     2678.99  &      213.54  &     6546.57  &     3016.95  &     3037.51  &      291.49  &        9.07  &        0.04  &        8.95  &        0.16  &        9.27  &        0.14  &        9.25  &        0.12  &        9.15  &        0.14 \\
J1021-0027  &     4912.88  &      628.93  &     3260.01  &      159.14  &     9085.38  &     1675.61  &     8817.56  &      719.31  &     2790.07  &      342.82  &     2409.75  &      157.98  &     5060.65  &      525.81  &     4593.35  &      183.54  &        8.29  &        0.13  &        8.52  &        0.05  &        9.15  &        0.21  &        9.29  &        0.09  &        9.43  &        0.08 \\
J0038-0019  &     4332.98  &      333.19  &     2656.77  &      189.51  &     3557.91  &      284.34  &     3303.65  &       118.2  &     3397.61  &      353.28  &     2036.88  &      227.39  &     5494.81  &       663.5  &     2146.07  &      272.75  &        8.26  &        0.08  &        8.35  &        0.07  &        8.33  &         0.1  &        8.39  &        0.04  &        8.34  &        0.04 \\
J0912-0040  &     4884.79  &      331.59  &     3859.93  &      251.59  &      5803.7  &     3022.45  &     4746.34  &      227.74  &     3198.91  &      193.24  &     2002.02  &      195.75  &     2164.52  &         213.84  &     2015.46  &      581.81  &        8.37  &        0.07  &        8.65  &        0.07  &        8.71  &        0.73  &        8.66  &        0.07  &        8.65  &        0.07 \\
J1048-0019  &     4755.88  &       649.8  &     2901.43  &      155.43  &     2278.84  &       350.3  &     2966.74  &     1312.75  &     3167.81  &      288.13  &     2260.65  &      203.08  &      967.48  &      372.04  &     1261.05  &         249.41  &        8.34  &        0.14  &        8.37  &        0.06  &        7.81  &         0.2  &        7.99  &       0.136  &        8.11  &        0.71 \\
J1045-0047  &     6027.95  &      520.11  &     2832.87  &      308.37  &     9706.63  &     3147.47  &      4724.3  &      403.71  &     3841.54  &      183.51  &     2202.59  &       227.6  &     6403.41  &     1425.91  &     2820.17  &      151.58  &        8.44  &         0.1  &        8.29  &        0.11  &         8.9  &        0.38  &        8.46  &        0.11  &         8.6  &         0.1 \\
J0042-0011  &     4197.93  &      217.44  &     1829.42  &       210.0  &     1578.09  &      122.63  &     1489.82  &       95.27  &     2929.25  &       161.1  &     1097.93  &       599.1  &     1701.43  &      475.43  &     1177.48  &      165.01  &         8.1  &        0.06  &         7.9  &        0.11  &        7.52  &        0.12  &        7.49  &        0.09  &        7.42  &        0.08 \\
J1046+0025  &      3971.3  &      481.38  &     3210.69  &      164.83  &     4319.81  &     1113.03  &      3382.7  &      584.84  &     2915.58  &       146.4  &     2207.23  &      226.84  &     1013.97  &      984.74  &     1003.89  &      196.21  &        8.11  &        0.12  &        8.35  &        0.06  &        8.26  &        0.33  &        8.16  &        0.24  &        7.81  &        0.24 \\
J0930-0018  &     6704.04  &      856.67  &     5058.85  &      416.66  &     7035.13  &      796.99  &     6079.72  &     7366.64  &     4590.25  &      423.08  &     3059.34  &      415.94  &      842.29  &         247.30  &     1508.94  &     1415.15  &        8.47  &        0.13  &        8.72  &        0.09  &        8.65  &       0.496  &        8.48  &       0.106  &        8.59  &       0.134 \\
J1108+0141  &      7468.9  &       410.2  &     2601.39  &        91.7  &     4093.15  &      591.92  &     4750.88  &       63.75  &     3350.98  &      125.47  &     2830.93  &      118.89  &     3947.92  &      434.39  &     4184.72  &      134.75  &        9.59  &        0.05  &        9.24  &        0.04  &        9.37  &        0.14  &        9.52  &        0.01  &        9.42  &        0.02 \\
J1005+0245  &     6095.34  &     1550.64  &     2475.74  &      144.98  &     3878.66  &      3284.9  &     5050.79  &      198.95  &     4574.16  &      821.83  &     1745.09  &      154.14  &     4291.27  &      196.04  &     4053.76  &       188.6  &        9.07  &        0.26  &        8.94  &        0.06  &        9.11  &        1.64  &        9.45  &        0.04  &        9.44  &        0.04 \\
\hline
\vspace*{0.05cm}\\
\multicolumn{19}{l}{$^{\rm a}$~~\MbhHa\ measurements obtained through \fwha\ and \Lsix.}\\
\multicolumn{19}{l}{$^{\rm b}$~~\MbhHa\ measurements obtained through \fwha\ and \Lha.}\\
\end{tabular}
\label{tab:line_params1}
\end{table}
\end{landscape}

}

\begin{figure*}
\centering
\includegraphics[width=0.495\textwidth,angle=0]{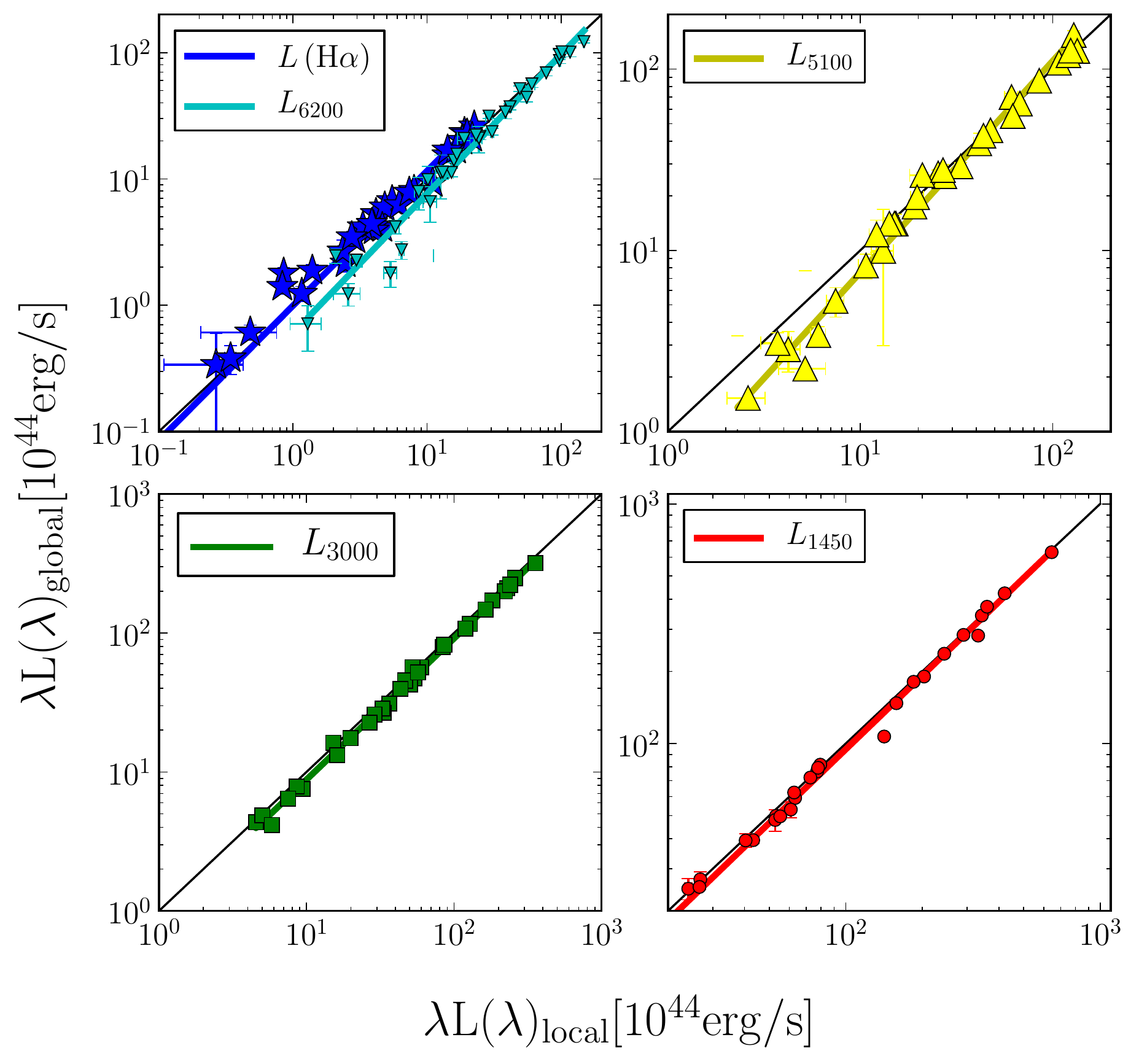}
\includegraphics[width=0.495\textwidth,angle=0]{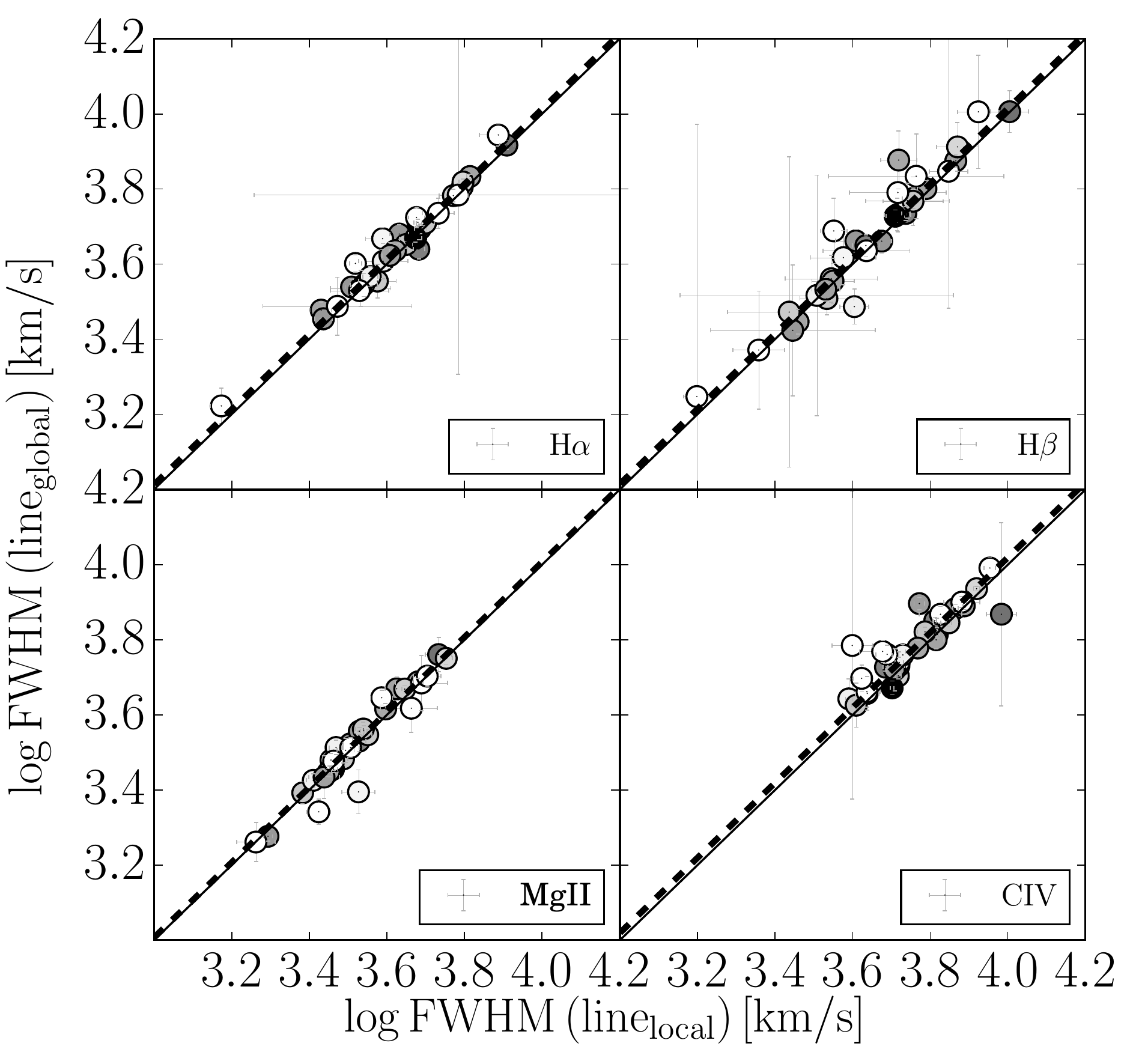}
\includegraphics[width=0.495\textwidth,angle=0]{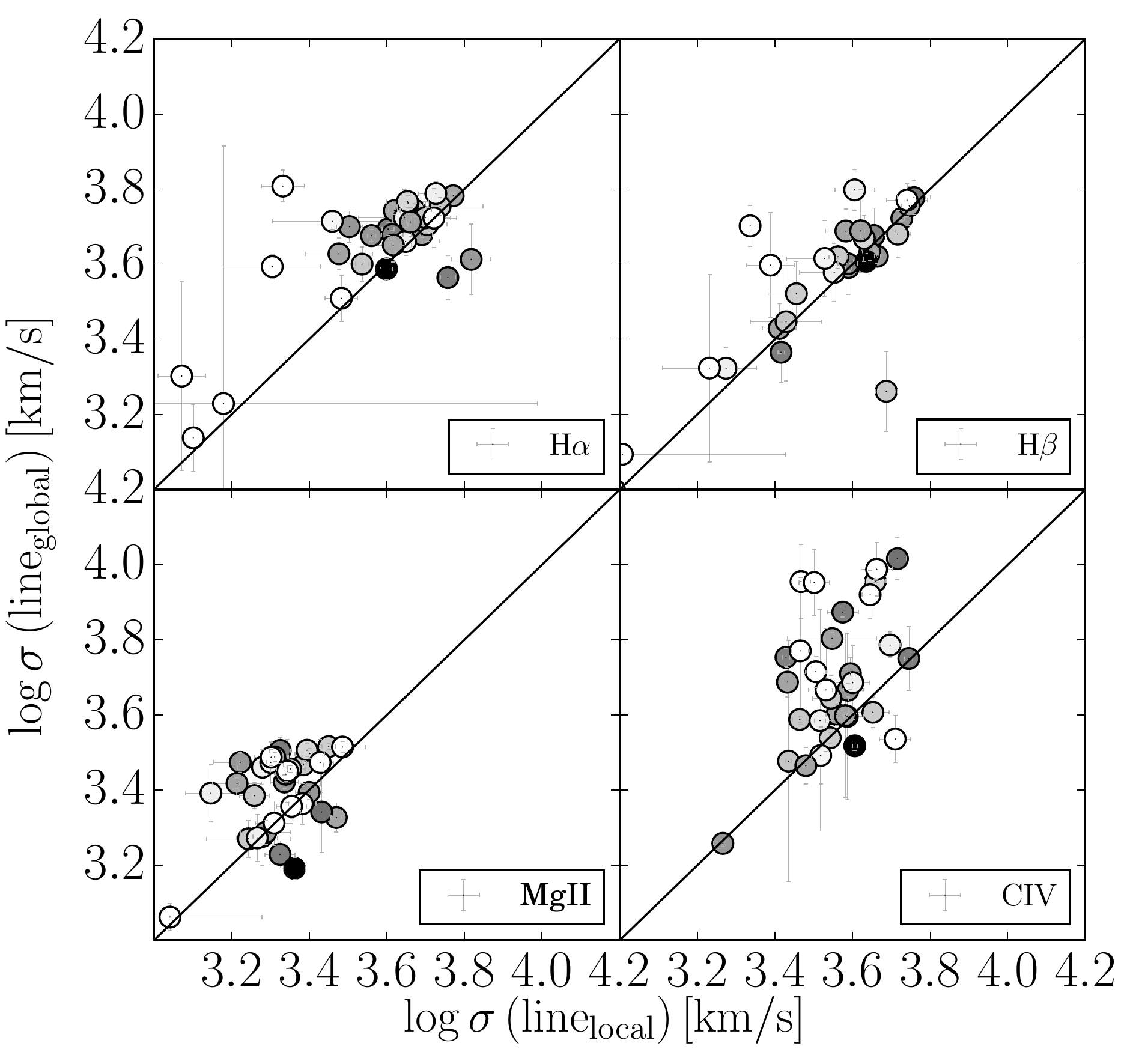}
\includegraphics[width=0.495\textwidth,angle=0]{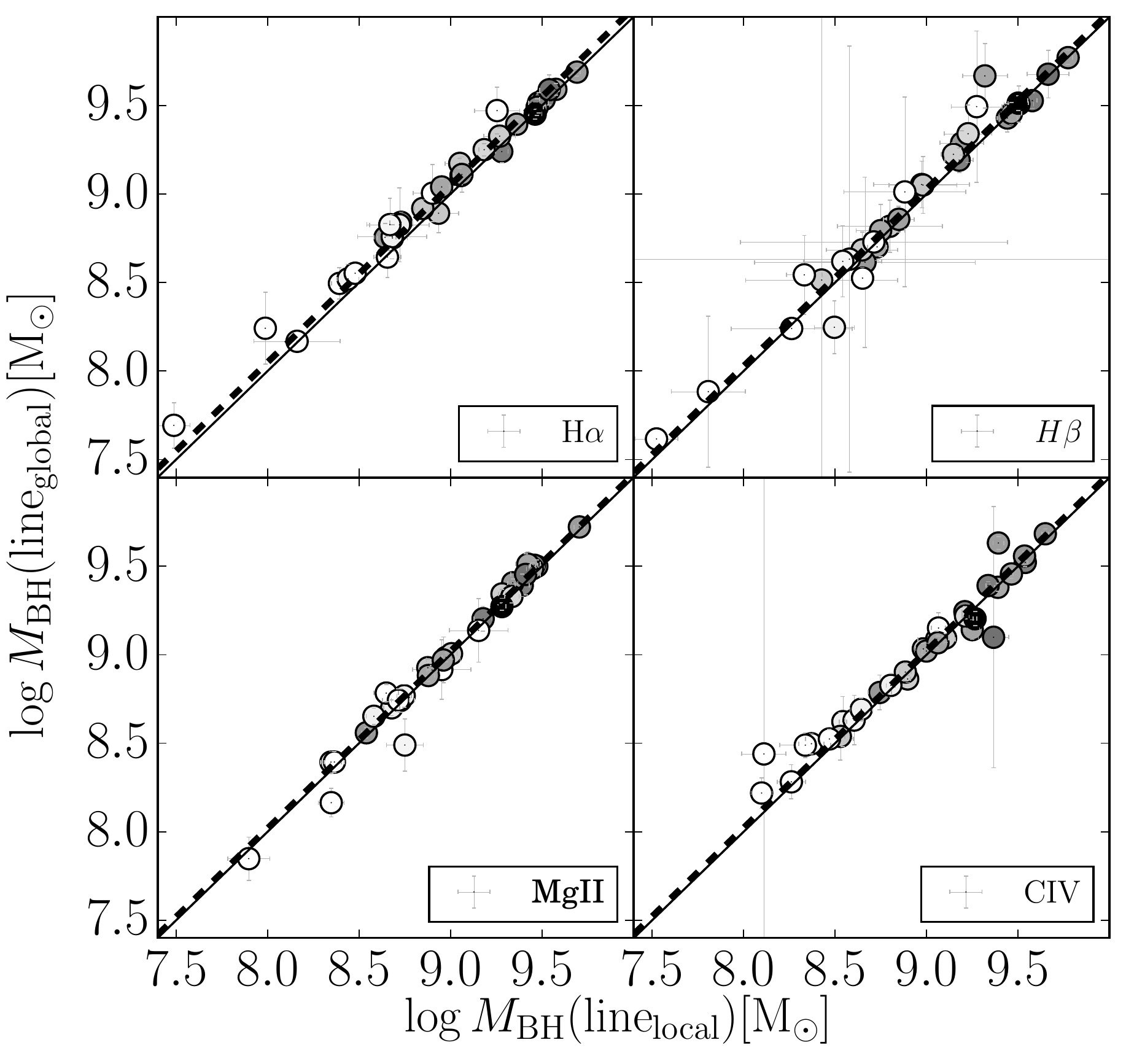}
\caption{Comparison of various line and continuum properties measured under the two general assumptions of \local (horizontal axis)  and \glob (vertical axis) SEDs. Top-Left panel: \Lha (blue stars), \Lsix (cyan triangles), \Lop\ (yellow triangles), \Lthree\ (green squares) and \Luv\ (red dots).  The colored solid lines  are the best linear fits to the corresponding data. Top-Right panel: \fwha\ (top-left), \fwhb\ (top-right), \fwmg\ (bottom-left) and \fwciv\ (bottom-right) lines measurements. Bottom-left panel: Same as Top-Right panel but comparing \sigline s instead of \fwhm s (note the much larger scatter in this case).  Bottom-right panel: \Mbh\ comparison between \local and \glob approaches. In the top-right and bottom-right panels black dashed lines represents the median offset between \glob and \local measurements. In all panels the black solid diagonal line represents the 1:1 relation. Points have  been  color-coded in gray scale by S/N where darker colors correspond to larger S/N.}
\label{fig:globalvslocal}
\end{figure*}

\begin{table}
\centering
\tabcolsep=0.25cm 

\begin{tabularx}{0.47\textwidth}{ccccl}
\hline
Line    & $\Delta\Mbh$           & $\Delta\fwhm$            & $\Delta L^{1}$          & \multicolumn{1}{c}{$\Delta L_{\text{line}}$} \\ 
        & (dex)                   & (dex)                     & (dex)                   & \multicolumn{1}{c}{(dex)}                    \\  \hline \\
\Halpha & $ 0.03^{+0.06}_{-0.06}$ & $0.015^{+0.020}_{-0.015}$ & $0.05^{+0.04}_{-0.06}$  & $0.05^{+0.04}_{-0.06}$                       \\ \\
\Hbeta  & $0.04^{+0.09}_{-0.05}$ & $0.020^{+0.035}_{-0.025}$ & $-0.01^{+0.03}_{-0.05}$ & $0.03^{+0.12}_{-0.05}$                      \\ \\
\mgii   & $0.01^{+0.03}_{-0.02}$ & $0.010^{+0.015}_{-0.010}$ & $-0.03^{+0.03}_{-0.02}$ & $0.03^{+0.03}_{-0.03}$                       \\ \\
\civ    & $ 0.05^{+0.06}_{-0.03}$ & $0.020^{+0.025}_{-0.020}$ & $-0.02^{+0.02}_{-0.03}$ & $0.07^{+0.06}_{-0.07}$    \\ \hline                                       
\end{tabularx}
\caption{Median induced offsets when the local approach is used instead of the \glob approach.}
\label{tab:offsets}
\end{table}

\section{Results and Discussion}
\label{sec:results}

\subsection{Local versus global continuum measurements }
\label{subsec:globalvslocal}

In this subsection we compare the \local and \glob  continuum approaches in order to quantify the possible biases that are introduced when the real underlying shape of the continuum cannot be accurately established. As we will describe below in detail, our main conclusion is that \local continuum measurements of  \fwhm s, continuum luminosities and, consequently, black hole masses present very small but systematic offsets with respect to the corresponding  \glob continuum measurements.

\subsubsection{Continuum biases}
In figure \ref{fig:globalvslocal} we present the comparison between \Llocal\ and \Ltdc\ (top-left panel) for different chosen wavelengths. We generally find small but systematic offsets between  quantities derived via the \local versus \glob\ approach. We find that the \Luv, \Lthree, \Lop\ and \Lsix\   median offsets   ($ \Delta  L \equiv \log \left(\Ltdc/\Llocal \right)$)  are typically small ($\lesssim \abs{-0.05}$ dex, see Table \ref{tab:offsets} for details). These offset are  consistent with a very subtle overestimation of the continuum emission when the \local approach is adopted (see Fig. \ref{fig:local} for a particular example).

\subsubsection{Line width biases}
The  \textit{ systematic  continuum overestimation} that we found coming from adopting the \local instead of the  \globapp will naturally lead to \textit{systematical} \fwhm\ \textit{underestimation} as can be also seen in Figure \ref{fig:globalvslocal} (top-right panel). 
Indeed, all the relevant line width measurements present small median offsets  ($ \Delta \fwhm \equiv \log$ $\fwtdc/\fw$) smaller than $\lesssim 0.02$ dex as can be seen in Table. \ref{tab:offsets}. As mentioned in \S\ref{sec:data} the measurements of \fwhb\ are  more challenging for low S/N and/or objects where most iron emission is suppressed by telluric absorption.  This explains the outliers and large uncertainties for some objects in the \fwhb$_{\text{local}}$-\fwhb$_{\text{global}}$ plot.  Except for these few outliers, the \fwhm$_{\text{local}}$ measurements of all the emission lines are  proportional to, and  systematically but slightly smaller than the \fwhm$_{\text{global}}$ measurements. 

When we perform the same analysis on the velocity dispersion \sigline\ (see the bottom-left panel in Fig. \ref{fig:globalvslocal}) we find a large scatter ($\sim 0.14$ dex)  and  usually weak, if any, correlations ($P>0.01$)  between the \local and \glob measurements in \Halpha, \mgii\ and \civ. On the other hand, the \Hbeta\ line shows a much tighter correlation ($r_{\rm s}<0.78$, $P=2\times10^{-9}$) but the scatter is still very large ($\sim 0.12$ dex). These results indicate a strong and perhaps non-linear dependency between the measured \sigline\ and the level of its local continuum.  As a result  \sigline-based determinations of \Mbh\ may be unreliable for data of limited spectral coverage. In particular,  such estimates may suffer from higher systematic uncertainties compared to those based on \fwhm.

\subsubsection{Black hole mass biases}
In \S\S\ref{subsec:mass} we will describe in detail the methods that we follow for  \Mbh\ calibration using the \local and \glob approaches. 
The form of the virial mass estimator (see Eqn. \ref{eqn:virial_eqn}) indicates that biases in \Mbh\ determinations are mainly driven by  the (small) line width biases. 
This is not the case for the continuum luminosity  since one can, in principle, re-calibrate the $\RBLR-L$ relations to use either one of the \local or \glob measured continuum luminosities, thus completely eliminating the systematic biases.

After following the procedure described in \S\S\ref{subsec:mass} and the strict virial assumption ($ \Mbh \propto \fwhm^2$) we found that the \Mbh\ median offsets ($\Delta \Mbh= \log M_{\text{BH, global}} / M_{\text{BH, local}}$)  are in very good agreement with our predictions, as can be seen in the bottom-right panel of Figure \ref{fig:globalvslocal} and are smaller than $\lesssim$0.04 dex (see Table \ref{tab:offsets}). 
From the values listed in Table \ref{tab:offsets} and from a visual inspection of Fig.~\ref{fig:globalvslocal}  one can conclude that \Halpha, \mgii, and \civ\ are consistent (within the scatter) with   $\Delta \Mbh$ being   independent of \Mbh. \Hbeta\ is again a bit more complicated, due to the difficulties we mentioned above. However, after removing the low-quality outliers we eventually find \Hbeta\ to be consistent with  $\Delta \Mbh$ being independent of \Mbh. 
Among all the lines  considered here, we find \mgii\ to be the one showing the smallest biases when following the \localapp.
This is somewhat surprising, given the several important spectral features (BC, FeII lines) that are influencing this spectral region.

\subsubsection{Line luminosity biases}

Line luminosities are  more sensitive to continuum placement than the other quantities we examined. Indeed, we found line luminosity median offsets ($\Delta L_{\text{line}}= \log L\left({\text{line}}\right)_{\text{global}}/ L\left({\text{line}}\right)_{\text{local}}$) of  $0.06^{+0.08}_{-0.08}$ dex , $0.03^{+0.06}_{-0.04}$ dex, $0.02^{+0.09}_{-0.08}$ dex  and $0.05^{+0.04}_{-0.06}$ dex  for \civ, \mgii, \Hbeta\ and  \Halpha, respectively. 
Furthermore, we find that the large scatter that is generally found in $\Delta L_{\civ}$, $\Delta L_{\mgii}$, $\Delta L_{\Hbeta}$  and $\Delta L_{\Halpha}$ is due to the fact that these quantities are anti-correlated with continuum luminosity. 
In particular, the relations between these line luminosity biases and \Lop\ show $r_{\rm s}$ correlation coefficients of  -0.38, -0.44, -0.65 for \civ, \mgii, \Hbeta\ and  \Halpha, respectively.  
This implies that using the local approach to estimate line luminosities generally leads to an underestimation of the latter, and its effect is larger for low luminosity objects (up to 0.14 dex, or 38\%, in the case of \civ).

In summary, the impact of using the \localapp  to estimate the local luminosities, lines widths and black hole masses when the \glob  continuum is unknown is found to be  small ($< 0.06$ dex). However, the impact using the \localapp to estimate \textit{line luminosities} is found be luminosity dependent, being stronger  for low luminosity objects. The
median values of $\Delta\fwhm$, $\Delta L$, $\Delta\Mbh$, and $\Delta L_{\text{line}}$ that we found are summarized  in Table~\ref{tab:offsets}.
Based on the general good agreement between \local and \glob measurements and in order to provide the community with strategies more applicable to observations with limited wavelength coverage, in \emph{the analysis that follows is based only on the \local measurements}, unless otherwise stated.

\begin{figure}
\centering
tf\includegraphics[width=0.45\textwidth,angle=0,trim=1cm 0cm 1cm 1cm]{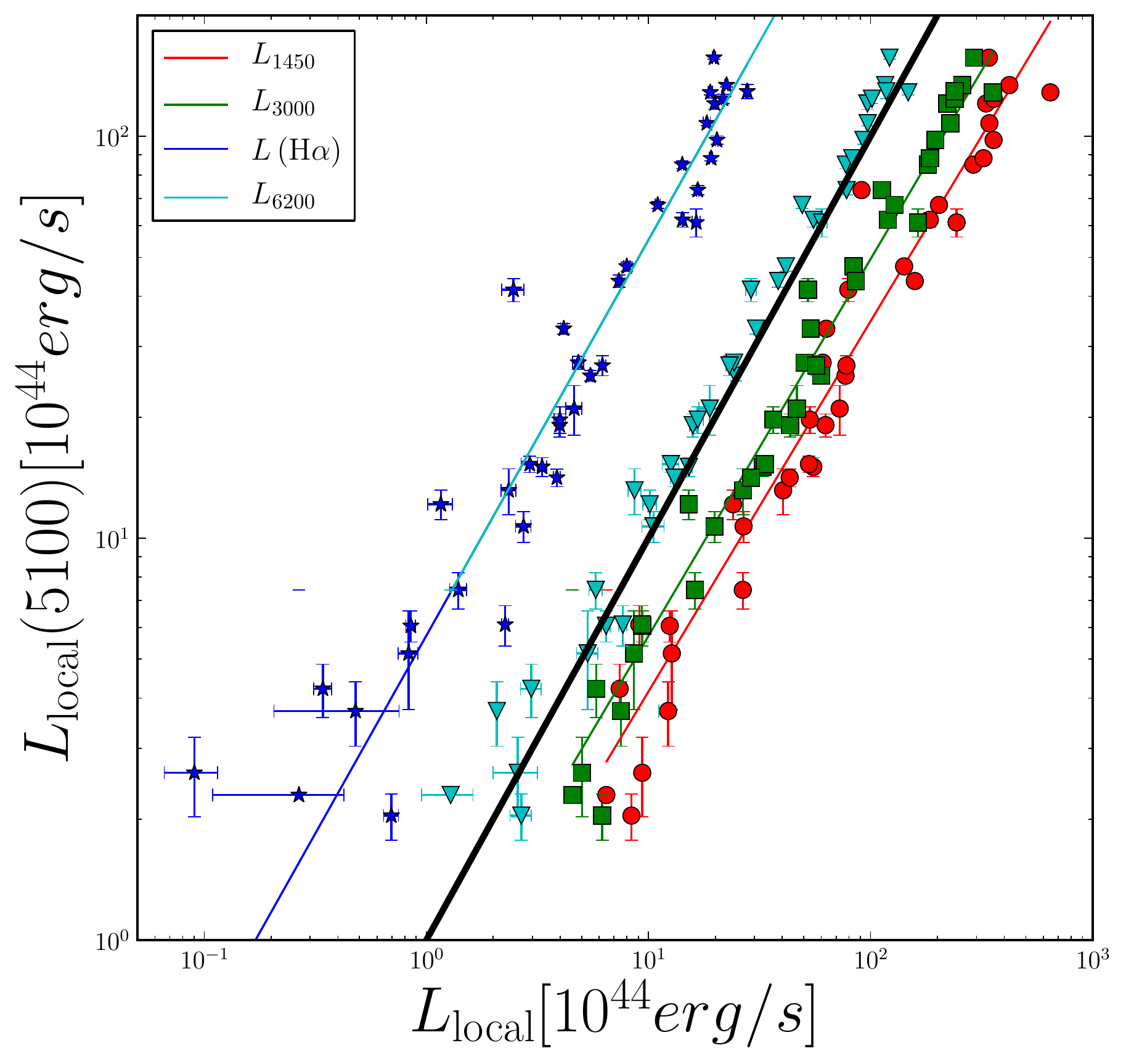}
\caption{  
\local \Halpha\  line luminosity (blue), \Lsix\ (cyan) \Lthree\ (green) and \Luv\ (red) vs  $\Lop^{\text{local}}$. The color solid lines represent the best linear fits to the corresponding data.  Black solid line represents 
the 1:1 relation.}  
\label{fig:Lscaling}
\end{figure}

\begin{figure*}
\centering
\includegraphics[width=0.49\textwidth,angle=0]{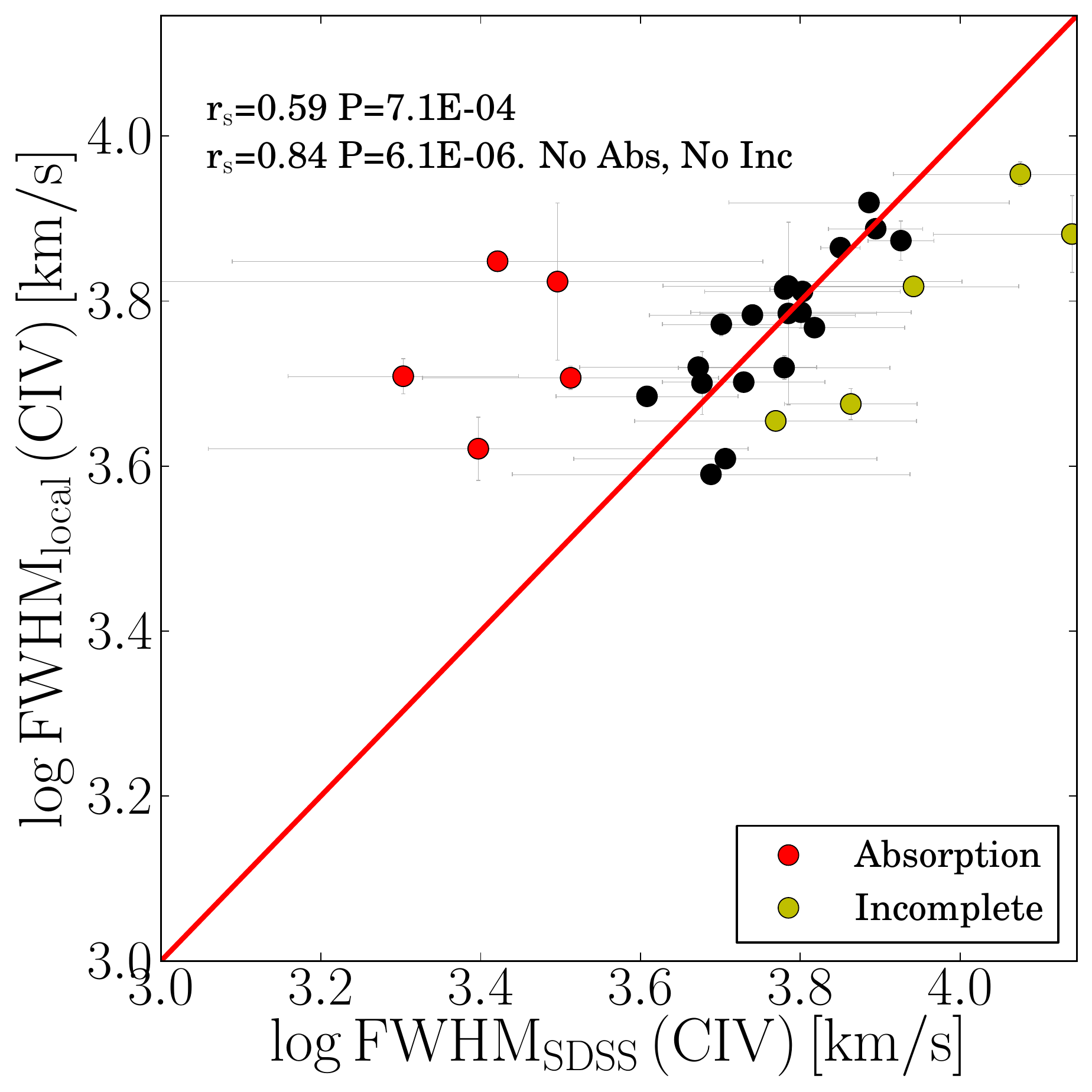}
\includegraphics[width=0.49\textwidth,angle=0]{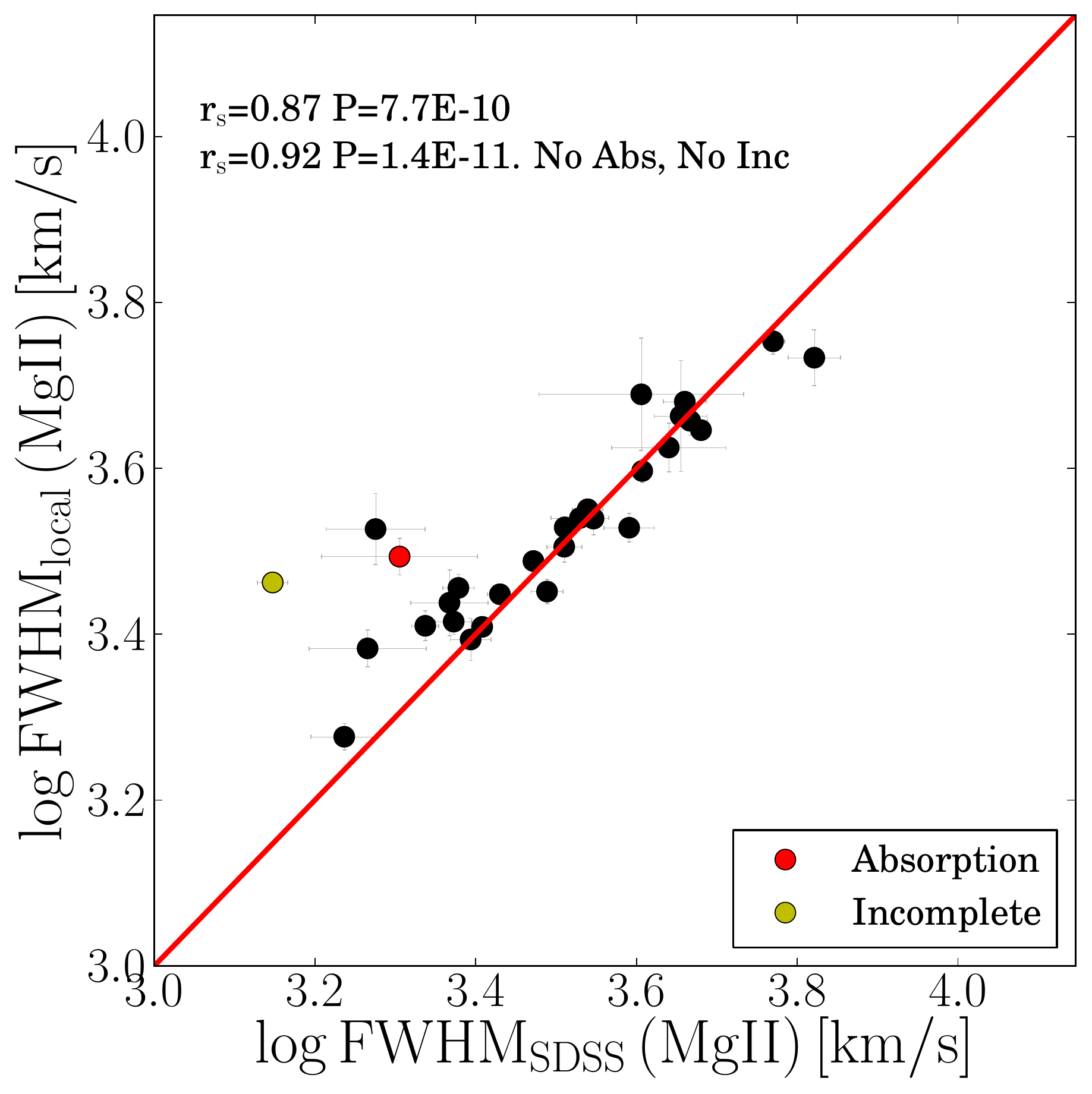}
\includegraphics[width=0.49\textwidth,angle=0]{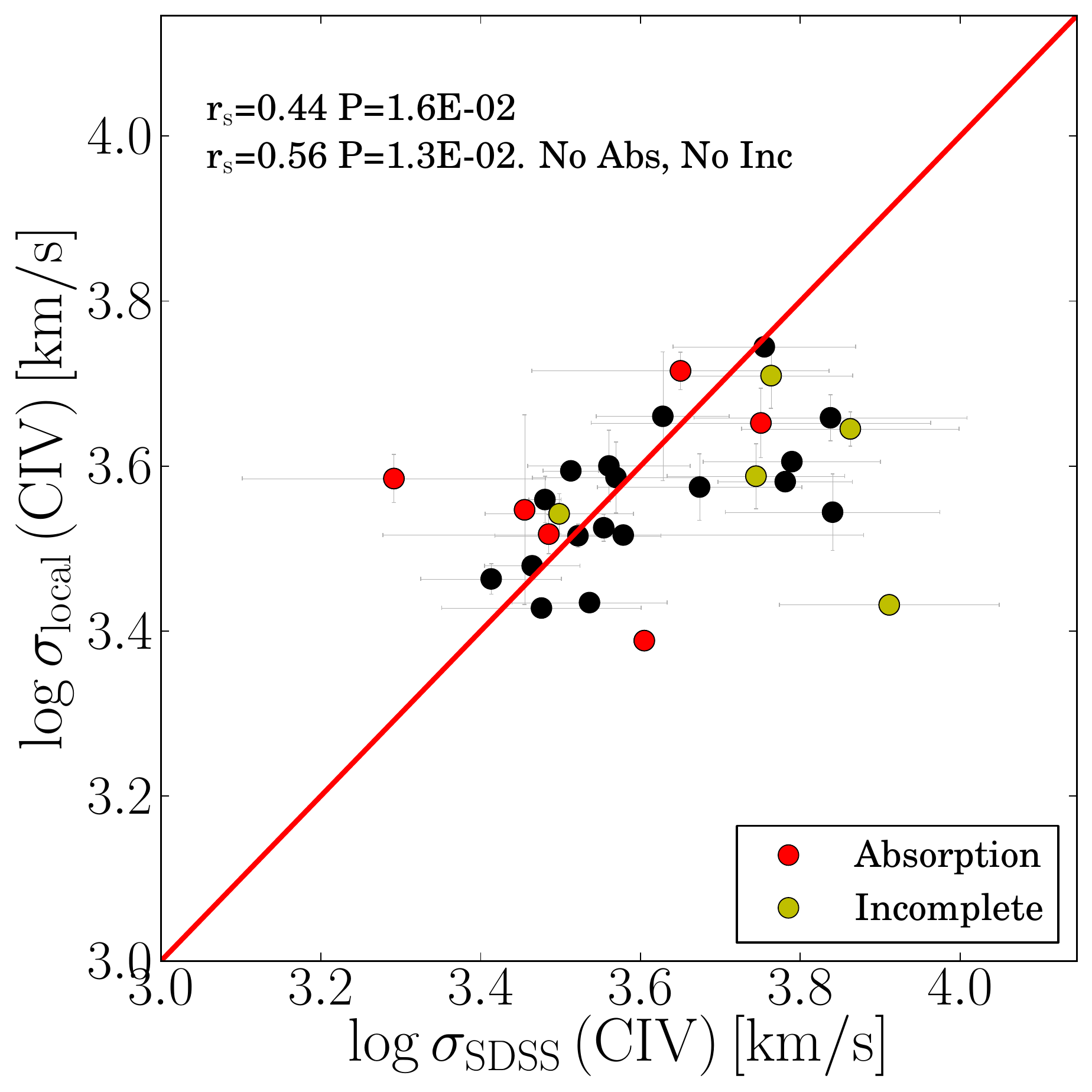}
\includegraphics[width=0.49\textwidth,angle=0]{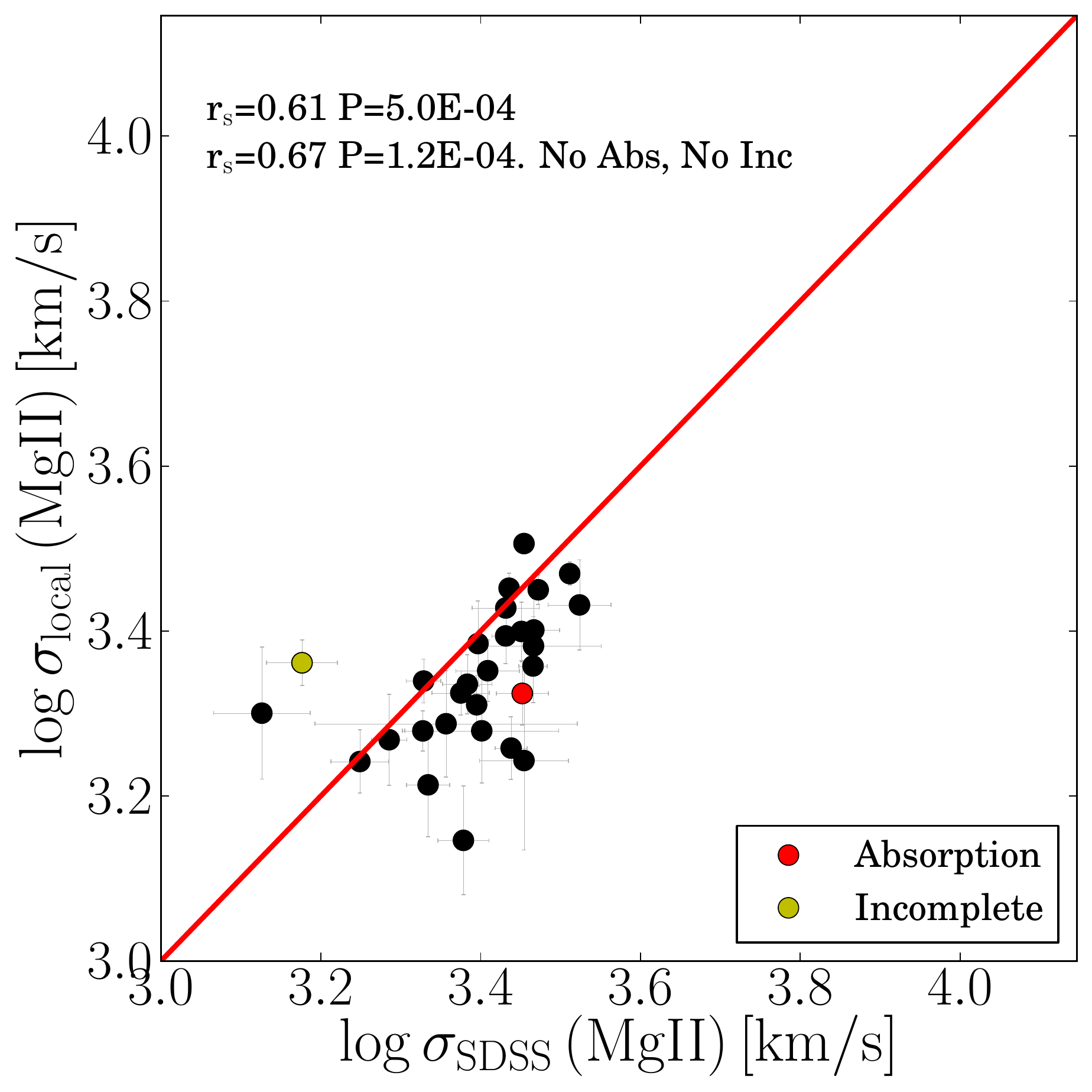}
\caption{ \fwhm s (top-panels) and \sigline s (bottom-panels) of the \civ\ (left) and \mgii\ (center) 
profiles found with low quality SDSS data versus those measured in our high quality X-Shooter 
using the \local continuum approach. 
The right panels shows \civ\ vs \mgii\ line widths measured with SDSS data. 
Red solid line shows the 1:1 relation and red dashed line
represents $\fwciv=\sqrt{3.7}\,\fwmg$. 
Red dots represent objects with noticeable absorption features while 
yellow dots are objects with SDSS incomplete profiles.  It can be 
seen that \civ\ profiles with strong absorption features artificially 
populate the zone where  $\fwciv<\fwmg$. }
\label{fig:CIVabs}
 \end{figure*}

\subsection{Luminosity Correlations}
\label{subsec:L}

Figure \ref{fig:Lscaling} presents a comparison between \Lop\ and  the luminosity indicators most commonly used in the context of \Mbh\ estimates.  
 The best-fit parameters of all the correlations can be found in Table \ref{tab:LLtable}. 
These relations provide us with the links necessary to connect  each luminosity indicator and $\RBLR\left(\Hbeta\right)$, through the $\RBLR-\Lop$ relation obtained from reverberation mapping experiments \citep{Kaspi2000,Kaspi2005,Bentz2009,Bentz2013}. 
For the purposes of the present work, we use the same calibration as  in TN12, which is appropriate for sources with $\Lop \gtrsim 10^{44}\, \ergs$:

\begin{equation}
\RBLR \left(\Hbeta\right) = 538 \left(\frac{\Lop}{10^{46}\,\ergs}\right)^{0.65}\,\,\text{lt-days} 
\label{eqn:RL}
\end{equation}

As shown in Fig. \ref{fig:Lscaling}  the  \Lha-\Lop\ relation  shows a larger scatter than those involving UV continuum luminosities (\Luv-\Lop\ and \Lthree-\Lop). This may therefore contribute to an increased uncertainty in \Lha-based determinations of \Mbh. 
This is not surprising, given the expected range of conditions in the BLR.
Consequently, we also investigate use  of \Lsix\ as an alternative to \Lha. 
As can be seen in Fig. \ref{fig:Lscaling} (cyan inverted triangles) the \Lsix-\Lop\ relation shows an even smaller scatter than \Luv\ and \Lthree. 
This is particularly the case for objects with $\Lop \gtrsim 10^{45}\,\ergs$,  where host galaxy contribution is negligible.

\Luv-\Lop\ and \Lthree-\Lop\ luminosity correlations are supra-linear, in the sense of showing $L \propto \Lop^{\beta}$ and $\beta>1$ (see first column of Table \ref{tab:LLtable} and note that $\beta=\gamma^{-1}$). This has been noted earlier by \citet{VandenBerk2004} but is in contrast to \citet{ShenLiu2012} who found consistency with $\beta=1$ in the sample of high luminosity quasars ($\Lop\left[\ergs \right] >10^{45.4}$).

While there are various correlations with \MbhHa\ and \LLedd\ (measured from \Halpha)  that can, perhaps, explain these differences, it is important to note that our sample is by no means complete. It was chosen to sample the high-L z$=$1.55 AGN population by giving equal weight to a group of sources with the same \Mbh\ and \LLedd\ (see paper I). Hence, the relationship found here should be checked in a larger and complete sample that represents the entire AGN population.

\begin{table}
\centering
\tabcolsep=0.12cm 
\begin{tabularx}{0.47\textwidth}{ccccccccc}

\hline
	&    \multicolumn{2}{c}{$\Lop^{\text{local}}$ vs $L_{\text{local}}$ $^a$}   &  \multicolumn{2}{c}{$\Lop^{\text{local}}$ vs $L_{\text{global}}$ $^b$}   &   \multicolumn{2}{c}{$L_{\text{global}}$ vs $L_{\text{local}}$ $^c$}     \\ 
\hline    
	& $\gamma$ & A & $\gamma$ & A  & $\gamma$ & A   \  \\
	L(\Halpha) & 1.04 & 4.73 & 1.09 & 3.59 & 0.94 & 1.32  \\ 
    \Lsix & 0.98 & 1.23 & 0.94 & 1.57 & 0.96 & 1.28  \\ 
	\Lop & 1 & 1 &  0.89 & 1.61 & 1.15 & 0.60 \\
	\Lthree & 0.92 & 0.67 &  0.91   & 0.77  & 1.01 & 0.88 \\
	\Luv & 0.88 & 0.56 &  0.87 & 0.64  & 1.02 & 0.87\\ \hline
\end{tabularx}
\caption{ Best fit power law parameters to the following relations: $^a$\ $\Lop^{\text{local}} = A L_{\text{local}}^{\gamma}$, $^b$\  $\Lop^{\text{local}} = A L_{\text{global}}^{\gamma}$,$^c$\ $L_{\text{global}} = A L_{\text{local}}^{\gamma} $. }
\label{tab:LLtable}
\end{table}

\begin{figure*}
\includegraphics[width=0.92\textwidth,angle=0,trim=0cm 0.2cm 0cm 0.5cm]{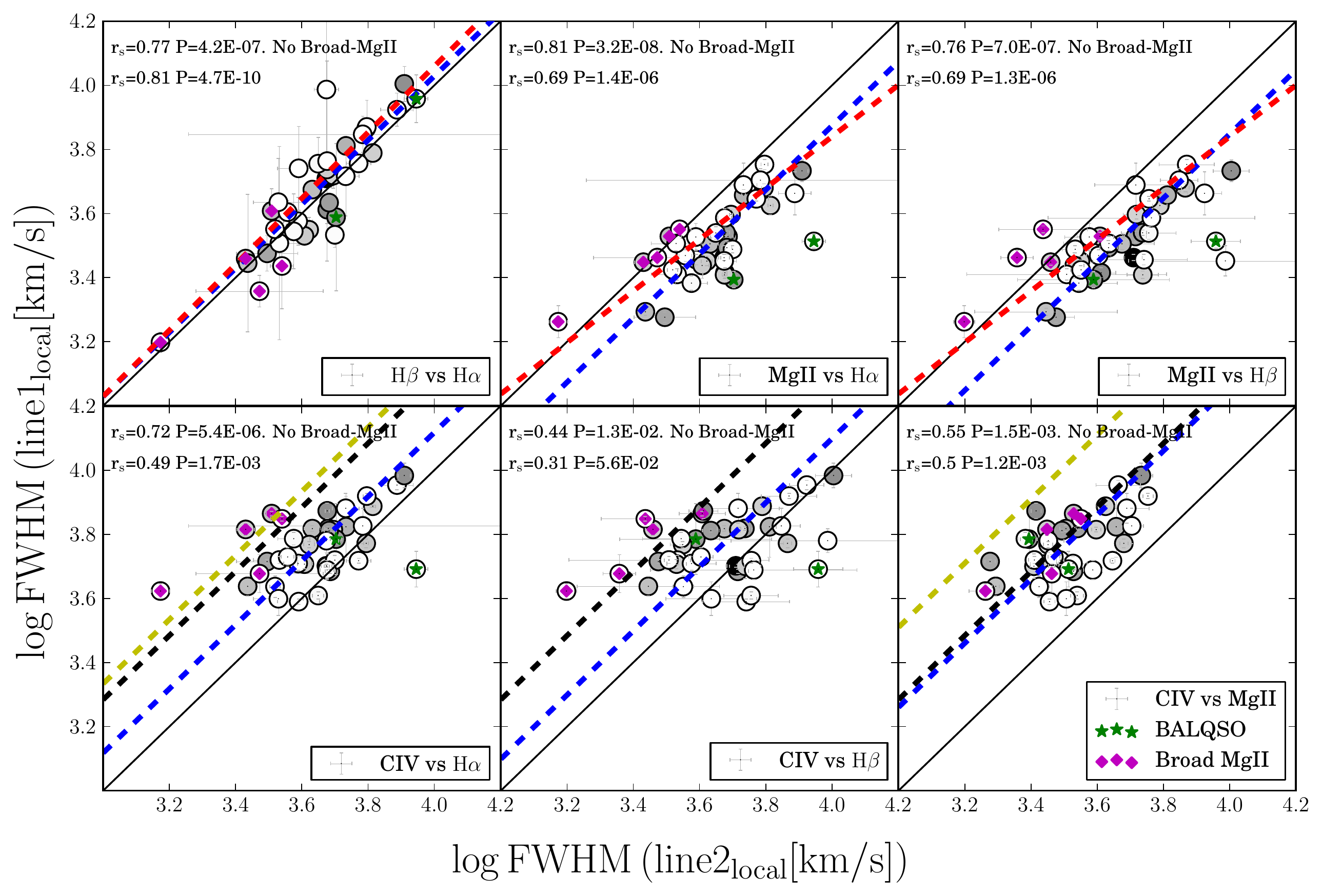} \\
\includegraphics[width=0.92\textwidth,angle=0,trim=0cm 0.5cm 0cm 0.2cm]{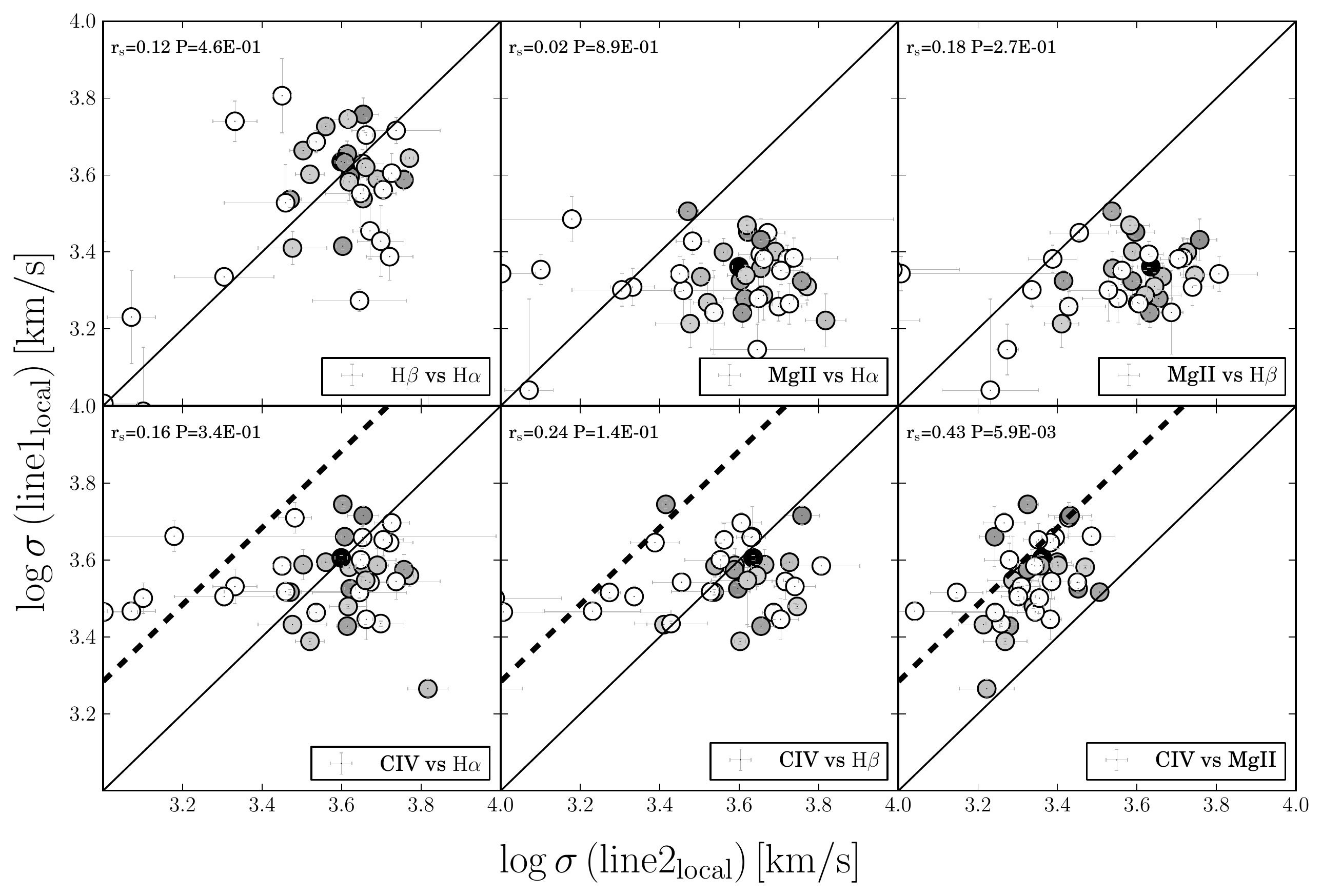}
\caption{ \fwhm\ (top) and $\sigma$ (bottom) comparisons between different lines in the {\bf local} continuum approach as indicated in the inserts of each panel (line1 vs line2).  The black solid lines represents the 1:1 relation.  The black dashed line represents $\fwciv = \sqrt{3.7}\fwhm\left(\text{line} \right)$. The yellow dashed line represents $\fwciv = \sqrt{3.7}\fwhb$ after rescaling the \fwhm\ of each line to \fwhb\ using the median value of $\fwhb/\fwhm\left(line\right)$. Red dashed lines represent previous scaling relations (\fwhb\ vs \fwmg\ from TN12, and \fwhb\ vs \fwha\ from \citet{GreeneHo2005}).  Blue dashed lines represent the best fit after assuming \fwhb$\propto$\fwha$\propto$\fwmg$\propto$\fwciv. Points are color-coded  in gray scale by the S/N of the continuum bands around \Hbeta\  where darker colors translates into larger S/N. Broad absorption lines quasars (BALQSO, green stars) and the broad-\mgii\ objects (magenta diamonds,  see \S \ref{subsec:broadmg})   are the main sources of discrepancies of the \civ\ and \mgii\ \fwhm s when compared  to the \Halpha\ and \Hbeta\ \fwhm s.  }
\label{fig:FWHMlocal}
\end{figure*}

\subsection{Line widths and line offsets}
\label{subsec:fwhm}

\subsubsection{Comparison with SDSS data}
\label{subsec:civabs}

At the redshift range of our sample, the archival SDSS spectroscopy covers both the \civ\ and \mgii\ lines in 29 out of 39 objects.\footnote{For the remaining 10 objects, the only archival spectroscopy available is from the 2SLAQ survey, which is of limited $S/N$ and is not flux calibrated.} {  In  Figure. \ref{fig:XshvsSDSS} we show an example of the SDSS and X-Shooter spectra  in the overlapping region.}
Comparing SDSS and X-shooter data allow us to test the effects of having only survey-grade data, with limited S/N and spectral resolution, on the measurement of line widths.
To this end, we used our \civ\ and \mgii\ fitting code for the lower quality archival SDSS DR7 spectra. 
In  Fig. \ref{fig:CIVabs} we compare  the \fwhm\ (top-panels) and \sigline\ (bottom-panels) values of the \civ\ and \mgii\ lines obtained from the SDSS data, with those  obtained from our higher quality spectra under the \local approach.\ {  We also show the Spearman correlation coefficients and corresponding  $P$-values  in each panel. }

We find that SDSS-based \fwciv\  for objects with absorption features which are unresolved in the SDSS data (4 out of 29 objects, red symbols), or those with partially-observed profiles because of the limited SDSS wavelength coverage  (5 out of 29, yellow symbols) result in \fwhm\ measurements which are systematically  different from those obtained from the higher quality data. 
Specifically, while \emph{unresolved absorption} features are likely to result in a systematic \emph{underestimation} of \fwciv, by about  $50\pm10\%$, \emph{incomplete} profiles are likely to lead to a systematic \emph{overestimation} of \fwciv, by about  $40\pm20\%$.  
This result was found in previous works \citep[e.g.][]{Denney2013,Park2013,Tilton2013} and could  explain, to some extent,  the over-population of narrow \civ\ objects that is reported in TN12. 
The \mgii\ line  does not  generally show strong absorption features. Indeed, we find  that the SDSS-based \fwmg\ measurements are generally consistent with our higher quality \fwmg\ measurements with the exception of five objects. Of these 5 objects, three have very low S/N, one has an incomplete profile, and one shows signs of absorption.

Looking into the corresponding comparison with \sigline\ (bottom panels of Fig. \ref{fig:CIVabs}), we generally find that sources with absorption features or incomplete profiles do not stand out from the ``normal'' population. 
The entire sample shows considerable scatter when comparing the SDSS {  and X-Shooter line measurements and show less signifcant correlations  than the   \fwhm (top panels of Fig. \ref{fig:CIVabs})}. 
For $\sigma\left(\mgii \right)$, we find the SDSS measurements to be systematically broader than our $\sigma_{\text{X-Shooter}}\left(\mgii \right)$ estimations, and the scatter is larger than the one in the \fwhm\ comparison.
For $\sigma\left(\civ\right)$, there is a large dispersion (0.2 dex) between  SDSS and X-Shooter measurements, that could be caused by the high sensitivity of \sigline\ measurements to continuum placement.

We conclude that the usage of \sigline\ to measure line width in data of limited quality introduces significant scatter. For such data, the use of \fwhm\ is preferred, especially for the \mgii\ line. In addition,   the absorption features often seen in the \civ\ line necessitate the use of high-quality spectra, in order to resolve and properly account for these features, even if one uses \fwhm\ instead of \sigline.

\begin{table*}
\centering
\caption{Line width ratios and correlations.\\
For each pair of lines, we list median values and scatter of  ${\rm Q}\equiv \log \left( \fwhm\left( \text{line1} \right)/\fwhm\left( \text{line2} \right) \right)$ and the Spearman correlation coefficients between$\fwhm\left( \text{line1} \right)$ and $\fwhm\left( \text{line2} \right)$. 
We tabulate these quantities for both the complete sample (under the \localapp), and after excluding the five broad-\mgii\ and the two BALQSO.}
\label{tab:fwhm}
\tabcolsep=0.12cm
\begin{tabular}{ccccccccccccccccccc}
\hline
 & \multicolumn{6}{l}{-------------------------\Halpha-------------------------} & 
\multicolumn{6}{c}{-------------------------\mgii-------------------------}  & 
\multicolumn{6}{c}{-------------------------\civ-------------------------} \\ 
 & 
\multicolumn{3}{c}{-------All objects $^{\rm a}$-----} & 
\multicolumn{3}{c}{----No Broad-\mgii $^{\rm b}$--} & 
\multicolumn{3}{c}{-------All Objects $^{\rm a}$-----} & 
\multicolumn{3}{c}{----No Broad-\mgii $^{\rm b}$--} & 
\multicolumn{3}{c}{-------All Objects $^{\rm a}$-----} & 
\multicolumn{3}{c}{----No Broad-\mgii\ $^{\rm b}$--} \\ 
line1 & 
\multicolumn{1}{c}{Q} & 
\multicolumn{1}{c}{scatter} & 
\multicolumn{1}{c}{$r_{\text{s}}$} & 
\multicolumn{1}{c}{Q} & \multicolumn{1}{c}{scatter} & \multicolumn{1}{c}{$r_{\text{s}}$} & \multicolumn{1}{c}{Q} & \multicolumn{1}{c}{scatter} & \multicolumn{1}{c}{$r_{\text{s}}$} & \multicolumn{1}{c}{Q} & \multicolumn{1}{c}{scatter} & \multicolumn{1}{c}{$r_{\text{s}}$} & \multicolumn{1}{c}{Q} & \multicolumn{1}{c}{scatter} & \multicolumn{1}{c}{$r_{\text{s}}$} & \multicolumn{1}{c}{Q} & \multicolumn{1}{c}{scatter} & \multicolumn{1}{c}{$r_{\text{s}}$} \\ \hline
\Halpha\ &  ...      & ...    & ...    & ...    & ...               & ...      & 0.13 & 0.08 & 0.69 & 0.13 & 0.07 & 0.81                 & -0.12 & 0.14 & 0.48 & -0.11 & 0.10 & 0.72 \\ 
\mgii\      & -0.13 & 0.08 & 0.69 & -0.13 & 0.07 & 0.81  & ...     & ...      & ...      & ...      & ... & ... & -0.26 & 0.10 & 0.50 & -0.25 & 0.10 & 0.55 \\ 
\Hbeta\  & 0.03 & 0.07 & 0.81 & 0.04 & 0.07 & 0.77 & 0.15 & 0.10 & 0.69 & 0.16 & 0.08 & 0.88                                & -0.10                  & 0.17                         & 0.31                                & -0.09                  & 0.13                         & 0.44                                \\  
\hline\\
\end{tabular}
\end{table*}

\subsubsection{Line Offsets}
\label{subsec:lineoff}

We  measured  the line offsets with respect to the laboratory wavelengths of \Halpha, \Hbeta\ and \mgii. Their absolute values ($\abs{\Delta v}$) are found to be  (within the 16\% and 84\% percentiles) smaller than  600\,\kms, 550\,\kms\ and 250\,\kms\ respectively. 

Many of the observed \civ\ lines show large negative velocity offsets  ($\Delta v \simeq -1200 \pm 1000\ \kms$) suggesting non virial equilibrium of the \civ\ emitting clouds. This has been noted in numerous earlier publications, \citep[e.g.][]{Shang2007,Wang2009,ShenLiu2012,TrakhtenbrotNetzer2012,Runnoe2013,Brotherton2015}.  Moreover, the \civ\ velocity offsets are anti-correlated with \LLedd\ ($r_{\rm s}=-0.53$, $P=0.0004$), i.e., higher \LLedd\ will translate into bluer line centers \citep[e.g.][]{Marziani2006,Sulentic2007}. 
We also find that the much smaller velocity offsets of the \mgii\ lines  are also anti-correlated with \LLedd\ ($r_{\rm s}=-0.49$, $P=0.001$) which is also in agreement with \citet{Marziani2013a}. 
 We repeated the analysis using the normalized accretion rate 
($\dot{m}\equiv L_{\rm model}/L_{\rm Edd}\left[\Mbh^{\rm model }\right]$) taken from the best-fit AD models (to be presented in paper~III; see paper~I for details). 
We find that our measured $\dot{m}$, too, is anti-correlated with \civ\ velocity offsets ($r_{\rm s}=-0.49$, $P=0.001$), however the analogous correlation with \mgii\ velocity offsets becomes insignificant  ($P=0.07$).
{These results suggest that \LLedd\ is playing an important role in the line offsets of the \civ\ profile, while \mgii\ velocity offsets may involve additional parameters.  As explained earlier, the way we selected our sample makes it difficult to make strong statements regarding the entire population of AGN.}
When the same analysis is done with the Balmer lines,  we find no correlation between neither \LLedd\ nor $\dot{m}$ and  the Balmer line velocity offsets ($P=0.26$ and  $P=0.90$, for \Halpha\ and \Hbeta, respectively). 
We further confirm earlier results \citep[e.g.,][]{Corbin1990,Richards2011} of a significant anti-correlation between the \civ\ blueshifts and  the \civ\ line strength, EW(\civ)  ($r_{\rm s}=0.43$, $P=0.006$), but not with EW(\mgii)  ($r_{\rm s}=0.25$,$P=0.12$).

Several studies investigated the possibility that broad emission lines are gravitationally red-shifted by few hundred  to few thousand \kms\ \citep[e.g.][]{Netzer1977,ZhengSulentic1990,Popovic1995,Muller2006,Tremaine2014}.  
This effect is enhanced  in very broad emission line components  ($\fwhm\gtrsim 7000\,\kms$)  that are formed close to the BH. In this work we made no attempt to include this in the modeling of the line profiles since we are mainly after the line \fwhm\ which is insensitive to such small variations. We verified, however, that line offset due to this effect are smaller than the general uncertainty and scatter associated with our measurements of the line center velocity. We  come back to this issue in paper IV (Mejia-Restrepo et al, in preparation).

\subsubsection{Line width correlations}

Figure \ref{fig:FWHMlocal} presents a comparison between the widths of some of the broad emission lines in our X-shooter observations, in terms of FWHM (top panel) and line dispersion (\sigline; bottom panel).  For reference, we also illustrate the 1:1 relation (black solid line), and a constant scaling of  $\fwciv = \sqrt{3.7}\,\fwhm\left(\Halpha,\Hbeta,\mgii \right)$ (black dashed line).
The latter scaling is motivated by the typical ratio of the corresponding BLR sizes for \Hbeta\ and \civ, as measured in RM experiments, and under the virialized BLR assumption (see detailed discussion in TN12). 
We have  plotted in yellow a dashed line that represents  $\fwciv = \sqrt{3.7}\,\big \langle \frac{\fwhb}{\fwhm\left(\Halpha,\mgii \,\right)} \big \rangle  \,\fwhm\left(\Halpha,\mgii \right)$  to account for  the median \fwhm\ ratio between  \fwhb\ and the \fwhm\ of \Halpha\ and \mgii. 
Finally, we have color coded the points  in gray scale by the S/N of the continuum bands around \Hbeta\  where darker colors translates into higher S/N.  In Figures \ref{fig:app:profile_comp1} and \ref{fig:app:profile_comp2} of the  Appendix \ref{app:profile_comp}  we show the normalized \Halpha, \Hbeta,  \mgii\ and \civ\ observed line profiles in velocity space to provide the reader with a direct visual comparison of the most prominent emission lines. 
The large error bars in the \Hbeta\ line widths are due to the low signal to noise and the difficulty of constraining the iron emission around \Hbeta, because of the telluric absorption (see \S\ref{sec:data}). 

We generally find very good agreement between the \fwhm s of \Hbeta\ and \Halpha\ (Fig. \ref{fig:FWHMlocal} top-left panel). 
On average, \fwhb\ is broader than \fwha\ by  0.03 dex (see blue dashed line in Fig. \ref{fig:FWHMlocal}), with a scatter of about 0.08 dex. 
This result is in good agreement with several previous studies, as well as with the scaling relation reported in \citet{GreeneHo2005}  (see red dashed line in Fig. \ref{fig:FWHMlocal}).
 
We also find that objects with $ \log \fwhb\left[\kms\right] \leq 3.6$ ($\sim 4000\, \kms$)   show \fwhb\  slightly narrower than the median trend (i.e. below the blue dashed line in Fig. \ref{fig:FWHMlocal}) by about 0.04 dex (10\%). These  objects are however fainter and their values are less accurate because of the difficulties with \Hbeta\ measurements. This results is in agreement with \citet{Denney2009b} where they found that the estimated \fwhb\ in low quality  data ($S/N \lesssim 20$) is not reliable.

From Fig. \ref{fig:FWHMlocal} we can also see  that there are significant  correlations between the \fwhm s of:  1) \Halpha\ and \mgii\ (scatter of $\sigma_\Delta=0.08$ dex),  2) \Hbeta\ and \mgii\ ($\sigma_\Delta=0.10$ dex)  and  3) \Halpha\ and \Hbeta\  ($\sigma_\Delta=0.07$ dex)  in agreement with several previous works  \citep[e.g.][]{GreeneHo2005,Shang2007,Wang2009,ShenLiu2012,TrakhtenbrotNetzer2012,Marziani2013}. 
Also, \fwmg\  is  proportional to and narrower than \fwhb\ by 0.16 dex (30\%), with a scatter of about 0.08 dex and no dependence on \fwhb.  There are however some outliers in these general trends: The  two BALQSOs  (green dots in Fig. \ref{fig:FWHMlocal}) and 5 objects that show $\fwmg\gtrsim\fwhb$ and have high  \LLedd\ ($>0.17$, hereafter broad-\mgii objects,  magenta diamonds in Fig. \ref{fig:FWHMlocal}). These 7 objects and their implications in the \fwmg-Balmer lines correlations are further discussed  in \S \ref{subsec:broadmg}.

From the discussion above  it is reasonable to assume that the emissivity  weighted \mgii\ region is more distant from the central BH than the corresponding regions for the \Halpha\ and \Hbeta\ lines. On the other hand,   both Balmer lines seem to come from the same part of the BLR. As a consequence and based on the \fwhm\ linear correlation among \Halpha,\Hbeta\ and \mgii, assuming  virialization of \Hbeta\ would reasonably imply virialization of \mgii\ and \Halpha.

The correlations of \fwciv\ with the measured \fwhm\ of the other lines are weaker, occasionally insignificant (i.e. P$>$0.01) and non-linear: 1) \Halpha\  ($r_{\rm s}=0.48$, P$=0.02$, $\sigma_\Delta=0.14$ dex), 2)  \Hbeta\ (insignificant, P$=0.05$) and  3)\mgii\ ($r_{\rm s}=0.50$, P$=0.001$, $\sigma_\Delta=0.10$ dex). This would mean that  \fwciv\ is not linearly proportional to the \fwhm\ of \Halpha, \Hbeta\ and \mgii. For example, $\fwciv\propto \fwha^{1.41\pm0.50}$.   
Moreover, when combining the results of  the RM experiments \citep[e.g.][]{Kaspi2007}  with the virial assumption, it is expected that the \civ\ line would be broader than 
\Hbeta, by a factor of about $\sqrt{3.7}$. 
\footnote{The scaling factor is somewhat luminosity dependent. See TN12 for a discussion of this issue.} 
In contrast,  the vast majority of sources in our sample (35/39; 90\%) show $\fwciv<\sqrt{3.7}\,\fwhb$ and  one third of the sources have \fwhb$>$\fwciv.
These results indicate either a non-virialized \civ\ emission region, or a very different ionization structure for objects with low and high \fwhb.

Finally, when we compare the velocity dispersion  (\sigline) between the lines of interest  (bottom panels of Fig. \ref{fig:FWHMlocal}) we  only find one significant  correlation between \fwmg\ and \fwciv\ ($r_{\rm s}=0.43$, P$=0.005$) in the \local approach. However, even this correlation does not hold under the \globapp ($P=0.36$).
Due to the fact that the correlations between the \fwhm\ of different lines are much tighter  than  the \sigline\ correlations (under both continuum  approaches), and  the fact that \sigline\ is  strongly affected by flux in the line wings, we  choose to use the \fwhm\  to estimate \Mbh\ in the analysis that follows.

\subsubsection{Broad-\mgii\ and BALQSO objects}
\label{subsec:broadmg}

As  discussed in \S\ref{subsec:fwhm} we  found that \mgii\ profiles are generally and systematically narrower than \Halpha\ and \Hbeta\ profiles. However, the top right and top center panels of Fig. \ref{fig:FWHMlocal} show that  around $ \log \left( \fwhb \right)$ and $\log \left( \fwha \left[\kms\right]  \right)  \lesssim 3.6$ ($\leq 4000 \kms$) there are a handful of objects (magenta diamonds) that show $\fwmg\ \gtrsim \fwhm\left(\Halpha,\Hbeta\right)$ and were noted earlier as ``broad-\mgii\ objects''.

\citet{Marziani2013a} and \citet{Marziani2013} presented a thorough Eigen-vector 1 analysis of the \mgii\ and \Hbeta\  profiles following \citet{Sulentic2002} from an SDSS selected sample of 680 quasars. Their classification is based on the location of type-I AGN in the $R_{\text{op}}$-\fwhb\ plane where   $R_{\text{op}}=L\left(\FeII\left( 4750\AA \right) \right)/L\left(\Hbeta \right)$. They claimed that the so called ``Broad-\mgii\ objects'' belong to the extreme population A category (A3 and A4 according the their classification, see Fig. 8 in \citet{Marziani2013}) and represents about 10\% of the total population of high luminosity AGN. 
These  extreme population A objects  have narrow \Hbeta\ profiles ($\leq 4000\, \kms$) and the highest $R_{\text{op}}$ values. They are also among the objects with the highest Eddington ratios and largest velocity offsets. 
Unfortunately, our difficulties to properly measure the \FeII\ emission around \Hbeta\ do not allow us to  measure $R_{\text{op}}$ and test their assumptions. can however compare their \LLedd\ estimates to our \Halpha-based \LLedd\  estimates by applying a bolometric  correction as described in TN12. {   As can be seen in Figure. \ref{fig:MLLedd}  all  these objects occupy the top 20 percentile of the \LLedd\  distribution in our sample ($\LLedd\geq0.20$) in agreement with \citet{Marziani2013}. The Broad-\mgii\ objects in our sample also show  relatively large \civ\ and \mgii\ velocity blue-shifts (top 20\%,  $\Delta v_{\text{Broad-\mgii}}\left(\civ\right) \lesssim -2200\,\kms$,  $\Delta v_{\text{Broad-\mgii}}\left(\mgii\right)\lesssim -200\, \kms$) which is also in agreement with \citet{Marziani2013}. We note however that broad-\mgii\ objects are not the only ones that meet the mentioned conditions. } 

As can be seen in Figure \ref{fig:FWHMlocal}, the BALQSOs in our sample show exactly  the opposite behavior. They show  narrower  \mgii\  profiles than usual. Unfortunately, it is impossible to draw any conclusion based on only two sources. 

In Table \ref{tab:fwhm} we present the median values and corresponding scatter  of  Q$\equiv \log \left( \fwhm\left( \text{line1} \right)/\fwhm\left( \text{line2} \right) \right)$ as well as the Spearman correlation coefficient between the \fwhm\ of the listed lines under two cases: a) including \textit{all} objects in the analysis and b) excluding  the broad-\mgii\ and the BALQSOs from the analysis. It can be seen  in Table \ref{tab:fwhm} and Fig. \ref{fig:FWHMlocal} that after removing these outliers the \fwhm\ correlations becomes tighter (i.e.   $r_{\text{s}}$  increases) and the Q factors remain almost unchanged. We emphasize that this result is also true for \fwciv while the correlations between \fwciv\ and the \fwhm\ of the Balmer lines approach to linearity after removing such 7 objects. Consequently, for the following \Mbh\ analysis we exclude both the 5 Broad-\mgii\ objects and the two BALQSOs.

\begin{figure*}
\centering
\includegraphics[width=0.99\textwidth,angle=0]{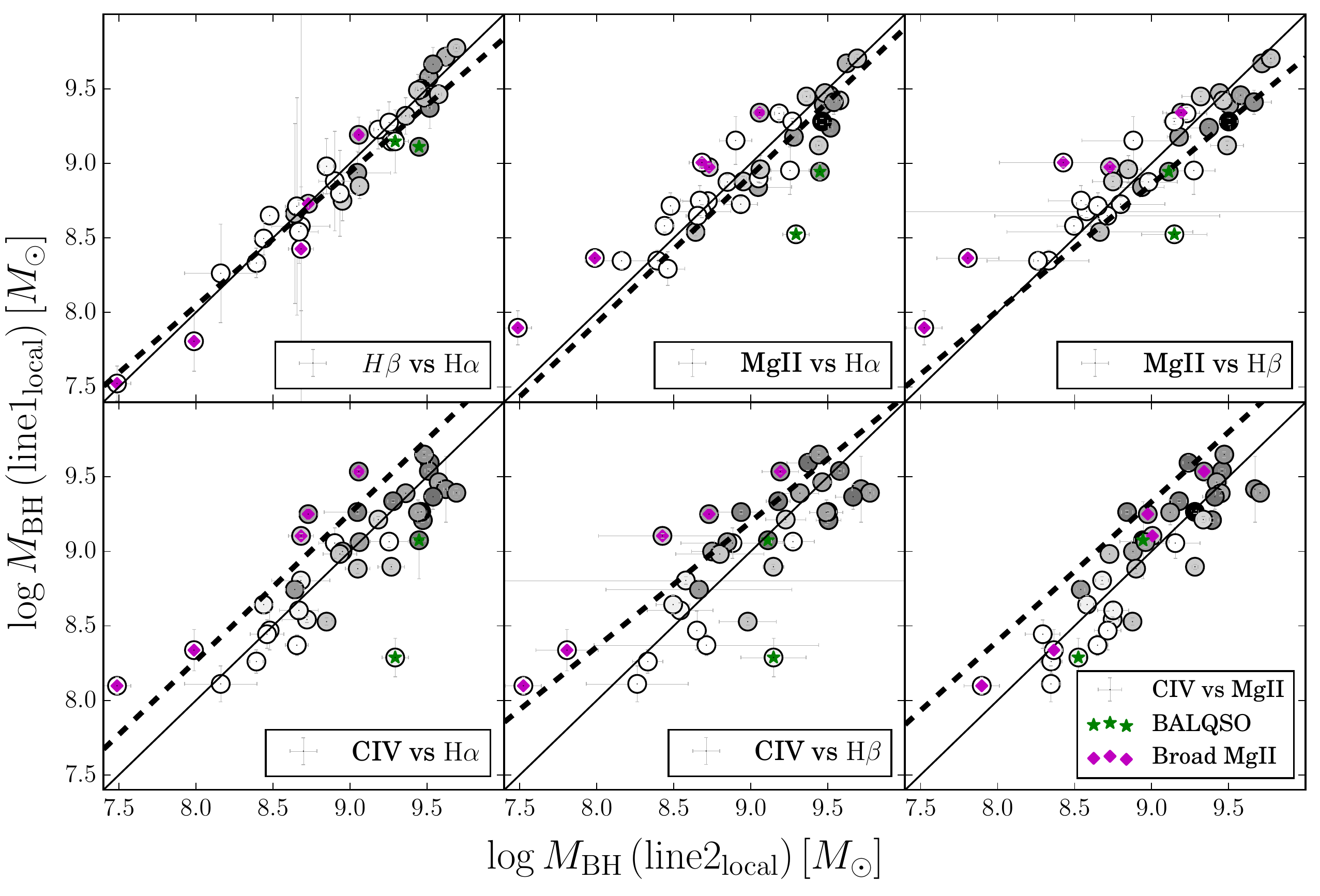}
\caption{
Comparisons between  different  \Mbh\ estimates that are derived from different lines as indicated in the inserts of each panel. The black solid line represents the 1:1 relation. \Halpha\ values were derived using \Lsix. The dashed black line represents the best fit to previous black hole mass estimators.  Points are color-coded  in gray scale by the S/N of the continuum bands around \Hbeta\  where darker colors translates into larger S/N. BALQSO  and the broad-\mgii\ objects (see \S \ref{subsec:broadmg})  are the labeled by green stars and magenta diamonds respectively.
}
\label{fig:mass}
\end{figure*}

\subsection{Black Hole Mass estimators}
\label{subsec:mass}

In this subsection we present the procedure we use to obtain, and compare, different \Mbh\ estimates using the different line and continuum measurements. Our starting point, and the basis for all the following correlations,  is the sub-sample of 32 AGN obtained by removing from the original sample 5 sources showing large discrepancy between \fwhb\ and \fwmg\ (see \S \ref{subsec:broadmg}) and the two BALQSOs in the sample. A major aim is to find a practical strategy that will allow the identification of sources that are not suitable for accurate mass determination based on single line and continuum measurement. 

\subsubsection{$\Lop-\RBLR\left(\Hbeta\right)$ relation and \Hbeta }

Most present-day  single epoch mass measurements are  based on the 
$\RBLR \left(\Hbeta \right)$-$\Lop^{\text{local}}$ relation,  established through RM experiments (see \S \ref{sec:intro} and Eq.\ref{eqn:RL}).
In this case $\Lop^{\text{local}}$ is a \local estimation of the continuum and \RBLR\ is  obtained from the time lag of the response of the \Hbeta\ line to (optical) continuum variations. 
This lag is assumed to properly represent the emissivity weighted radius of the broad \Hbeta\ line. 
\Mbh\ is obtained  from equation \ref{eqn:virial_eqn} where both  $\fwhb_{local}$ and
$\Lop^{\text{local}}$ are obtained using \textit{ local} continuum measurements.  These values can be used to obtain the "local" BH mass estimate, \Mbh$_{local}$. We can then use the expressions derived in \S \ref{subsec:L}, and the various biases between the \local and \glob \Lop\ and \fwhm, to derive a \glob expression for \MbhHb. 

We start by using the local \MbhHb\ expression obtained by TN12. This expression is most appropriate for our intermediate and high luminosity AGN:
{\small
\begin{equation}
\Mbh\left(\Hbeta\right)_{\text{local} }=5.26 \times 10^{6}\,\Msun\,\left( \frac{\Loploc}{10^{44}\,\ergs}\right)^{0.65} \left( \frac{\fwhbloc}{10^{3}\,\kms}\right)^{2} \,\,  ,
\label{eqn:mhblocal}
\end{equation}
}
Obtaining the equivalent \glob expression is not trivial since  we need first to find a relation between \RBLR\ measured from RM and  $\Lop^{\text{global}}$ and not simply use the recipe that connects local measurements. 
However, we do not know   $\Lop^{\text{global}}$  for the objects targeted by RM campaigns  and we have to rely on the scaling relation between $\Lop^{\text{local}}$ and  $\Lop^{\text{global}}$ that we find in this work (see table \ref{tab:LLtable}). Substituting in Eq.~\ref{eqn:RL} we get:  
{\small
\begin{equation}
\Mbh\left(\Hbeta\right)_{\text{global}}=7.17\times 10^{6}\,\Msun\,\left( \frac{\Lopglob}{10^{44}\,\ergs} \right)^{0.58} \left( \frac{\fwhbglob}{10^{3}\,\kms}\right)^{2} \,\, ,
\label{eqn:mhbglobal}
\end{equation}
}
It is important to note  that we have simply re-scaled the empirical $\RBLR \left(\Hbeta \right)$ vs $\Lop^{\text{local}}$ relation to a  $\RBLR \left(\Hbeta \right)$ vs $\Lop^{\text{global}}$ relation that is  adjusted to  predict the \emph{same} \RBLR\ measurements. 
Consequently, we do not expect any systematic bias in \Mbh\ measurements coming from intrinsic $\Lop^{\text{global}}$-$\Lop^{\text{local}}$ biases. The bias between \Mbh$_{\text{local}}$ and \Mbh$_{\text{global}}$ are simply the results of  the intrinsic differences between the $\fwhm_{\rm local}$ and $\fwhm_{\rm global}$  (see \S \ref{subsec:globalvslocal}). The small \Mbh\ biases that we found are shown in the bottom right set of panels in Figure \ref{fig:globalvslocal}. 

\begin{table*}
\centering
\tabcolsep=0.4cm 
\begin{tabularx}{\textwidth}{cccccccccc}
\hline
                      & \multicolumn{3}{c}{-----------------Local$^{a}$-----------------}                        & \multicolumn{3}{c}{-----------------global$^{b}$-----------------}                          & \multicolumn{3}{c}{-----------------Local$_{\rm corr}^{b}$-----------------}           \\ 
                      & $\log{K}$ & $\alpha$ & scatter               & $\log{K}$ & $\alpha$ & scatter            & $\log{K}$ & $\alpha$ & scatter             \\
                      &           &          & \multicolumn{1}{c}{(dex)} &           &          & \multicolumn{1}{c}{(dex)} &           &          & \multicolumn{1}{c}{(dex)} \\ \hline
\fwha, \Lop           & 6.779     & 0.650    & 0.16                      & 6.958     & 0.569    & 0.19                      & 6.845     & 0.650    & 0.16                      \\ 
\fwha, $L_{\rm 6200}$ & 6.842  &  0.634        &    0.16                 & 7.062     & 0.524    & 0.22                           &    6.891       &   0.634         &     0.16                       \\ 
\fwha, \Lha           & 7.072     & 0.563    & 0.18                      & 7.373     & 0.514    & 0.23                      & 7.389     & 0.563    & 0.18                      \\ 
\fwhb, \Lop           & 6.721     & 0.650    & 0.00                      & 6.864     & 0.568    & 0.00                      & 6.740     & 0.650    & 0.00                      \\ 
\fwmg, \Lthree        & 6.906     & 0.609    & 0.25                      & 6.955     & 0.599    & 0.29                      & 6.925     & 0.609    & 0.25                      \\ 
\fwciv, \Luv          & 6.331     & 0.599    & 0.33                      & 6.349     & 0.588    & 0.38                      & 6.353     & 0.599    & 0.33                      \\     \hline                                      
\end{tabularx}
\caption{Virial BH mass calibrations of Equation \ref{eqn:virial_eqn} ($\Mbh = K (L_{\lambda} )^{\alpha} \fwhm^{2}$)  based on different line width and luminosity combinations for  32/39 objects in our sample, calibrated against the  \Hbeta\ virial mass calibration given in Equation. \ref{eqn:mhblocal}. $^{a}$ \Mbh\ calibration based on \local measurements. $^{b}$ \Mbh\ calibration based on \glob measurements. $^{c}$  \local \Mbh\ calibrations corrected  for the  \emph{small} systematic offsets that we found with respect to \glob \Mbh. Note that the values in this table are valid for L in units of $10^{44}$ \ergs\ and \fwhm\ in units of $1000$ \kms. {  For these calibration we assume $f=1$  which is appropriate for \fwhm\ \Mbh\ estimates.}}
\label{tab:mass}
\end{table*}

\begin{figure*}
\includegraphics[width=0.49\textwidth,angle=0]{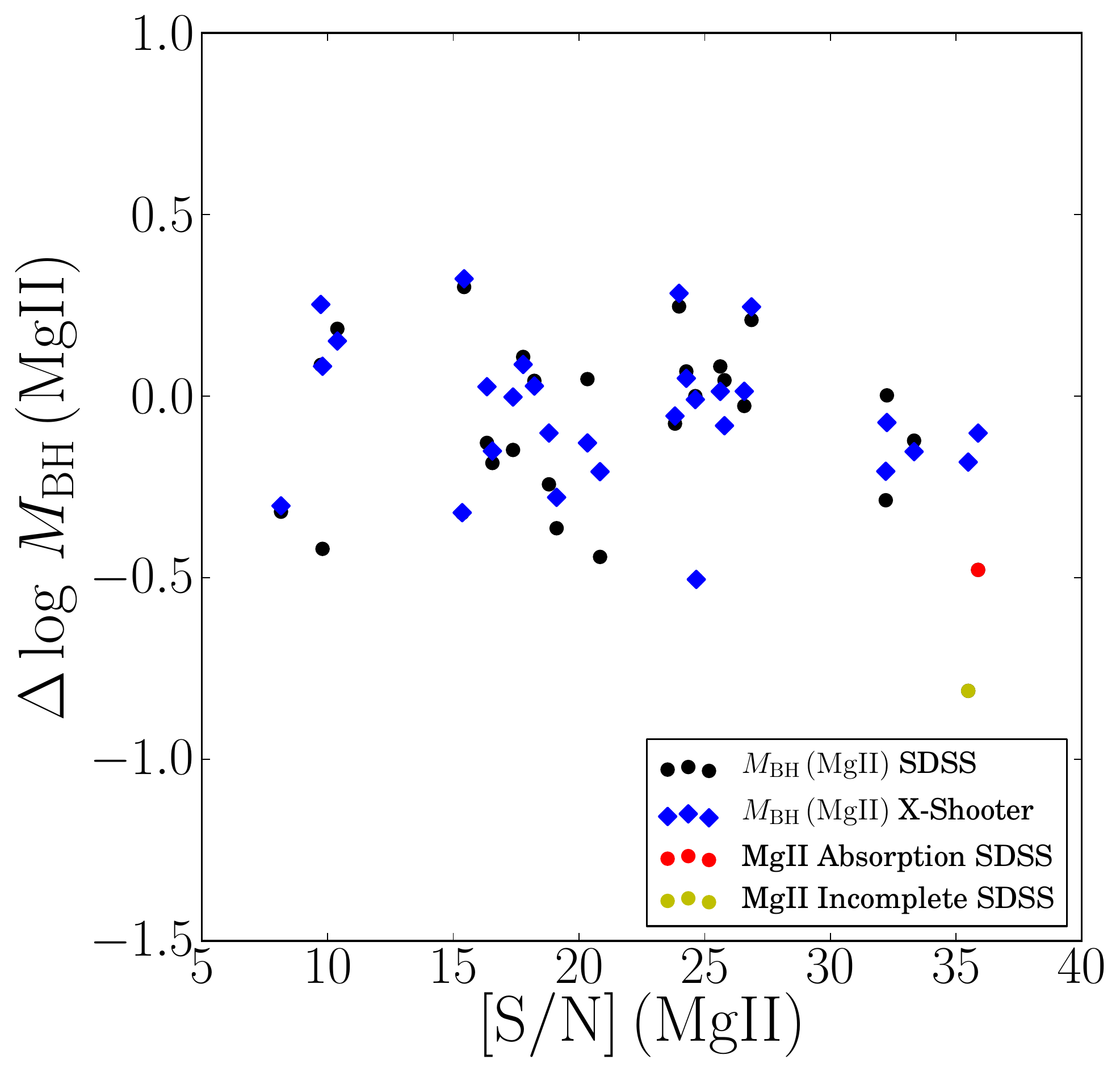}
\includegraphics[width=0.49\textwidth,angle=0]{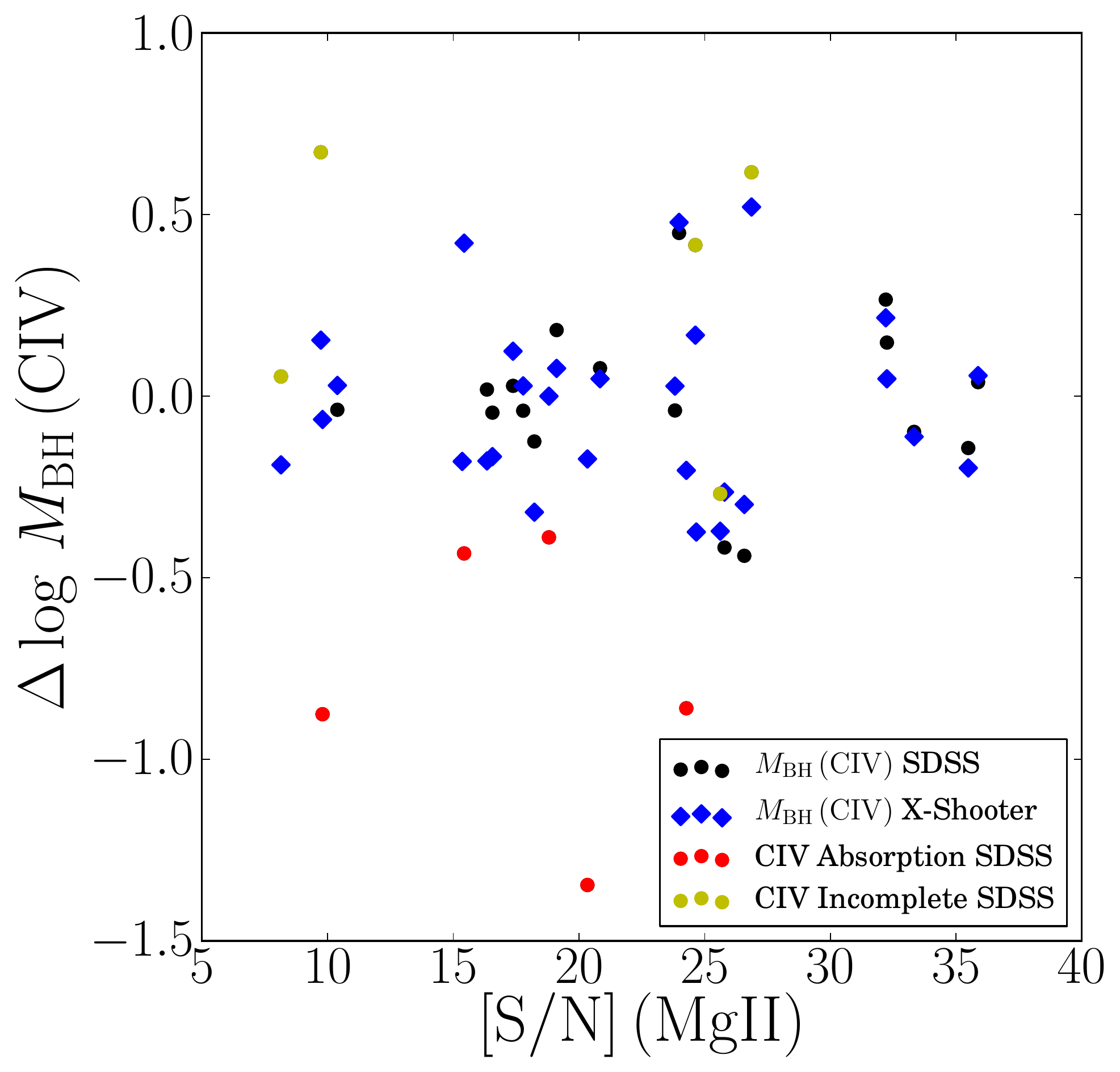} 
\caption{A comparison of \Mbh\ estimates from X-Shooter and SDSS spectra. We show the offsets in mass estimates, $\Delta\,\log\,\Mbh \equiv  \log\left( \Mbh\left(\text{line}\right)/\MbhHa\right)$, vs.  the S/N of the continuum around \mgii\ ($\left[\rm{S/N}\right]\left(\mgii\right)$) using SDSS (black dots) and X-Shooter (blue diamonds) data for the \mgii\ (left panel)  and \civ\ (right panel) lines. 
SDSS data with unresolved absorption features (red dots) or incomplete line profiles (yellow dots) are also shown.}
\label{fig:MBHsdss}
\end{figure*}

\subsubsection{Other lines}
\label{subsubsec:otherlines}

In order to calibrate \Halpha, \mgii\ and \civ\ line measurements to match the \MbhHb\ predictions  we follow standard procedures \citep[e.g.][]{MclureDunlop2004,VestergaardPeterson2006,TrakhtenbrotNetzer2012} that basically  rescale  $\RBLR \left(\Lop \right)$ to $\RBLR \left( L_{\lambda} \right)$   (see Eqn. \ref{eqn:RL})  and then rescale $\mu \left(\lambda \right)$ to $\Mbh \left(\Hbeta \right)$  where 
$ \mu(\lambda)=G^{-1} \RBLR \left(L_{\lambda} \right)  \fwhm\left(\text{line}\right)^{2}$. 

This approach  assumes that \Mbh\ scales as $ \fwhm^{2}$,  which follows from a virialization of  the line emitting  region. 
According to the direct proportionality that we found  between \fwhb, \fwha\ and  \fwmg\ (see \ref{subsec:fwhm}), it will be enough to assume virialization of the \Hbeta\ emitting region. 
We note that several  previous studies have instead allowed total freedom to the dependence of  \Mbh\ on  \fwhm\ \citep[e.g.,][]{ShenLiu2012}, instead of assuming a virial relation. 
However, there is no physical motivation for this approach (except perhaps for \civ) apart from the attempt to minimize residuals with regard to \MbhHb. 
We focus on identifying those sources which appear to represent  the largest deviation from virial equilibrium, and excluding them from the analysis. As explained in \S \ref{subsec:broadmg}, these are the five sources with the largest deviations between \fwhb\ and \fwmg, that are mostly small width ($\fwhb<4000\,\kms$), high \LLedd\ ($\gtrsim0.17$) sources , and the two BALQSOs. In such cases \MbhHa\ and \MbhHb\ are the only methods providing reliable \Mbh\ determination.

The results of the rescaled  single epoch \Mbh\ estimators  based on \Halpha, \Hbeta, \mgii\ and \civ\ in 32/39 sources are summarized in Table\, \ref{tab:mass} and shown in Figure  \ref{fig:mass} where the black solid lines represent the 1:1 relations. We also show the 7 removed sources; BALQSOs in green and objects with discrepant \fwhb\ and \fwmg\ in magenta.
 
 Figure \ref{fig:mass} shows that the main sources of scatter in all the \Mbh\ relationships in the original sample are the above 7 sources.  Removing these objects leaves almost perfect correlations ($r_{\rm{s}}>0.85$, $P<10^{-12}$) between mass estimates based on \Halpha, \Hbeta\ and \mgii\ and even \civ. In fact, the scatter in \MbhHa-\MbhMg\ and  \MbhHa-\MbhC\   is reduced from 0.23 to 0.15 dex  and from 0.29 to 0.16  respectively for the \Lsix-\fwha\ estimates. Unfortunately, it is not easy to identify and remove such objects from a sample where only the \civ\ line region is observable. We come back to this issue later in the paper.

The use of \Lha\ in \citet{Xiao2011}, as well as other studies \citep[e.g.,][]{GreeneHo2005}, is motivated by the possibility of host-light contribution to \Lsix, especially in low luminosity (low-redshift) AGN. 
However, as previously mentioned (\S \ref{sec:fit}), most of our objects have negligible host galaxy contamination,  and we have accounted for it in the few objects where it is relevant. Thus,  we can safely use \Lsix\ for \Halpha-based \Mbh\ estimates. In table \ref{tab:mass} we present both \Lsix-\fwha\  and \Lha-\fwha\ \Mbh\ calibrations.

{  In Figure \ref{fig:mass} we also present the best-fit relations that compare our new mass prescriptions with previously published ones (black dashed lines). 
Particularly we compared our new calibrations with the TN12 \mgii-based calibration, the \citet{Xiao2011} \Halpha-based calibration (an updated version of \citet{GreeneHo2005}) and the \citet{VestergaardPeterson2006} \civ-based calibration. We note that these are somewhat simplified comparisons, as a proper analysis of the deviation from each \Mbh\ calibration is not straightforward, due to the usage of different f factors; different $\RBLR-L$ relations; assumed cosmology; and even of fitting procedures. 
Nevertheless, it is evident from the diagram that the deviation from the earlier mass estimates based on \civ\ are the largest among the three (bottom panel of Fig. \ref{fig:mass}).
}

\subsubsection{X-Shooter versus SDSS \Mbh\ estimates}

{
In Fig. \ref{fig:MBHsdss} we compare the \mbh\ estimations using  the (lower-S/N) SDSS spectra and (higher S/N)  X-Shooter spectra, by plotting $\Delta\,\log\,\Mbh \equiv  \log\left( \Mbh\left(\text{line} \right)/\MbhHa\right)$ for \mgii\ and \civ\ versus the S/N of the continuum around \mgii ($\left[\rm{S/N}\right]\left(\mgii\right)$). We  note that the typical difference between the data sets is $\left(S/N\right)_{XSh}\simeq 4\times\left(S/N\right)_{\rm SDSS}$.  }
As expected (see \S \ref{subsec:fwhm}), objects with unresolved absorption features or incomplete line profiles generally show the largest  offsets in mass. {  Apart from these objects,  the scatter in \MbhC\ and \MbhMg\ estimates is \emph{independent} of the S/N. 
This is not surprising because of the good agreement between X-Shooter- and SDSS-based \fwciv\ measurements (see \S \ref{subsec:civabs}).
We conclude that the scatter in \mgii- and \civ-based mass estimates is dominated by intrinsic differences between \fwmg-\fwciv\ and \fwha\ as well as between \Lop-\Lthree\  and \Lsix.}%by the physics of the BLR and variability effects rather than the uncertainties due to S/N of the data.} 

\begin{table*}
\centering
\begin{tabular}{lllllllllll}
\hline
               & \multicolumn{5}{c}{------$\log\left(L_{\text{P}}\left(\SiOIV\right)/L_{\text{P}}\left(\civ\right)\right)$------} & \multicolumn{5}{c}{------$\log\left(L_{\text{P}}\left(\CIII\right)/L_{\text{P}}\left(\civ\right))\right)$------} \\
               & $r_{\rm s}$        & P       & scatter(dex)       & $\beta$       & C           & $r_{\rm s}$       & P(\%)       & scatter (dex)       & $\beta$       & C          \\ \hline
$\log\left(\fwciv/\fwha\right)$ & 0.36               & 0.02           & 0.35               & 0.76          & -0.51       & 0.34              & 3           & 0.34                & 0.72          & -0.55      \\
$\log\left(\fwciv/\fwhb\right)$ & 0.44               & 0.003         & 0.32               & 0.55          & -0.31       & 0.47              & 0.2         & 0.30                & 0.57          & -0.33      \\
$\log\left(\fwciv/\fwmg\right)$ & 0.51               & 0.003         & 0.28               & 0.69          & -0.72       & 0.57              & 0.02        & 0.19                & 0.52          & -0.52 \\ \hline
$\log\left(\fwciv/\fwhb\right)$ from R13  & 0.64               & $3\times10^{-9}$         & 0.26               &  0.57        & -0.36       & - - -              & - - -         & - - -                & - - -          & - - -       \\ \hline
\end{tabular}
\caption{Spearman correlation coefficients, probability, scatter, and the best fit parameters ($\log \fwhm_{ratio} = \beta\,\log L_{P}^{ratio}+C$) between the listed quantities.}
\label{tab:rehabciv}
\end{table*}

\begin{figure*}
\centering
\includegraphics[width=0.95\textwidth,angle=0]{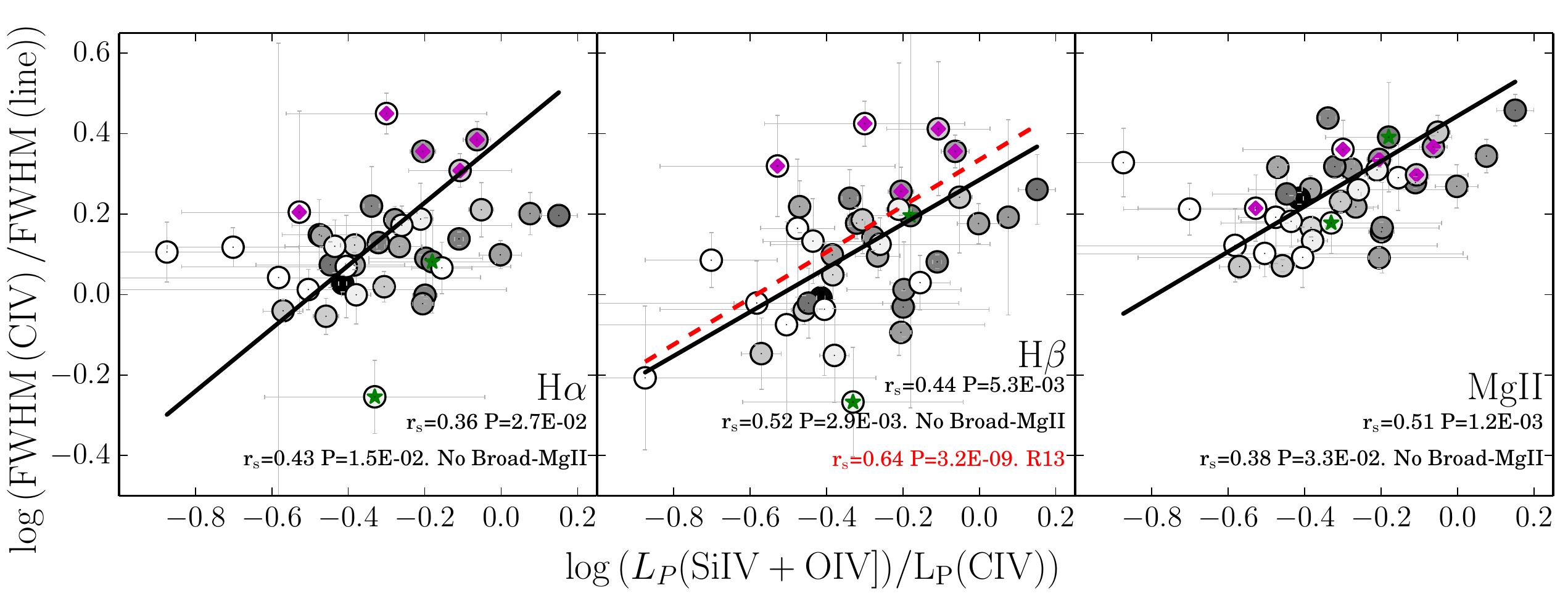}
\includegraphics[width=0.95\textwidth,angle=0]{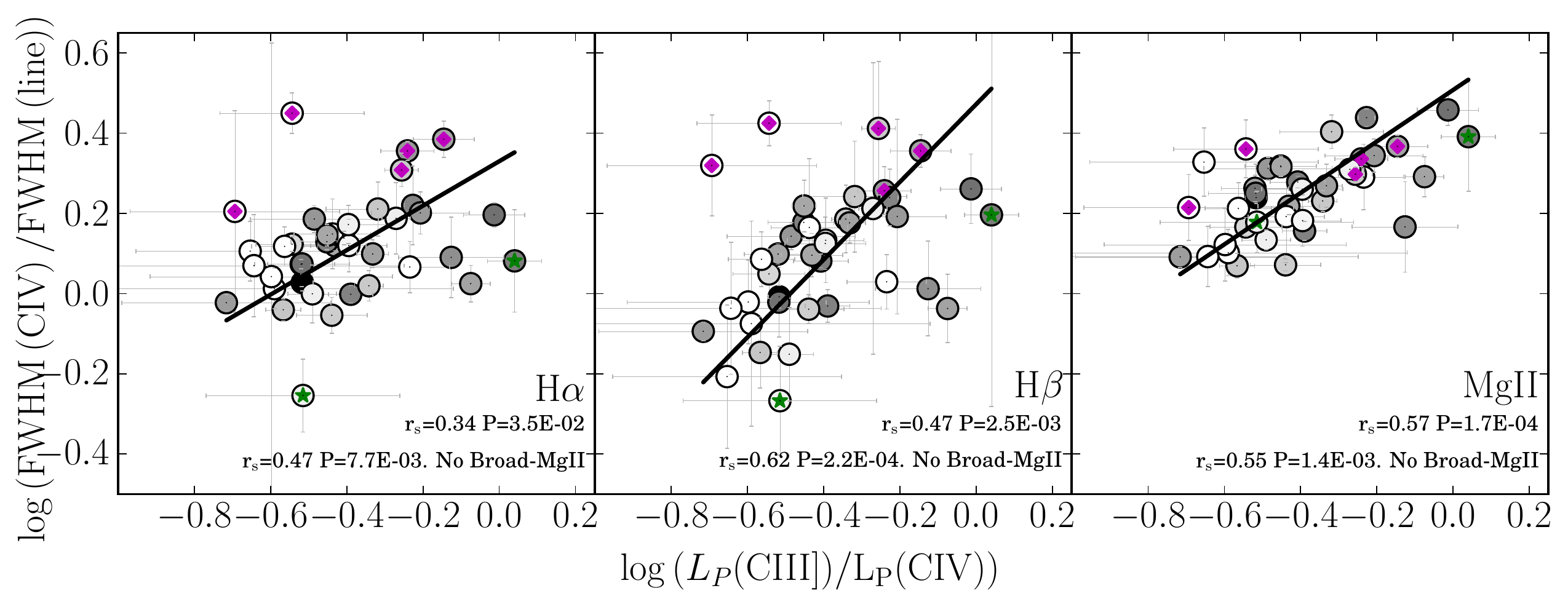}
\includegraphics[width=0.95\textwidth,angle=0]{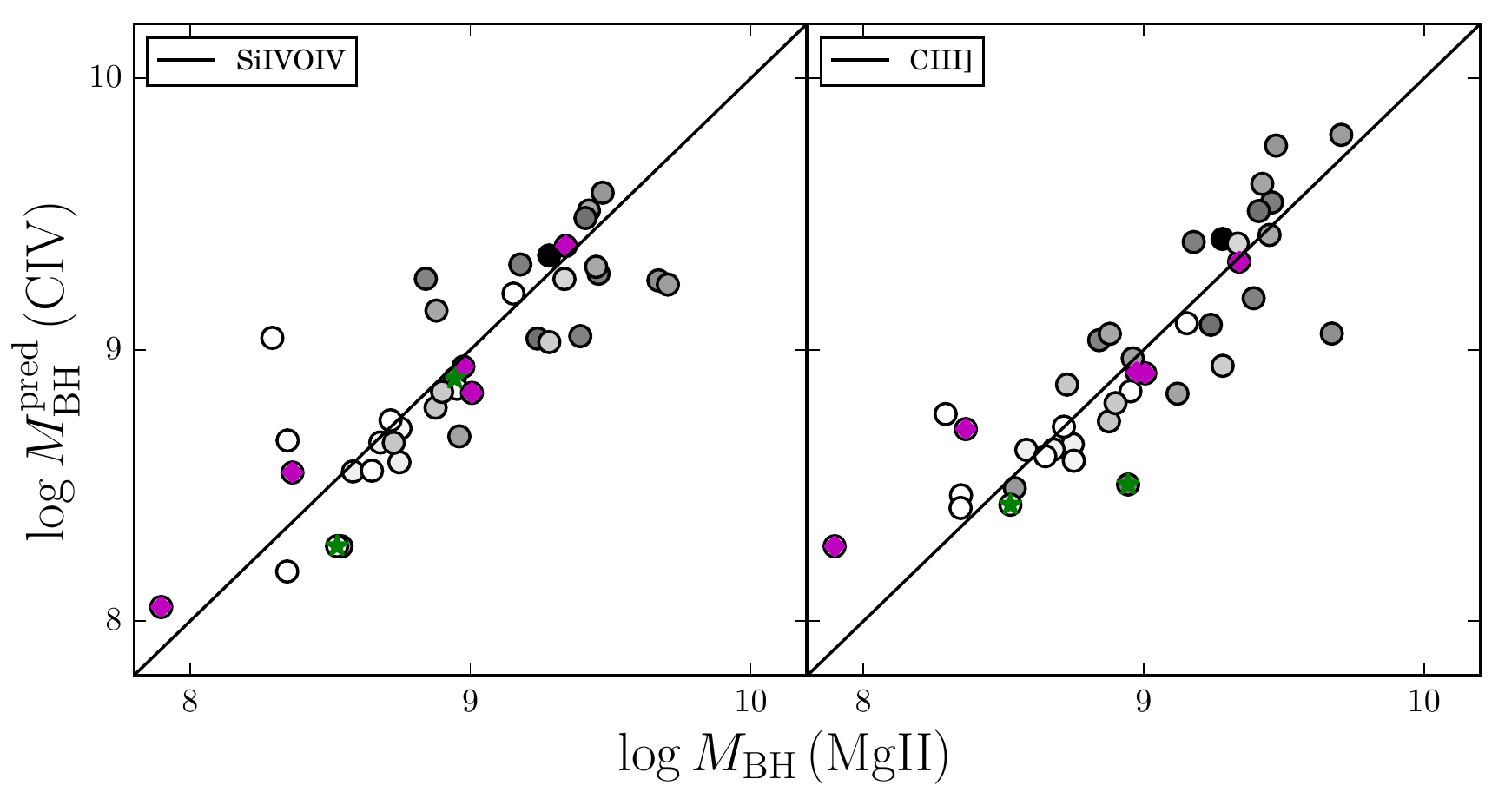}
\caption{ Top panel:  Comparison of the  \Halpha-\civ\ (left), \Hbeta-\civ\ (middle) and \mgii-\civ\ (right) \fwhm\ ratios with the \sioiv - \civ\ 
line peak ratio.The Red dashed line represents the best-fit relation reported by \citet{Runnoe2013} and the black solid lines represent our best fit relation. Middle panel: same as top panel but this time we compare with the \ciii - \civ\ line peak ratio. Bottom panel: Predicted \Mbh\ masses using the correlations of  the  \mgii-\civ\  \fwhm\ ratio  with the \sioiv - \civ\ (bottom-left) and the \ciii - \civ\ (bottom-right) line peak ratios. Green stars represent Broad absorption lines quasars (BALQSO) and magenta points represent the broad-\mgii\ objects. The black solid lines represent the 1:1 relation.}
\label{fig:Rehab_CIV}
\end{figure*}
  
\subsection{The \civ\ line as a Black Hole mass estimator}
\label{subsec:civ}

As can be seen in Figure \ref{fig:FWHMlocal}  and also  mentioned in \S\S\ref{subsec:fwhm}, the width of \civ\ shows only weak correlations (if at all) with the widths of the other lines  we study in this paper. 
This result  together with the significant blue-shifts observed in the \civ\ line center ($\Delta v = -1200 \pm 1000$) {  make mass estimates based on the CIV line significantly more uncertain.} 
However, In high-z objects ($2 \lesssim z \lesssim 5$) \civ\ is the only prominent broad emission line that lies within the optical window. It is therefore important to explore possibilities to improve \Mbh\ determination by means of \civ. There have been already some attempts in this direction.  For instance,  \citet{Runnoe2013}  and  \citet{Brotherton2015} claim a correlation between the line peak ratio $L_{P}\left(\sioiv\right)/L_{P}\left(\civ\right)$ and the \fwhm\ ratio $\fwciv/\fwhb$ driven by Eigenvector 1 \citep{BorosonGreen1992} that would help to reduce the scatter in \Mbh\ from 0.43 dex to 0.30 dex, \citet{Denney2013} propose that having high quality spectra and using the velocity dispersion  of the line (\sigline ), instead of \fwhm, will lead to accurate \Mbh\ estimations. However,   \citet{Denney2013} sample is limited to only 6 objects and our larger, high quality sample does not show  any correlation between  $\sigline\left(\Hbeta\right)$ and $\sigline\left(\civ\right)$. 

In the following section we test the  \citet{Runnoe2013} suggested relation as well as other relationships that can be used to improve the \civ-based mass determination method.

\subsubsection{Rehabilitating \civ?}
\label{subsec:Rehab_CIV}

In Table \ref{tab:rehabciv}  we show the correlation coefficient, correlation probability and scatter between $L_{\text{P}}\left(\sioiv\right)/L_{\text{P}}\left(\civ\right)$ and   $\fwciv/\fwhb$ as well as several other similar line peak and \fwhm\ ratios that are listed in the table. In Figure \ref{fig:Rehab_CIV} we compare such quantities.

As can be seen in table \ref{tab:rehabciv} and Fig. \ref{fig:Rehab_CIV}, we confirm the correlation reported by \citet{Runnoe2013}, however with a lower level of significance and larger scatter. 
{  These differences may be attributed to the the smaller size of our sample (39 objects here vs.\ 85 in R13), and the somewhat lower S/N in the \Hbeta\ region for the fainter sources in our sample, compared with R13. 
We can also see in Table \ref{tab:rehabciv} and Fig. \ref{fig:Rehab_CIV}  that our best fit relation between $L_{\text{P}}\left(\sioiv\right)/L_{\text{P}}\left(\civ\right)$ and   $\fwciv/\fwhb$ (black solid line in top-middle panel) is in very good agreement with the one presented in R13 (red dashed line in top-middle panel).}

We also find that $L_{\text{P}}\left(\ciii\right)/L_{\text{P}}\left(\civ\right)$  correlations are slightly stronger than 
 the analogous $L_{\text{P}}\left(\sioiv\right)/L_{\text{P}}\left(\civ\right)$ correlations. At the same time the strongest correlations are those involving these line peak ratios and $\fwciv/\fwmg$.  These relationships can be used to derive ``corrected''\ \Mbh\ estimates in cases where the relevant line peak ratios can be observed.
 
Below we present  the corrected \Mbh\ that can be derived from  \civ\ and \sioiv\ measurements:

{\small
\begin{eqnarray}\nonumber
\Mbh\left(\mgii\right)_{\text{pred}} &=& 1.13\times 10^{6}\left( \frac{L_{1450}}{10^{44}} \right)^{0.57} \times \left( \frac{\fwciv}{10^{3}\,\kms}\right)^{2} \\
&& \times \left(\frac{L_{P}\left(\sioiv\right)}{L_{P}(\civ)}\right)^{-1.66}
\label{eqn:mhbsioiv}
\end{eqnarray}
}
and from \civ\ and \ciii\ measurements:
{\small
\begin{eqnarray}\nonumber
\Mbh\left(\mgii\right)_{\text{pred}}  &=& 5.71\times 10^{5}\, \left( \frac{L_{1450}}{10^{44}\,\ergs} \right)^{0.57}\times \left( \frac{\fwciv}{10^{3}\,\kms}\right)^{2} \\
&& \times \left(\frac{L_{\text{P} }\left(\ciii\right)}{L_{\text{P}}\left(\civ\right)}\right)^{-2.09}\,\,\, .
\label{eqn:mhbciii}
\end{eqnarray}
}

{  
The confirmation of the \citet{Runnoe2013} correlation, and the new correlations reported here, should assist in rehabilitating \civ\ for more reliable \Mbh\  measurements, by relying on the nearby \sioiv\ and/or \ciii\ emission lines. 
Even for those combinations of observables which do not significantly reduce the scatter in \Mbh\ determinations, they provide an improvement in the \emph{accuracy} of rest-frame UV-based \Mbh\ estimations since these prescriptions compensates the effect of \LLedd\ in the \civ\ profile. }

\section{Summary and Conclusions}
\label{sec:conclusions} 

This paper uses a unique sample of 39 type-I AGN observed by X-Shooter and covering, uniformly, the $\Mbh-\LLedd$ plane at $z=1.55$ down to $i_{\rm AB}\sim21$ mag. 
Our sample allows for a comprehensive comparison between different luminosity probes and emission line measurements, for the prominent broad emission lines \Halpha, \Hbeta, \mgii\ and \civ, which are commonly used for virial BH mass estimates. 
Thanks to the broad spectral coverage we were also able to test two approaches for continuum fitting and test for possible biases in \Mbh\ determinations: a physically-motivated approach based on fitting an accretion disc model to each spectrum; and a more practical approach which treats the continuum around each prominent line as an independent power-law.

In summary, the main findings of this work are:

\begin{enumerate}

\item
Comparing the two continuum fitting approaches, we  find only small (although systemic) offsets in the derived line luminosities, local continua luminosities, and line FWHMs, and consequently in \Mbh\ determinations ($<0.05$ dex).  This implies that a precise modeling of the continuum emission is \emph{not} crucial for \Mbh\ determinations.

\item
Line dispersion measurements (\sigline)  are highly sensitive to continuum modeling, and cannot be safely used for \Mbh\ determination, even for the well-studied Balmer lines and/or when high-quality spectra of broad UV lines are available.

\item  We corroborate that both the  \Halpha\ and \Hbeta\  lines  show very similar FWHMs and  can  be consistently used for estimating \Mbh\ based on the virial assumption. 

\item The \mgii\ line width is found to follow that of \hb, and, generally, can  be safely  used for \Mbh\ estimations. Our new observations show that the MgII line is about 30$\pm$15\% narrower than \Hbeta (in FWHM). We also found that about 10\% of the objects show atypically \emph{broad} \mgii\ lines, with $\fwmg\gtrsim \fwha$. These \mgii\ profiles are also systematically blue-shifted, probably due to non-virial dynamics, and further shown to be not suitable for reliable  \Mbh\ estimation (see \S \ref{subsec:broadmg} ).  We note that broad-\mgii\ objects can only be identified using additional information  from one of the Balmer lines, which would in turn eliminate the necessity to identify them. Without any additional information, such sources may be present in any sample of AGN. 

\item We find that \fwhm\ measurements for \civ\ in low-S/N spectra are systematically underestimated, for objects with partially resolved or unresolved \civ\ absorption features. We also find  and that the FWHMs  of \mgii and the FWHMs of non-absorbed-\civ-profiles  are consistent in low- and high-S/N data sets.  On the other hand, the line dispersion measurements (\sigline) for both \civ\ and \mgii\ profiles differ significantly (a scatter of $\sim0.2$ dex).

\item We find  better agreement and  lower dispersion between \Lsix\ and \Lop\ than between \Lha\ and \Lop, especially for high luminosity objects ($\Lop>10^{45}\ergs$), and recommend to use the \Lsix-\fwha\  black hole mass calibration (Table~\ref{tab:mass}) for objects with an AGN-dominated continuum in this luminosity range.

\item The considerable uncertainties associated with \civ-based determination of \Mbh\  are not solely due to insufficient spectral resolution and/or S/N. They are more likely related to the physics of the BLR. Our results are in agreement with some earlier findings about the systematic uncertainties associated to \civ. We found that the \LLedd\ is strongly correlated with $\fwciv/\fwha$ and with the velocity offset of the \civ\ line. We stress, however, that these correlations show large scatter and cannot practically assist in improving \MbhC\ estimates.

\item {  We confirm the result of \citet{Runnoe2013}, finding a significant correlation between the  \sioiv/\civ\ line peak ratio and \fwciv/\fwhb, which may in principle assist rehabilitating \civ-based \Mbh\ determinations. Moreover, we find even stronger correlations associated with the \ciii/\civ\ line peak ratio. Although these empirical correlations do \emph{not} significantly reduce the scatter in \MbhC\ estimates, we propose that their application, whenever possible, would improve the accuracy of \civ-based \Mbh\ determinations.}

\item \LLedd\ seems to  affect the dynamics of the \mgii-emitting region, especially in objects with extreme accretion rates (as pointed out by  \citet{Marziani2013}).  

\item We provide new single epoch calibrations for \Mbh, based on the \fwhm\ of \Halpha, \Hbeta, \mgii\ and \civ.  

\item We constructed a new (UV) iron template that aims to improve on previous templates \citep{VestergaardWilkes2001,Tsuzuki2006}, particularly in the region of  $\sim2200-3650$ \AA.
\end{enumerate}

\section*{Acknowledgments} 
{  
We thank the anonymous referee for the detailed and constructive feedback, which helped us in improving the paper.
}
JM acknowledges ``CONICYT-PCHA/doctorado Nacional para extranjeros/2013-63130316'' for their PhD scholarship support , Fondecyt Project \#1120328  for their support in computing facilities and travel to Tel Aviv University, where an important part of the work  was done. 
JM is also grateful for the hospitality and support of Tel Aviv university during two visits. 
Funding for this work has been provided by the Israel Science Foundation grant 284/13.\

%\begin{thebibliography}{}
%   \input{mass_paper.bbl}
%\end{thebibliography}
\bibliographystyle{mn2e}

\appendix

{
\section{Demonstrating the quality of X-Shooter spectra} 
\label{app:xs_sdss_example}

\begin{figure}
\centering
\includegraphics[scale=0.46]{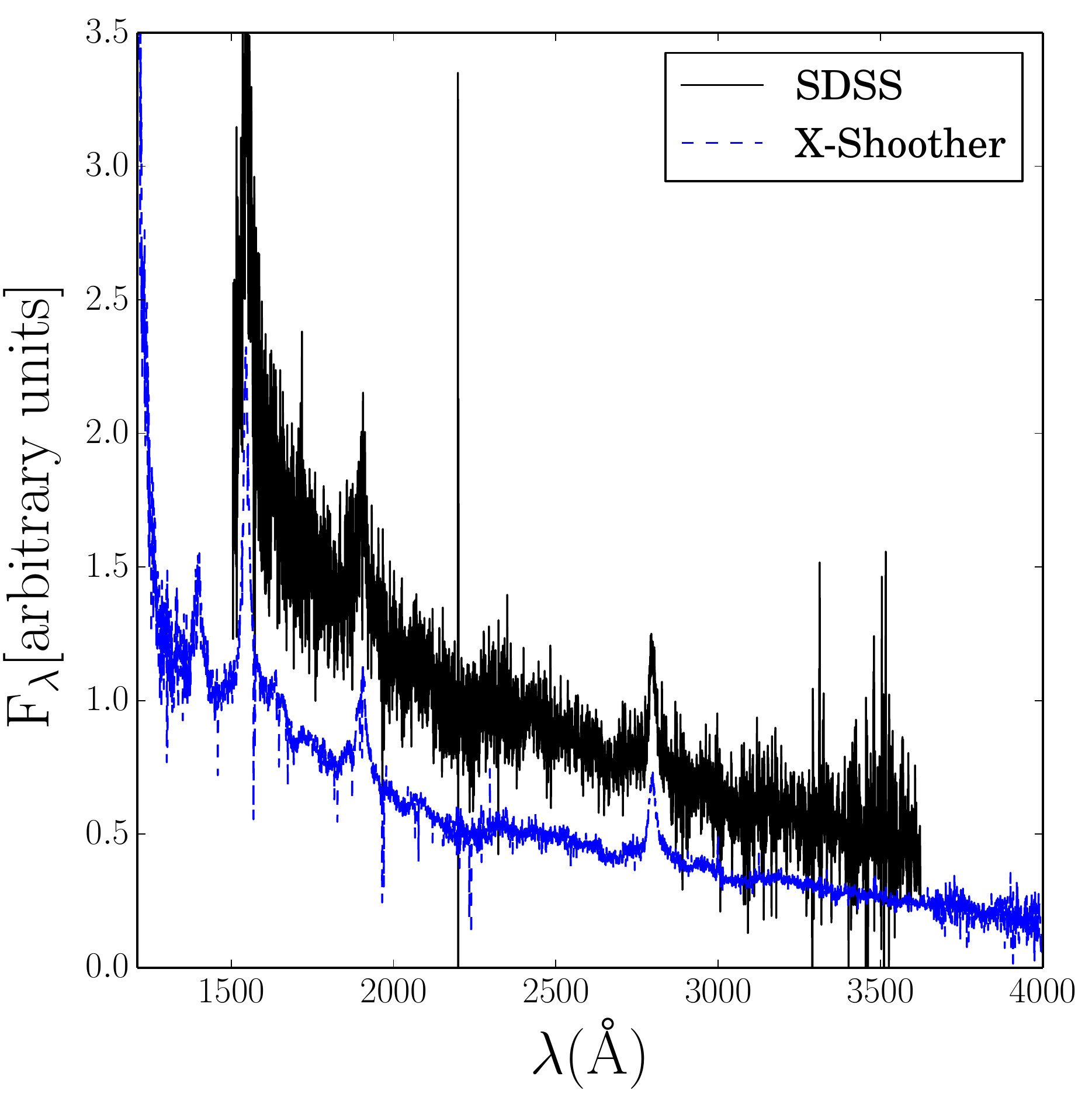}
\caption{{  SDSS  and X-Shooter spectra of J0143-0056. Both spectra have been rescaled to avoid overlapping.}}
\label{fig:XshvsSDSS}
\end{figure} 

Figure~\ref{fig:XshvsSDSS} compares the newly obtained X-Shooter spectrum  (UVB+VIS arms) to the publicly available SDSS spectrum, for J0143--0056 - the source shown in Figs.~\ref{fig:global} and \ref{fig:local}. 
This source has a $S/N\simeq25$ at 2000\AA\ which lies in the middle of the $S/N$ range for the entire sample.
Both spectra are presented \emph{without} any binning or smoothing, including the residual sky and/or instrumental features.
We note the significant improvements to $S/N$ and spectral resolution, as evident from the minor absorption feature on the blue wing of the \CIII\ line. 
The broader spectral coverage allows for a much more robust determination of the continuum level next to the \civ\ and \mgii\ emission lines (i.e., $L_{1450}$ and $L_{3000}$. 
Obviously, the NIR arm of X-Shooter includes the \hb\ and \ha\ spectral regions (not shown here), which are unavailable in the SDSS data.

}

\section{New UV iron emission template} 
\label{app:fe_template}

\begin{figure}
\centering
%\hspace{-6mm}
\includegraphics[trim=2cm 0cm 3cm 2cm, width=0.49  \textwidth,angle=0]{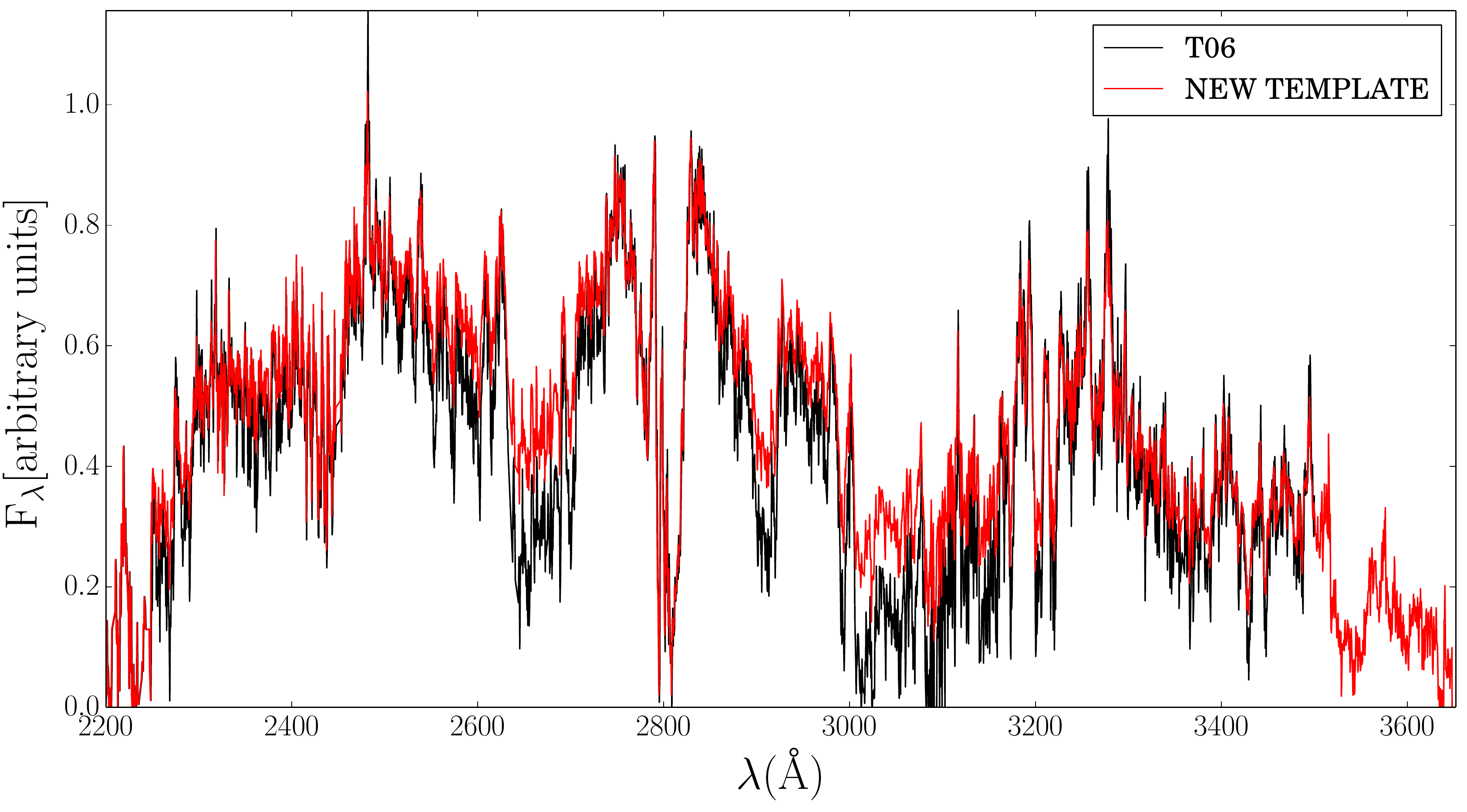}
\caption{ Comparison of our new template (red) 
and \citet{Tsuzuki2006} template (black).}
\label{fig:feIItemplate}
\end{figure} 

In  figure \ref{fig:feIItemplate}  we compare our new UV iron template with the template of T06. The new template, covering 2200-3646\AA\ and with an intrinsic width of 900 \kms, can be downloaded from \url{http://www.das.uchile.cl/~jemejia/feII_UV_Mejia-Restrepo_et_al_2015\_2200-3646AA.data}

We prefer the use of our new template motivated by the following three reasons:

\begin{itemize}

\item The T06 template  severely underestimates the continuum emission around 2100\AA.

\item T06 modeled the BC continuum as a modified Black Body  following \citep{Grandi1982}. This does not provide a good approximation to Balmer emission and we prefer templates based on photo-ionization calculations.

\item  The T06 template only extends between 2200\AA\ and  3500\AA. However,  there is still a remaining weaker but still non-negligible contribution from iron emission up to the Balmer limit (3647\AA). The correct estimation of iron emission in this regions (3500\AA\ to 3647\AA) is crucial for estimating the emission by iron lines and to prevent overestimation of the BC.

\end{itemize}

We constructed the template  following T06 and VW01 procedures and using our own estimations of the accretion-disk emission and Balmer continua. We redefined the accretion-disk-continuum by  manually selecting the continuum windows   at  $\sim$2100\AA\ and $\sim$4200\AA\ which account for the region where we require to obtain the new iron template (2100-3647\AA) . The Balmer continuum model that we use  is described in section \ref{sec:fit}. 

Our template provides stronger iron emission, particularly in the range of 2620-3500 \AA, which is crucial for \mgii\ measurements. 
This could be explained by  our different Balmer continuum approach and disk continuum windows.

\begin{figure*}
\centering
%\hspace{-6mm}
\includegraphics[trim=0.5cm 0.5cm 0.5cm 0.7cm,width=22cm,height=17cm,angle=90]{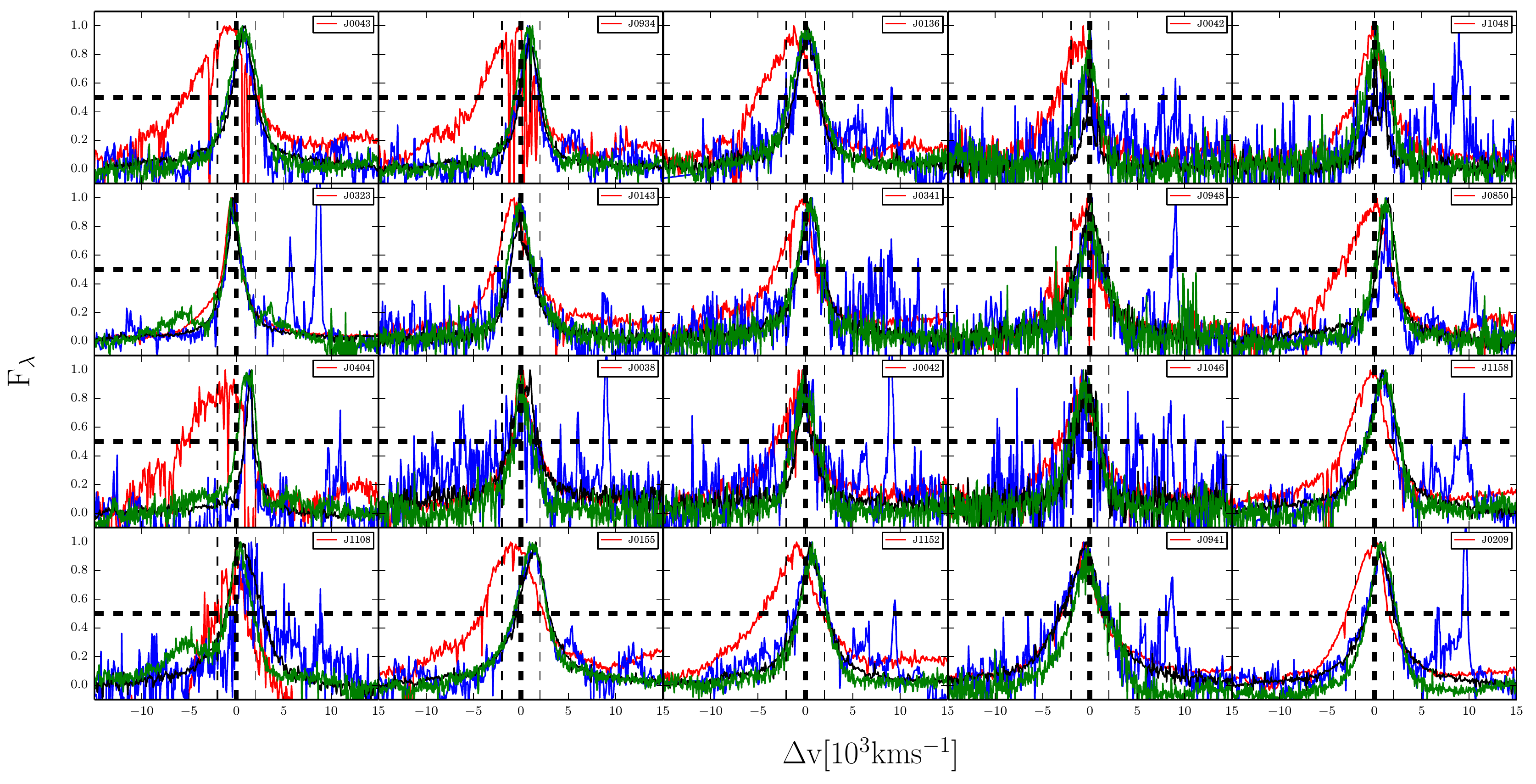}
\caption{Comparison of  the observed \Halpha\ (black), \Hbeta\ (blue), \mgii\ (green) and \civ\ (red) line profiles in the velocity space for the objects in the sample with satisfactory Thin disk continuum fits. 
All profiles have been normalized relative to the peak flux density of the line. It is important to remark that both the \mgii\ and \civ\ profiles are doublets and their decomposed profiles are narrower than shown here. In the top row we show the five broad-\mgii\ objects.
}
\label{fig:app:profile_comp1}
\end{figure*}

\begin{figure*}
\centering
%\hspace{-6mm}
\includegraphics[trim=0.5cm 0.5cm 0.5cm 0.7cm,width=22cm,height=17cm,angle=90]{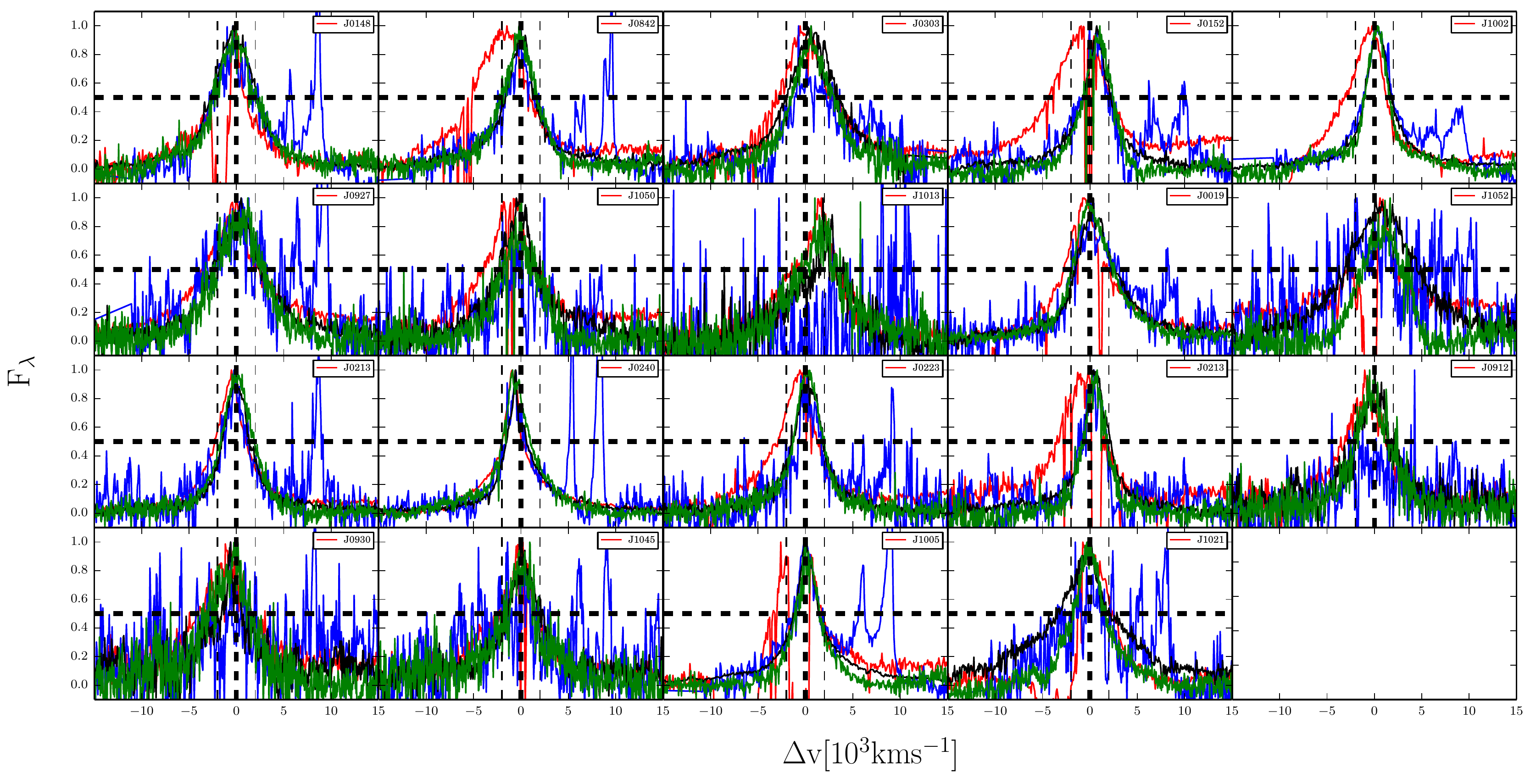}
\caption{
- continued. Comparison of  the observed \Halpha\ (black), \Hbeta\ (blue), \mgii\ (green) and \civ\ (red) line profiles in the velocity space for the objects in the sample with satisfactory Thin disk continuum fits. 
All profiles have been normalized relative to the peak flux density of the line. It is important to remark that both the \mgii\ and \civ\ profiles are doublets and their decomposed profiles are narrower than shown here. To the left of the bottom row we show the two BALQSO objects.
}
\label{fig:app:profile_comp2}
\end{figure*}

\section{Comparison of Observed Emission Line Profiles} 
\label{app:profile_comp}

In Figures \ref{fig:app:profile_comp1} and \ref{fig:app:profile_comp2} we show the normalized  profiles of the \Halpha, \Hbeta, \mgii\ and \civ\  emission lines, in velocity space. 
In most but not all sources, the  \civ\ profiles (red) are  broader and blue shifted with respect to the \Halpha\ and \Hbeta\ line profiles as discussed in \S\ref{subsec:fwhm}. 
The low ionization lines, \Halpha, \Halpha\ and \mgii, show  similar shape profiles.  \Hbeta\ is generally slightly broader than \Halpha. \mgii\ is, on  average, 30\% narrower than \Hbeta.  The five broad-\mgii\ objects (top-row) show \mgii\ that are broader than \Halpha\ and \Hbeta. These \mgii\ profiles are also slightly blue-shifted (about 300 \kms) relative to the \Hbeta\ line. The two BALQSOs are  the last two objects of the bottom row on the second set.

\section{Emission line constraints}
\label{app:fitting_params}

Table \ref{tab:line_const} lists the constraints on the emission line modeling for each of the components in our fitting procedure following \citet{Shang2007} and \citet{VandenBerk2004}.

\begin{table*}
\begin{tabular}[t]{  l  l l  l  l l l  l  }
\hline
	ID & LINE & $\lambda$ & GAUSSIAN COMPONENT & Flux & Center & FWHM & FLUX RATIO \\ \hline
	Si\,{\sevensize IV} + O\,{\sevensize IV}] Region &  &  &  &  &  &  &  \\ \hline
    1 & Si\,{\sevensize IV} & 1396.75 & Broad & Free & Free & Free & Free \\ 
	2 &  &  & Narrow & Free & 1 & Free & Free \\ 
	3 & O\,{\sevensize IV}] & 1402.34 & Broad & Free & 1 & Free & Free \\ 
	4 &  &  & Narrow & Free & 2 & Free & Free \\ \hline
	C\,{\sevensize IV}  Region&  &  &  &  &  &  &  \\ \hline
	1 & N\,{\sevensize IV}] & 1486.5 &  & Free & Free & Free &  \\ 
	2 & C\,{\sevensize IV} & 1548.2 & Narrow & Free & Free & Free & Free \\ 
	3 &  &  & Broad & Free & Free & Free & Free \\ 
	4 & C\,{\sevensize IV} & 1550.77 & Narrow & Free & 2 & 2 & 1 \\ 
	5 &  &  & Broad & Free & 3 & 3 & 1 \\ 
	6 & He\,{\sevensize II} & 1640.72 & Narrow & Free & Free & Free &  \\ 
	7 &  &  & Broad & Free & 6 & Free &  \\ 
	8 & O\,{\sevensize III}] & 1660.8 &  & Free & 1 & Free & 0.29 \\ 
	9 &  & 1666.14 &  & 8 & 8 & 8 & 0.71 \\ 
    10 & N\,{\sevensize IV} & 1718.75 &  & Free & Free & Free & Free \\ \hline
    C\,{\sevensize III} Region &  &  &  &  &  &  &  \\ \hline
    11 & C\,{\sevensize III}] & 1908.73 & Narrow & Free & Free & Free &  \\ 
	12 &  &  & Broad & Free & 13 & Free &  \\ 
	13 & Si\,{\sevensize III}] & 1892.03 &  & Free & 11 & Free &  \\ 
	14 & Al\,{\sevensize III} & 1854.72 &  & Free & 13 & Free & 1 \\ 
	15 &                     & 1862.78 &  & 14 & 14 & 14 & 1 \\ 
    16 &  Si\,{\sevensize II}  & 1818.17 &  & Free & 11 & Free &  \\ 
    17 &  Fe\,{\sevensize II}  & 1788.73 &  &   16 & 16   & 16   &  \\ 
    18 & N\,{\sevensize III}] & 1748.65 &  & 13 & 13 & 13 & 0.41 \\ 
	19 &  & 1752.16 &  & 18 & 18 & 18 & 0.14 \\ 
	20 &  & 1754.00 &  & 18 & 18 & 18 & 0.45 \\ \hline
	Mg\,{\sevensize II} Region &  &  &  &  &  &  &  \\ \hline
	1 & Mg\,{\sevensize II} & 2795.53 & Narrow & Free & Free & Free & 2 \\ 
	2 &  &  & Broad & Free & 1 & Free & 2 \\ 
	3 & Mg\,{\sevensize II} & 2802.71 & Narrow & 1 & 1 & 1 & 1 \\ 
	4 &  &  & Broad & 2 & 2 & 2 & 1 \\ 
	5 & Fe & Template &  & Free & Free & Free &  \\ \hline 
	\Hbeta\ Region &  &  &  &  &  &  &  \\ \hline
	1 & \Hbeta & 4861.32 & Narrow & Free & Free & Free &  \\ 
	2 &  &  & Broad & Free & Free & Free &  \\ 
	3 &  &  & NLR & Free & 4 & 4 &  \\ 
	4 & [O\,{\sevensize III}] & 5006.84 &  & Free & Free & Free & 3 \\ 
	5 &  & 4958.91 &  & 4 & 4 & 4 & 1 \\ 
	6 & He\,{\sevensize II} & 4685.65 &  & Free & Free & Free &  \\ 
	7 & Fe\,{\sevensize II} & s &  & Free & ... & Free &  \\ \hline
	\Halpha\ Region &  &  &  &  &  &  &  \\ \hline
	1 & \Halpha & 6562.8 & Narrow & Free & Free & Free &  \\ 
	2 &  &  & Broad & Free & Free & Free &  \\ 
	3 &  &  & NLR & Free & Free & 4 &  \\ 
	4 & [N\,{\sevensize II}] & 6548.06 &  & Free & 4 & [O\,{\sevensize III}] width & 1 \\ 
	5 &  & 6583.39 &  & 4 & 4 & 4 & 3 \\ 
	6 & [S\,{\sevensize II}] & 6716.47 &  & Free & 4 & 4 & 1 \\ 
	7 &                       & 6730.85 &  & 6 & 6 & 6 & 1 \\ \hline
\end{tabular}
\caption{Line regions and adopted constraints. Under the \globapp the \civ\ and \ciii\ line regions are fitted simultaneously}
\label{tab:line_const}
\end{table*}
\end{document}